%% file: righigks_rev1_arxiv.tex
\documentclass[%
preprint,%
 amssymb, amsmath,%
 ,aip,
 pof,
numeric,
longbibliography
]{revtex4-1}

\usepackage{docs}%
\usepackage{bm}%
\usepackage{color}
\expandafter\ifx\csname package@font\endcsname\relax\else
 \expandafter\expandafter
 \expandafter\usepackage
 \expandafter\expandafter
 \expandafter{\csname package@font\endcsname}%
\fi
\usepackage{natbib}%
\usepackage{graphicx}%
\usepackage{enumitem}%

\def\bvec{\left\{\begin{array}{c}} 
\def\evec{\end{array}\right\}}

\newcommand\solidrule[1][1cm]{\rule[0.5ex]{#1}{.4pt}}

\newcommand\dotdashedrule{\mbox{%
  \solidrule[4mm]\hspace{0.8mm}\solidrule[0.3mm]\hspace{0.8mm}\solidrule[4mm]}}
\newcommand\verythindashedrule{\mbox{%
  \solidrule[0.65mm]\hspace{0.65mm}\solidrule[0.65mm]\hspace{0.65mm}\solidrule[0.65mm]\hspace{0.65mm}\solidrule[0.65mm]\hspace{0.65mm}\solidrule[0.65mm]\hspace{0.65mm}\solidrule[0.65mm]\hspace{0.65mm}\solidrule[0.65mm]}}
\newcommand\thindashedrule{\mbox{%
  \solidrule[1mm]\hspace{1mm}\solidrule[1mm]\hspace{1mm}\solidrule[1mm]\hspace{1mm}\solidrule[1mm]\hspace{1mm}\solidrule[1mm]}}
\newcommand\fewdots{\mbox{%
   \hspace{0.25mm} \texttt{o} \hspace{0.25mm}}}

\begin{document}

\title{A Gas-Kinetic Scheme For The Simulation Of Compressible Turbulent Flows}%

\author{Marcello Righi}%
\email{marcello.righi@zhaw.ch}
\affiliation{School of Engineering, Zurich University of Applied Sciences, \\ Technikumstrasse 9, 8401 Winterthur, Switzerland}%

\date{February 2013}%

\begin{abstract}

The kinetic theory of gases 
has suggested the idea of viscosity to model the effect of thermal fluctuations on the resolved flow. Supported by the assumed analogy between molecules and the eddies in a turbulent flows, the idea of an eddy viscosity has been put forward in the pioneering work by Lord Kelvin and Osborne Reynolds.   
In over hundred years of turbulence modeling, the numerical schemes adopted to simulate turbulent flow - with the  exception of the Lattice Boltzmann methods - have never 
exploited this analogy in any other way. 
In this work, a gas-kinetic scheme is modified to simulate turbulent flow; the turbulent relaxation time is deduced from assumed turbulent quantities. The new scheme does not adopt an eddy viscosity, yet it relies even more strongly on the analogy between thermal and turbulent fluctuations, as turbulence dynamics is mathematically modeled by the Boltzmann equation.  
In the gas-kinetic scheme, a measure of the degree of rarefaction is introduced, as the ratio between unresolved and resolved time scales of motion. 
At low rarefaction, the turbulent gas-kinetic scheme deviates negligibly from a conventional Navier-Stokes scheme. However, as the degree of rarefaction increases, 
the  kinetic  effects become evident. 
This phenomenon is evident in the mathematical description of the turbulent stress tensor and also in numerical experiments. 
This study does not propose an innovative turbulence model or technique. It addresses the fact that the traditional coupling numerical scheme and turbulence modeling 
might improve the physical consistence of numerical simulations. 
In the proposed gas-kinetic scheme, the turbulent stress tensor is no longer modeled as a self-contained stress tensor, but originates from the complex, yet physically consistent, handling of dissipation from the underlying gas-kinetic theory. 
In order to gain experimental evidence for these arguments, a few flow cases have been selected among those which are particularly challenging for conventional schemes. 
The results of the simulations carried out with the turbulent gas-kinetic scheme are very encouraging.  
\end{abstract}

\maketitle



\section{Introduction}








It is well accepted that describing mathematically the effects of unresolved turbulent fluctuations on the resolved flow may still represent a challenging task, 
despite the progresses achieved by turbulence modeling over the last century and in particular by unsteady methods (Large Eddy Simulation and derivatives) over the last twenty years,  

Turbulent flow is almost invariably simulated on the basis of the Navier-Stokes equations.
Yet most turbulence models adopt the concept of eddy viscosity to generate the turbulent stress tensor, or at least its linear component. The association between Newtonian hydrodynamics and eddy viscosity may have an inherent contradiction: in turbulent flow the degree of rarefaction -- measured as the ratio of unresolved time or space scales to the resolved ones -- may raise well above the normally accepted limits of the continuum regime, 
and therefore of the limits of validity of the Navier-Stokes equations.  
As is well known, these limits are due to the decoupled evaluation of advective and dissipative fluxes. Borrowing {\sl rarefied gas dynamics} terminology, Navier-Stokes fluid mechanics lacks the effects of particle {\sl collisions} on the {\sl transport} properties of the flow.

The relevance of {\sl kinetic effects} issue has been the subject of discussions over the last decade in regard to the simulation of turbulent flow with the Lattice Boltzmann method in a number of papers by Chen and Succi {\sl et al.} \cite{chen2003extended,chen2004expanded,succi2002towards}, in which the merits of the kinetic theory of gases in modeling a distribution of eddies are identified. The motivation behind this paper stems from the fact that gas-kinetic schemes,  a class of numerical methods derived from the kinetic theory, may also be used to model the mechanics of fluids, including turbulence. 

A number of  gas-kinetic schemes were developed over the past  twenty years \cite{xu2001gas,may2007improved,mandal1994kinetic,chou1997kinetic,xu1994numerical}
mainly to improve shock-capturing in laminar flow. 
The scheme developed by Xu in 2001 \cite{xu2001gas} has achieved a significant level of validation; it has been tested in a number of laminar flow cases in the continuum regime, ranging from low-Reynolds subsonic to hypersonics \cite{xu2005multidimensional,li2010high,may2007improved,righi2012aeronautical}. 
This scheme was also the starting platform for the development of schemes for the transitional regime by Liao {\sl et al.} \cite{liao2007gas} and the rarefied regime by Xu {\sl et al.} \cite{xu2010unified}.  
In this work, the basic gas-kinetic scheme has been adopted to simulate turbulent flow.

Gas-kinetic schemes model the non-equilibrium thermal fluctuations in laminar flow as a relaxation process, where the relaxation time is related to viscosity and heath conduction coefficient of the gas.
Non-equilibrium effects in turbulent flow include the interaction between eddies, which unlike molecular dynamics, involve a variety of time and space scales.
These effects are modeled with a turbulent relaxation time, which can be derived from eddy viscosity or independently from turbulent quantities like turbulent kinetic energy and dissipation rate, as is discussed by Chen {\sl et al.} \cite{chen2003extended,chen2004expanded} and Succi {\sl et al.} \cite{succi2002towards} and references therein.

The consideration of kinetic effects is essential in rarefied flow when the degree of rarefaction reaches a given threshold. In turbulent flow, kinetic effects become important when the interactions between eddies  (also) involve large scales of motion; for instance in the presence of an interactions between a shock wave and a turbulent boundary layer; in a shocklayer the mean flow scale of motion is very small and definitely comparable to the scale of turbulent fluctuations.

The analysis of the turbulent gas-kinetic scheme as well as the numerical experiments presented in this paper suggest that the turbulent gas-kinetic scheme deviates from conventional schemes whenever the degree of rarefaction reaches a threshold of indicatively one thousandth. Navier-Stokes schemes correctly identify and capture the shock wave as a discontinuity in the solution but do so on the basis of a numerical and not a physical process. The gas-kinetic scheme may remain physically consistent also in shocklayers.

This paper is structured as follows: the derivation and analysis of Xu's gas-kinetic scheme is presented in section \ref{sec:gks}, the turbulent gas-kinetic scheme is analysed  in section \ref{sec:turbulentgks}, its implementation is described in section \ref{sec:implementation}, the numerical experiments are presented in section \ref{sec:experiments}, conclusions are exposed in section \ref{sec:conclusions}.

\section{Gas-kinetic schemes for compressible flow }
\label{sec:gks}

The kernel of a gas-kinetic scheme consists in modeling the fluxes of the conservative variables across computational cells according to the Boltzmann equation,  instead of relying on the Navier-Stokes equations.
The fluxes are calculated from the moments of a distribution function $f(x,v,t)$ defined in the phase space. 
At each interface, at the beginning of a time step, a function $f$ is introduced as a solution to the Boltzmann equation, with initial conditions consistent with the gas states at both sides of the interface. 
In practical calculations, the collision operator in the Boltzmann equation is linearized, normally consistently with the BGK model \cite{bhatnagar1954model}, which states: 

\begin{equation}
Df = \frac{f^{eq}-f}{\tau}.
\label{eq:bgk}
\end{equation}

In Eq. \ref{eq:bgk} the following notation and symbols are used: 

\begin{itemize}
\item $Df= \partial f / \, \partial t +  u_i \partial f / \, \partial x_i$, where summation convention holds, 
\item $f^{eq}$ is a Maxwellian distribution function, related to a gas in thermodynamic equilibrium: 
\begin{equation}
f^{eq} = \rho \left( \frac{\lambda}{\pi} \right)^{\frac{N+2}{N}} \exp \left[{ -\lambda
\left(
(u_i-v_i)^2+\xi^2 \right)} \right],
\label{eq:maxwellian}
\end{equation}

\noindent where summation convention holds, $\lambda = {m}/\left({2kT}\right) = \rho/2 p$, $m$ is the molecular mass, $k$ is the Boltzmann constant, and $T$ is temperature, $\rho $, $ v $ indicate density, and velocity.  $\xi$ indicates the $N$ effective degrees of freedom of the gas molecules, given by:  $N=\left({5-3\gamma}\right)/\left({\gamma-1}\right)+1$,
where $\gamma$ is the specific heat ratio. The total energy is $E = 1/2 \left( {u_i}^2  +\xi^2 \right)$. 

\item The conservative variables $w = [\rho\,\,\,\rho v_i \,\,\, \rho E]^T $ can be recovered by taking moments of the distribution function: 

\begin{equation}
w = 
\int\psi f d\Xi,
\label{eq:moment1} 
\end{equation}

\noindent where the infinitesimal volume in phase space is $d\Xi = du_1 du_2 du_3 \, d\xi$ (in three dimensions)
and $\psi$ is:

\begin{equation}
{\bf \psi} = 
\left[
1 \,\,\,\,
u_i \,\,\,\,
\frac 12 \left( {u_i}^2 +\xi^2 \right) 
\right]^T.
\end{equation}

\item $\tau$ is a relaxation time which characterize the return to equilibrium and is related to viscosity and heath conduction coefficients in a gas in the continuum regime.  
\end{itemize}

It is well known that the Euler and Navier-Stokes equations can be derived from the Boltzmann equation and also from the BGK model Eq. \ref{eq:bgk}. The complete derivation is beyond the scope of this paper; such a derivation has been proposed by Cercignani \cite{cercignani1988boltzmann} for a monoatomic gas and  by Xu \cite{xu1998gas} for a diatomic gas.
However, to the purpose of the analysis of the gas-kinetic scheme, it is useful to remind that such derivation may be obtained by means of the Chapman-Enskog method \cite{cercignani1988boltzmann}; 
by introducing the non-dimensional quantity $\epsilon = \tau / \widehat{\tau}$, where $\widehat{\tau}$ is a reference time in the flow, Eq. \ref{eq:bgk} is re-expressed in the form: 
 
\begin{equation}
f = f^{eq} - \epsilon \widehat{\tau} D f. 
\label{eq:chapman-enskog0}
\end{equation}

\noindent By substituting Eq. \ref{eq:chapman-enskog0} into the right hand side of the same equation, one obtains: 

\begin{equation}
f = f^{eq} - \epsilon \widehat{\tau} D f^{eq} + \epsilon^2 \widehat{\tau} D \left( \widehat{\tau} D f^{eq} \right) + \dots. 
\label{eq:chapman-enskog}
\end{equation}

\noindent The link to the Euler and Navier-Stokes equations is obtained by taking moments of Eq. \ref{eq:chapman-enskog} one obtains: 

\begin{equation}
\int f d\Xi = \int \left( f^{eq} - \epsilon \widehat{\tau} D f^{eq} + \epsilon^2 \widehat{\tau} D \left( \widehat{\tau} D f^{eq} \right) + \dots \right) d\Xi. 
\label{eq:moments-chapman-enskog}
\end{equation}

The expression in Eq. \ref{eq:chapman-enskog} provides the Euler equations if the terms ${\cal O} (\epsilon)$ are dropped, and the Navier-Stokes equations if the terms 
${\cal O} (\epsilon^2)$ are dropped, as can be found in the complete derivations (\cite{xu1998gas,cercignani1988boltzmann}). 
It is interesting to note that for a diatomic gas the conditions to recover the Navier-Stokes equations are $\mu = \tau/p$ (bulk viscosity is $2N/(3K+9) \mu$) and $Pr= \mu C_p / \kappa = 1$, the latter being  a known drawback of the BGK model. 

Moreover, it is also interesting to point out that the expansion in Eq. \ref{eq:chapman-enskog} truncated at the first order - i.e. neglecting ${\cal O} (\epsilon)$  terms - provides a distribution function at {\sl Euler} level: 

\begin{equation}
f^{Euler} = f^{eq}, 
\label{eq:chapman-enskog-euler}
\end{equation}

\noindent whereas the expansion truncated at the second order - i.e. neglecting ${\cal O} (\epsilon^2)$  terms - provides a distribution function at Navier-Stokes level: 

\begin{equation}
f^{NS}  = f^{eq} - \epsilon \widehat{\tau} D f^{eq}. 
\label{eq:chapman-enskog-ns}
\end{equation}

A closed-form solution of the BGK equation in a time interval $[0,t]$ is given by the integral form presented by Kogan \cite{kogan1969rarefied}:


\begin{equation}
f(x,y,z,t,u,v,w,\xi) = 
\frac{1}{\tau} \int_o^t f^{eq}(x',y',z',t,u,v,w,\xi) e^{-(t-t')/\tau}\,dt' + e^{-t/\tau} f_0 (x - ut,y - vt,z - w t), 
\label{eq:integralbgk}
\end{equation}
 
\noindent where $x' = x - u(t-t'),\,\,\,y'=y-v(t-t'),\,\,\,z'=z-w(t-t')$. 

%
%

For the sake of clarity, in the following equations the interface is assumed to be perpendicular to $x_1$ which is indicated as $x$ to reduce the number of indexes,
the microscopic velocity $u_1$ is indicated with $u$. 
The left and right states of the gas are indicated with the suffix $(l)$ and $(r)$. A third, fictitious  gas state representing the gas at the interface is indicated with the suffix $(i)$. The left and right values of the conservative variables $w_{(l)}$, $w_{(r)}$ and their gradients ${w_{(l)}}_{/x}$, ${w_{(r)}}_{/x}$ are obtained from a standard reconstruction scheme (e.g. MUSCL/TVD, ENO, WENO).
On both sides a Maxwellian is defined, from  $w_{(l)}$ and $w_{(r)}$:

\begin{equation}
f^{eq}_{(l)} = \rho_{(l)} \left( \frac{\lambda_{(l)}}{\pi} \right)^{\frac{N+2}{N}} \exp \left[{ -\lambda_{(l)}
\left(
(u_i-{v_{(l)}}_i)^2+\xi^2 \right)} \right],
\label{eq:lmaxwellian}
\end{equation} 

\begin{equation}
f^{eq}_{(r)} = \rho_{(r)} \left( \frac{\lambda_{(r)}}{\pi} \right)^{\frac{N+2}{N}} \exp \left[{ -\lambda_{(r)}
\left(
(u_i-{v_{(r)}}_i)^2+\xi^2 \right)} \right].
\label{eq:rmaxwellian}
\end{equation} 

The intermediate state is reconstructed from the left and right states:

\begin{equation}
w_{(i)} = \int_{u<0} h^l f^{eq}_{(l)} \psi d\Xi + \int_{u>0} h^r f^{eq}_{(r)} \psi d\Xi,
\label{eq:intermediatestate}
\end{equation}
\noindent where  $H$ is the Heaviside function, $h^l= H(U)$, $h^r = 1-H(u)$.

The distribution functions $f_0$ and $f^{eq}$ in Eq. \ref{eq:integralbgk} can be defined on the basis of these three states. The initial condition $f(x,v,0) = f_0$ is defined as a solution to the BGK model \ref{eq:bgk} at Navier-Stokes level. $f_0$ is obtained from Eq. \ref{eq:chapman-enskog-ns}, imposing a discontinuity between left and right states and linearizing the operator $Df$:    

\begin{equation}
f_0= \left\{\begin{array}{l}
{f^{eq}_{(l)}} \left( \left(1+{a_{(l)}} x \right)-\tau \left( {a_{(l)}} u+A_{(l)} \right) \right),\,\,\,x_1\leq 0, \\
{f^{eq}_{(r)}} \left( \left(1+{a_{(r)}} x \right)-\tau \left( {a_{(r)}} u+A_{(r)} \right) \right),\,\,\,x_1>0, \end{array} \right. 
\label{eq:initialf0}
\end{equation}

\noindent where $a_{(l)}$ and $a_{(r)}$ are the coefficients of spatial expansion in the phase space,
$A_{(l)}$ and $A_{(r)}$ are the first order coefficient of the temporal expansions. 
The coefficients $a_{(l)}$ and $a_{(r)}$  may be calculated from the gradients of the conservative variables. In practice, a multi-dimensional approach may be preferred, where the distribution function may be expanded along all axes and not simply along $x$. 
Although it is not shown here, $a_{(l)}$ and $a_{(r)}$ are not constant values but approximated as linear functions of all degrees of freedom of the gas (microscopic velocities $u_i$ and internal effective degrees of freedom $\xi$).
The coefficients $A_{(l)}$ and $A_{(r)}$  are not calculated from the past history of the flow, but from the compatibility condition  $\int ( f^{eq} - f ) \psi d\Xi = 0$ at $t=0$.



The equilibrium distribution approach by $f$ in Eq. \ref{eq:integralbgk} is expressed as: 

\begin{equation}
f^{eq} = \left\{
\begin{array}{l}
f^{eq}_{(i)} \left(1+\overline{ a_{(l)}} x-\overline{ A} t \right),\,\,\,x_1\leq 0, \\
f^{eq}_{(i)} \left(1+\overline{ a_{(r)}} x-\overline{ A} t \right),\,\,\,x_1>0, 
\end{array} 
\right. 
\label{eq:eqlgform}
\end{equation}

\noindent where the coefficient $\overline{ a_{(l)}}$ and  $\overline{ a_{(r)}}$ are obtained from fictitious gradients from the linear interpolation between $w_{(l)}$, $w_{(i)}$ and $w_{(r)}$. 
$\overline{A}$ are obtained from the compatibility condition integrated in a time interval. 

The substitution of Eqq. \ref{eq:initialf0} and \ref{eq:eqlgform} into Eq. \ref{eq:integralbgk} finally provides the solution to the BGK equation $f$:

\begin{eqnarray}
f  &=& \left[
       \left( 1 - e^{-t/\tau} \right) 
  + u \left( - \tau  + \tau  e^{-t/\tau}  +t\,e^{-t/\tau} \right) 
  \left( h_{(l)} \, \overline{a_{(l)}} +  h_{(r)} \, \overline{a_{(r)}} \right)    
  + \left( t - \tau  + \tau e^{-t/\tau}  \right) \overline{A} \right] f^{eq}_{(i)} 
  \nonumber \\
   &+& e^{-t/\tau}  \left[  h_{(l)} f^{eq}_{(l)} + h_{(r)} f^{eq}_{(r)} 
   - u \left( t + \tau\right)  \left(  a_{(l)} h_{(l)} f^{eq}_{(l)} + a_{(r)}^r h_{(r)} f^{eq}_{(r)}  \right) \right] \nonumber \\ 
   &-& \tau  e^{-t/\tau}  \left( A_{(l)} h_{(r)} f^{eq}_{(l)} + A_{(r)} h_{(r)} f^{eq}_{(r)} \right)    
.
\label{eq:resultingf}
\end{eqnarray}

In order to obtain a more compact formulation, the following distribution functions at Navier-Stokes level are introduced: 

\begin{eqnarray}
f_{(c)} &=& 
f^{eq}_{(i)} \left( 1 - \tau \left( h_{(l)} \overline{a_{(l)}} u + h_{(r)} \overline{a_{(r)}} u + \overline{A} \right) \right), 
\label{eq:fcentral} \\
f_{(u)} &=& 
  h_{(l)} f^{eq}_{(l)}  \left[ 1 - \tau \left(  a_{(l)} u + A_{(r)} \right) \right] 
+ h_{(r)} f^{eq}_{(r)}  \left[ 1 - \tau \left(  a_{(r)} u + A_{(r)} \right) \right]. 
\label{eq:fupwind}
\end{eqnarray}


\noindent $f_{(c)}$ is built from the fictitious state  introduced with Eq. \ref{eq:intermediatestate} and, by analogy with the terminology used for numerical schemes, may be considered {\sl central}. 
$f_{(u)}$ keeps into account  the left and right reconstructed variables, and may be related to the idea of {\sl upwind}. The use of the terms {\sl central} and {\sl upwind} only 
refers to the way the distribution function is defined. No analogy is implied with conventional schemes involving a discontinuous reconstruction and solution of a Riemann problem. 

When combined with Eqq. \ref{eq:fcentral} and \ref{eq:fupwind}, Eq. \ref{eq:resultingf} can be re-expressed:

\begin{equation}
f = f_{(c)} ( 1 + \bar{A} t ) +
e^{-t/\tau} \left( f_{(u)} - f_{(c)} \right) +
t e^{-t/\tau} \left( \widetilde{f_{(u)}} - \widetilde{f_{(c)}} \right),
\label{eq:resultingf2}
\end{equation}

\noindent where $\widetilde{f_{(c)}}$ and $\widetilde{f_{(u)}} $ only retain the spatial expansion coefficients: 

\begin{eqnarray}
\widetilde{f_{(c)}} &=& 
f^{eq}_{(i)} \left( 1 - \tau \left( h_{(l)} \overline{a_{(l)}} u + h_{(r)} \overline{a_{(r)}} u  \right) \right), 
\label{eq:fcentraltilde} \\
\widetilde{f_{(u)}} &=& 
  h_{(l)} f^{eq}_{(l)}  \left[ 1 - \tau \left(  a_{(l)} u  \right) \right] 
+ h_{(r)} f^{eq}_{(r)}  \left[ 1 - \tau \left(  a_{(r)} u  \right) \right]. 
\label{eq:fupwindtilde}
\end{eqnarray}


Eq. \ref{eq:resultingf2} reveals a combination of {\sl central} and {\sl upwind} distribution functions, whose coefficients depend on collision rate $\tau$ and time. 

The numerical fluxes are then obtained by integration over the duration $\Delta t$ of the time step:

\begin{equation}
{\cal{F}}= \int_0^{\Delta t} \int f \, \psi \, u \, d\Xi \, dt =   \int \int_0^{\Delta t} f \, \psi \, u \,  dt\,d\Xi 
\label{eq:fluxesf}
\end{equation} 

An average value of $f$, $\widehat{f}$, can be introduced: 

\begin{equation}
\widehat{f}= \frac{1}{\Delta t} \int_0^{\Delta t} f \, dt,  
\label{eq:averagefover dt} 
\end{equation}

\noindent so that Eq. \ref{eq:fluxesf} can be re-expressed: 

\begin{equation}
{\cal{F}}= \Delta t \, \int \widehat{f} \, \psi \, u \, d\Xi.
\label{eq:fluxesfaverage}
\end{equation}

The substitution of  Eq. \ref{eq:resultingf2} in Eq. \ref{eq:averagefover dt} results in: 

\begin{eqnarray}
\widehat{f}&=& 
f_{(c)} ( 1 + 1/2 \bar{A} \Delta t ) +
\frac{\tau}{\Delta t} \left( 1 - e^{-\Delta t/\tau} \right)                                                          \left( f_{(u)} - f_{(c)} \right) \nonumber \\
&+& 
\left[ \frac{\tau}{\Delta t}   \left( 1 - e^{-\Delta t/\tau} \right) - \frac{\tau}{\Delta t}  e^{-\Delta t/\tau} \right]   
\left( \widetilde{f_{(u)}} - \widetilde{f_{(c)}} \right).
\label{eq:resultingf3}
\end{eqnarray}

The time integration introduces in the expression of $\widehat{f}$ and in the numerical fluxes through Eq. \ref{eq:fluxesfaverage} the timescale $\Delta t$, 
which is representative of the resolved flow dynamics. 
Introducing now the dimensionless quantity $\epsilon = \tau/\Delta t$, and the functions $\alpha(\epsilon) = \epsilon \left( 1 - e^{-1/\epsilon} \right)$ 
and $\beta(\epsilon) = \epsilon  e^{-1/\epsilon}$, 
Eq. \ref{eq:resultingf3} can finally be re-arranged into the compact form: 


\begin{equation}
\widehat{f} = 
f_{(c)} ( 1 + 1/2 \bar{A} \Delta t ) +
\alpha  \left( f_{(u)} - f_{(c)} \right) +
\left(\alpha - \beta \right)  \left( \widetilde{f_{(u)}} - \widetilde{f_{(c)}} \right).
\label{eq:resultingf4}
\end{equation}

The dimensionless quantity $\epsilon$ is the ratio of unresolved time scales (thermal fluctuations) to the  ones of the resolved flow. 
It is a measure of the degree of {\sl rarefaction}, although  $\epsilon$ is a hybrid indicator, as it compares a physical quantity to a numerical one. 
The presence of $\Delta t$ in Eq. \ref{eq:resultingf3} is a consequence of the fact that the gas-kinetic scheme is also time-accurate. This property is absent in conventional schemes. 
An significant observation concerns the limit of Eq. \ref{eq:resultingf4}  for a vanishing $\epsilon$ (or {\sl hydrodynamic limit}): 

\begin{equation}
\lim_{\epsilon \mapsto 0} \widehat{f} = f_{(c)} ( 1 + 1/2 \bar{A} \Delta t ).
\label{eq:hydrodynamiclaminar}
\end{equation}

Eqq. \ref{eq:resultingf4} and \ref{eq:hydrodynamiclaminar} suggest that the gas-kinetic scheme generates time-accurate Navier-Stokes fluxes through $f_{(c)}$ as well as corrections depending on reconstruction values and degree of rarefaction $\epsilon$.
This implies the capability to generate a  physically consistent dissipation as a reaction to a discontinuity in the reconstruction. 


%

\section{Gas-kinetic schemes for compressible turbulent flow}
\label{sec:turbulentgks}

A gas-kinetic scheme for turbulent flow is trivially obtained from the laminar scheme, Eq. \ref{eq:resultingf4},  by replacing the relaxation time $\tau$  with another one, $\tau_t$, taking the dynamics of turbulence into account. 
A turbulent relaxation time can be derived from an assumed eddy viscosity $\mu_t$ by setting merely 
\begin{equation}
\tau_t = \mu_t /p,
\label{eq:tauttrivial}
\end{equation}

\noindent  by analogy with the relation $\tau=\mu/p$ used for laminar flow.
%
%
%
%
%
However, unlike conventional turbulence modeling, the effect of unresolved turbulence is expressed by a turbulent relaxation time and not by an eddy viscosity; $\tau_t$ can also be obtained in a more sophisticated and physically more meaningful way directly from assumed turbulent quantities like the turbulent kinetic energy $k$ and the turbulent dissipation rate $\varepsilon$. On the basis of a $k$-$\varepsilon$ RANS turbulence model and a systematic renormalization group procedure, Chen {\sl et al.}  \cite{chen2003extended} and Succi {\sl et al.} \cite{succi2002towards} proposed: 

\begin{equation}
\tau^{k\varepsilon}_t = \tau + C_{\mu}\frac{k^2/\varepsilon}{T\left( 1 + \eta^2 \right)^{1/2} } ,
\label{eq:tautchen}
\end{equation}

\noindent where $C_{\mu}$ is a numerical coefficient used in the $k$-$\varepsilon$ model, normally around $0.09$, $k$ is turbulence kinetic energy, $\varepsilon$ is turbulence dissipation rate and $\eta=S k/\varepsilon$, $S$ is a measure of the local velocity gradient. 
The argument used by Chen is that $\tau^{k\varepsilon}_t$ in  Eq. \ref{eq:tautchen} should express the dependence of $\tau_t$ from the variety of unresolved time scales.  

As an example, the consequence of adopting a turbulent relaxation time instead of an eddy viscosity would appear significant in an hypothetical application to Large Eddy Simulation; the corresponding Smagorinsky subgrid model \cite{smagorinsky1963general} would be 
$
\tau_{SGS} = C S \Delta t^2,
$ 
(where $C$ is a numerical factor and $S$ is the magnitude of strain rate). Eq. \ref{eq:tauttrivial} applied to the conventional Smagorinsky model would instead provide $\tau_{sgs} = C\, S\, \Delta^2 / p$, where $\Delta $ is the filter width. 
This implies the use of  a frequency instead of a wave length to characterize unresolved turbulence. A dynamic subgrid model (Germano  {\sl et al.} \cite{germano1992turbulence}) would be comparing the biggest resolved frequencies instead of the smallest resolved scales of motion: the information obtained would not be equivalent in all circumstances. 

A turbulent gas-kinetic scheme based on the $k$-$\varepsilon$ model is therefore represented by the following set of equations: 

\begin{eqnarray}
\widehat{f^{k\varepsilon}} &=& f^{(c)} ( 1 + 1/2 \bar{A} \Delta t ) +
\alpha^{k\varepsilon} \left( f^{(u)} - f^{(c)} \right) +
\left(  \alpha^{k\varepsilon} - \beta^{k\varepsilon} \right)   \left( \widetilde{f^{(u)}} - \widetilde{f^{(c)}} \right), 
\label{eq:fkepsilon} \\
\alpha^{k\epsilon} &=& \epsilon^{k\epsilon}  \left( 1 - e^{-1/\epsilon^{k\epsilon} } \right), \, \beta^{k\epsilon} = \epsilon^{k\epsilon} \,  e^{-1/\epsilon^{k\epsilon} },   
\label{eq:alphabetakepsilon} \\
\epsilon^{k\varepsilon} &=& \frac{1}{\Delta t} \left( \tau + C_{\mu}\frac{k^2/\varepsilon}{T\left( 1 + \eta^2 \right)^{1/2} } \right).
\label{eq:epsilonkepsilon}
\end{eqnarray}


The hydrodynamic limit of the turbulent gas-kinetic scheme in Eqq. \ref{eq:fkepsilon} - \ref{eq:epsilonkepsilon} is the conventional Navier-Stokes scheme (plus time expansion coefficient) with the $k$-$\varepsilon$ turbulence model.

\begin{equation}
\lim_{\epsilon \mapsto 0} {f^{k\varepsilon} } = {f}_{(c)} = f_{(NS)}^{k\varepsilon}.
\label{eq:hydroturbolimit}
\end{equation} 

The hydrodynamic limit of the scheme would change and differ form a Navier-Stokes scheme if the initial condition $f_0$ were based on a higher order expansion. A second-order expansion would provide a distribution function at the level of the Burnett equation (derivation for instance in Ohwada {\sl et al.}  \cite{ohwada2004kinetic}): 

\begin{equation}
f^{Burnett} = f^{eq} - \epsilon \widehat{\tau} D f^{eq} + \epsilon^2 \widehat{\tau} D \left( \widehat{\tau} D f^{eq} \right).  
\label{eq:chapman-enskog-burnett}
\end{equation}

The Burnett distribution function  
contains non-linear terms, which lead to a non-linear contributions to the turbulent stress tensor which are remarkably closer to the assumed terms in algebraic stress tensor, as pointed out by Chen  {\sl et al.} \cite{chen2004expanded}. This option remains out of the scope of this paper.

A difficulty with the scheme in Eqq.  \ref{eq:fkepsilon} - \ref{eq:epsilonkepsilon} 
emerges in practical simulations.  
%
The degree of rarefaction $\epsilon = \tau_t / \Delta t$ - albeit perfectly suitable for laminar flow - may assume higher values in turbulent flow on a stretched grid, cause the bad conditioning of the preconditioning operator and lead to grid-dependent  results or instability. In this study, the degree of rarefaction has been calculated according to the following expression: 

\begin{equation}
\epsilon = \tau_t / \widehat{\tau}, 
\label{eq:tautautilde}
\end{equation}

\noindent where $\widehat{\tau}$ is a time scale representative of the mean, resolved flow. In practice, $\widehat{\tau}$ can be estimated on the basis of the gradients of the mean flow, e.g. density: 

\begin{equation}
\widehat{\tau} = \frac{\rho}{D\,\rho }. 
\label{eq:tautilde}
\end{equation}

Eq. \ref{eq:epsilonkepsilon} is therefore replaced by: 

\begin{equation}
\epsilon^{k\varepsilon} = \frac{\tau^{k\varepsilon}}{\widehat{\tau}} = \frac{\tau + C_{\mu}\frac{k^2/\varepsilon}{T \left( 1 + \eta^2 \right)^{1/2} } }{\rho/D\rho}.
\label{eq:rarefactionkepsilon}  
\end{equation}

Eq. \ref{eq:rarefactionkepsilon}, unlike Eq. \ref{eq:epsilonkepsilon}, expresses a degree of rarefaction and is grid-independent. $\epsilon^{k\varepsilon}$, as defined in Eq. \ref{eq:rarefactionkepsilon}, can be seen as a ``turbulent Knudsen number'', based on the ratio of unresolved and resolved time scales. 
Like the laminar gas-kinetic scheme, the turbulent gas-kinetic scheme given by Eqq. \ref{eq:fkepsilon} - \ref{eq:epsilonkepsilon} 
generates kinetic effects, by taking into account the effect of {\sl collisions} (in this case, between eddies) on {\sl transport}. 
In turbulent flow, $\epsilon$ assumes larger values, up to a few thousandths or a few hundredths in shocklayers at high Mach number, which would correspond to a flow in {\sl transitional } regime. 
Despite the fact that gas-kinetic schemes are not developed for rarefied flow, they might be able to handle moderate rarefaction, provided the collisions are suitably modeled. 
A similar application has been done in laminar flow;  Liao {\sl et al.}  \cite{liao2007gas}  have used the same gas-kinetic scheme with a variable relaxation time in order to model the collisions in the transitional regime.

\section{Implementation of the turbulent gas-kinetic scheme in a finite-volume solver}
\label{sec:implementation}


%
%

In practical calculations, the relaxation time is modified with the addition of a contribution ($\tau_a$) which provides artificial dissipation in the proximity of discontinuities: 

\begin{equation}
\tau = \tau + \tau_t + \tau_a. 
\label{eq:relaxation time}
\end{equation}

\noindent The artificial dissipation time $\tau_a$, following Xu \cite{xu2001gas} is taken to be proportional to the pressure jump across the interface: 

\begin{equation}
\tau_a = C_a \frac{\left|p^r-p^l\right|}{\left|p^r+p^l\right|}\, \Delta t,
\label{eq:taua}
\end{equation}

\noindent where $p^l$ and $p^r$ are pressure values of the left and right states of the gas, $C_a$ is a coefficient whose value varies from $0$ to $1.$ in laminar flow, depending on grid and shock, and can be omitted in turbulent simulations without any significant impact on stability.  
Complying with the original formulation of the scheme by Xu \cite{xu2001gas} the heath flux must be corrected to account for a realistic Prantdl number. 
Like in a few other implementations (refer for instance to Xu \cite{xu2001gas} and May  {\sl et al.} \cite{may2007improved}) the distribution function is expanded only in the direction of the fluxes. A multi-dimensional version was also developed; on reasonably good-quality grids it provides comparable results at a slightly higher computational cost.

In the numerical experiments carried out in this work, the gas-kinetic fluxes have been implemented in a 2D finite volume steady-state solver (Righi \cite{righi2012aeronautical,righi2012rgd}). 
Well-known acceleration techniques (4-level multigrid and LU-SGS preconditioning in the form proposed by Jameson and his coworkers \cite{yoon1988lower,jameson1983solution}) have provided convergence properties comparable to more traditional Navier-Stokes schemes. 
The  reconstruction techniques include second and third order TVD/MUSCL  schemes and fifth order WENO, although the results shown are only second-order. The {\sl minmod} limiter has been used in all cases for both  conservative variables and their gradients. 

As a results of higher suitability with the chosen flow cases, the well-known $k$-$\omega$ turbulence model by Wilcox \cite{wilcox2006}  has been chosen instead of the  $k$-$\varepsilon$ model.
The  turbulent relaxation time and the degree of rarefaction are calculated from Eq. \ref{eq:tautchen}, re-expressed for the specific dissipation $\omega$: 

\begin{equation}
\tau^{k\omega}_t = \tau + \tau_a +\frac{k/\omega}{T\left( 1 + \eta^2 \right)^{1/2} }, 
\label{eq:tautchenadjusted}
\end{equation}

\begin{equation}
\epsilon^{k\omega} = \frac{\tau^{k\omega}_t}{\widehat{\tau}}.
\label{eq:epsilonkomega}
\end{equation}

Eq. \ref{eq:tautchenadjusted} also includes the aritifical dissipation time $\tau_a$. 
Eqq. \ref{eq:tautchenadjusted} and \ref{eq:epsilonkomega} are combined with Eqq. \ref{eq:fkepsilon} and \ref{eq:alphabetakepsilon} to generate a $\widehat{f^{k\omega}}$.
Two allied equations for the two turbulent quantities $k$ and $\omega$ are solved alongside the equations for the conservative variables. 

No-slip wall boundary conditions have been used in all flow case; no changes with respect to conventional CFD have been implemented, as the ``rarefaction'' effects introduced in the previous sections are not expected to extend to solid walls.

\section{Numerical experiments }
\label{sec:experiments}

The flow cases presented are summarized in Table \ref{tab:flowcases}. The size of the grids used in all flow cases are summarized in Table \ref{tab:gridsflowcases}.

\begin{table}
\begin{tabular}{lp{76mm}lrr}
\hline \hline
{}\footnote{Position in text}           & Case & Reynolds & Mach\footnote{Freestream} \\ \hline
\ref{subsec:rae2822} & RAE2822 Aerofoil Case 9 \cite{cook1979aerofoil}   & $6.3 \times 10^6$ & $0.73$ \\   
\ref{subsec:rae2822} & RAE2822 Aerofoil Case 10  \cite{cook1979aerofoil} &  $6.3 \times 10^6$ & $0.73$ \\   
\ref{subsec:naca0012} & NACA 0012 Aerofoil $\alpha = 2.26^\circ$ \cite{harris1981two}  & $9.0 \times 10^6$ & $0.80$ \\   
\ref{subsec:naca0012} & NACA 0012 Aerofoil $\alpha=4.86^\circ$ \cite{harris1981two}& $9.0 \times 10^6$ & $0.74$ \\   
\ref{subsec:delery} & D\'elery bump channel (Case C) \cite{delery1983experimental}   & $1.0 \times 10^6$  \footnote{Based on bump height}& $0.67$ \\   
\ref{subsec:rampsettles} & Supersonic compression corner $\alpha = 8^\circ$ \cite{settles1979detailed} &  $Re_\theta = 23\,000 $ & $2.85$ \\   
\ref{subsec:rampsettles} & Supersonic compression corner $\alpha = 16^\circ$ \cite{settles1979detailed} &  $Re_\theta = 23\,000 $ & $2.85$ \\   
\ref{subsec:rampsettles} & Supersonic compression corner $\alpha = 20^\circ$ \cite{settles1979detailed} &  $Re_\theta = 23\,000 $ & $2.85$ \\   
\ref{subsec:rampsettles} & Supersonic compression corner $\alpha = 24^\circ$  \cite{settles1979detailed} & $Re_\theta = 23\,000 \footnote{Simulations at lower and higher Reynolds number are also included} $ & $2.85$ \\   
\hline \hline
\end{tabular}
\caption{Summary of flow cases presented}
\label{tab:flowcases}
\end{table}

\begin{table}
\begin{tabular}{lllllr}
\hline \hline
Case & Type & Coarse & Medium & Fine & Wall res. \footnote{Wall resolution in wall units ($y^+= y u_\tau/\nu $) } \\ \hline
RAE2822 Case 9 and 10 & C-type& $368 \times 112$  &  $416\times 128$ & $560\times 176$ & $<1$ \\   
NACA 0012 all cases   & C-type& $384 \times 112$  &  $464 \times 160$  &  $608 \times 160$ & $<1$ \\
D\'elery bump channel & Channel & $200 \times 160$ & $312 \times 256$  &  $456 \times 288$ & $<1$ \\
Compression corner, all cases&  Channel & $240 \times 96$  & $304 \times 120$ & $528 \times 136$ & $<1$ \\
\hline \hline
\end{tabular}
\caption{Size of grids used in all flow cases }
\label{tab:gridsflowcases}
\end{table}

For comparison, results obtained from the same solver but with a conventional  Navier-Stokes scheme with Roe's approximate Riemann solver. 
All flow cases have been computed on different grids and the results shown in the following sections are reasonably grid-independent. 
All computational meshes are stretched to improve shock and boundary layer resolution, the latter is guaranteed by the placement of the first layer of cells within the laminar sublayer.

\subsection{Transonic flow around a RAE 2822 airfoil in supercritical conditions}
\label{subsec:rae2822}

With reference to the experimental campaign carried out by Cook \cite{cook1979aerofoil} around the RAE 2822 airfoil, two flow cases have been considered. 
In case 9  the flow at $M=0.73$ and  $3.19^\circ$ angle of attack, generates a strong normal shock at around $55$ \% of the airfoil chord without causing the separation of the boundary layer. In case 10, the flow at $M=0.745$ and   $3.19^\circ$ angle of attack,   generates a stronger shock which causes an incipient separation, a more significant thickening of the boundary layer and a displacement upstream of the shock. The latter feature appears to be very challenging to capture for conventional schemes and linear two-equation models. Reynolds number is slightly above 6 million in both cases. 
In the experiments, the boundary layer has been tripped at $3\%$ of the airfoil chord on the upper and lower sides; the calculation is fully turbulent downstream of this point. 

As no special freestream conditions have been applied to keep into account the vorticity created, the domain has been extended to $50$ chords and the angle of attack has been chosen in order to match the pressure distribution upstream the shock. In all cases the angle of attack has resulted to be slightly smaller than the measured value. 

\begin{figure}[h]
  \begin{center}
    \input{cpcase9_}
    \input{cfcase9_}
  \end{center}
  \caption{Wall pressure (a) and skin friction (b) coefficients for the RAE2822 airfoil (Case 9, $Re = 6.3\times 10^6$, $M=0.73$,  angle of attack $\alpha =3.19^\circ$).
   (\solidrule) Gas-kinetic scheme (GKS) on finest grid,
   (\protect\thindashedrule) GKS on medium grid, 
   (\protect\verythindashedrule) GKS on coarsest grid,  
   (\protect\dotdashedrule) Navier-Stokes (Roe's approximate Riemann solver) on finest grid,
   (\protect\fewdots): experimental data from Cook \cite{cook1979aerofoil}. }
\label{fig:cpcfcase9}
\end{figure}
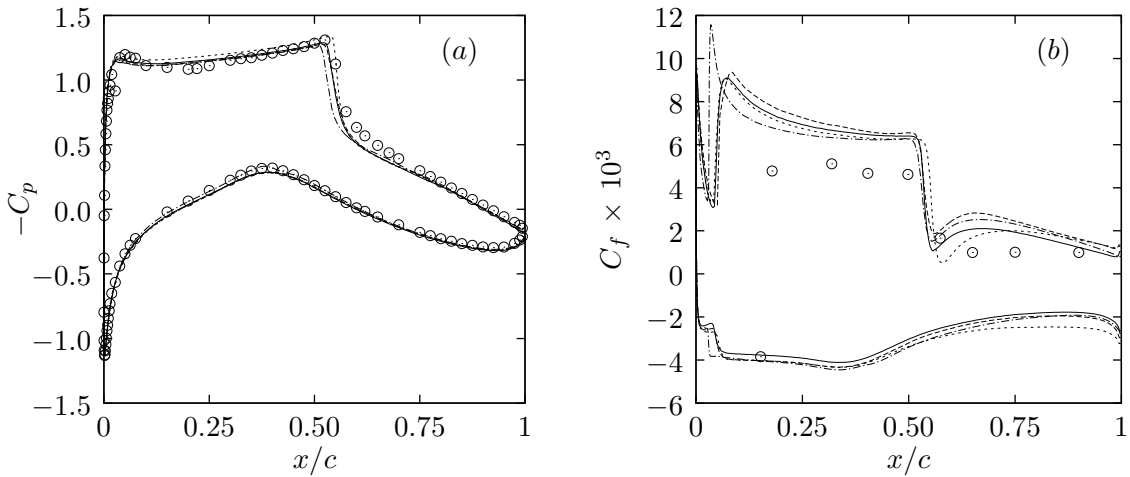

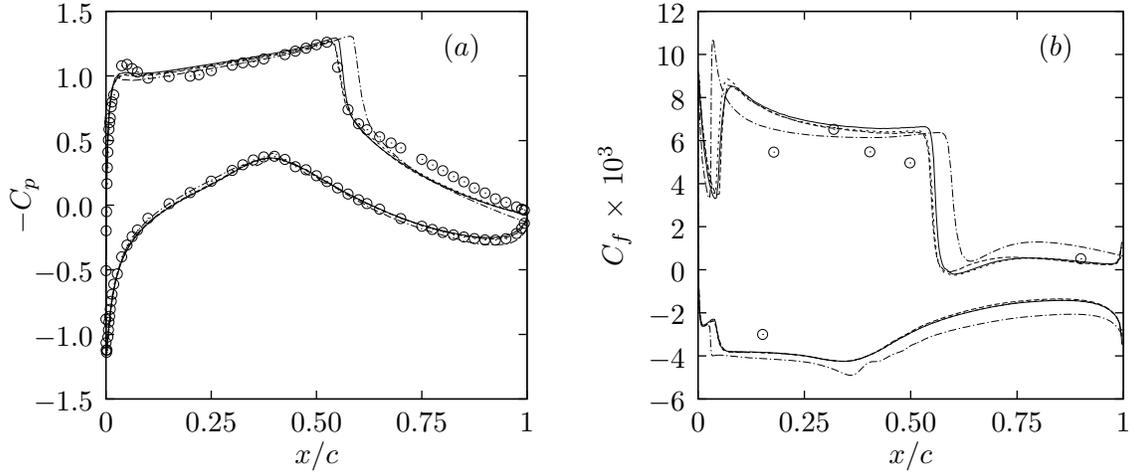
\begin{figure}[h]
  \begin{center}
    \input{cpcase10_}
    \input{cfcase10_}
  \end{center}
  \caption{Wall pressure (a) and skin friction (b) coefficients for the RAE2822 airfoil (Case 10 $Re = 6.2\times 10^6$, $M=0.745$,  angle of attack $\alpha =3.19^\circ$).
   (\solidrule)  Gas-kinetic scheme (GKS) on finest grid,
   (\protect\thindashedrule) GKS on medium grid, 
   (\protect\verythindashedrule) GKS on coarsest grid,  
   (\protect\dotdashedrule) Navier-Stokes (Roe's approximate Riemann solver) on finest grid,
   (\protect\fewdots): experimental data from Cook \cite{cook1979aerofoil}. }
\label{fig:cpcfcase10}
\end{figure}

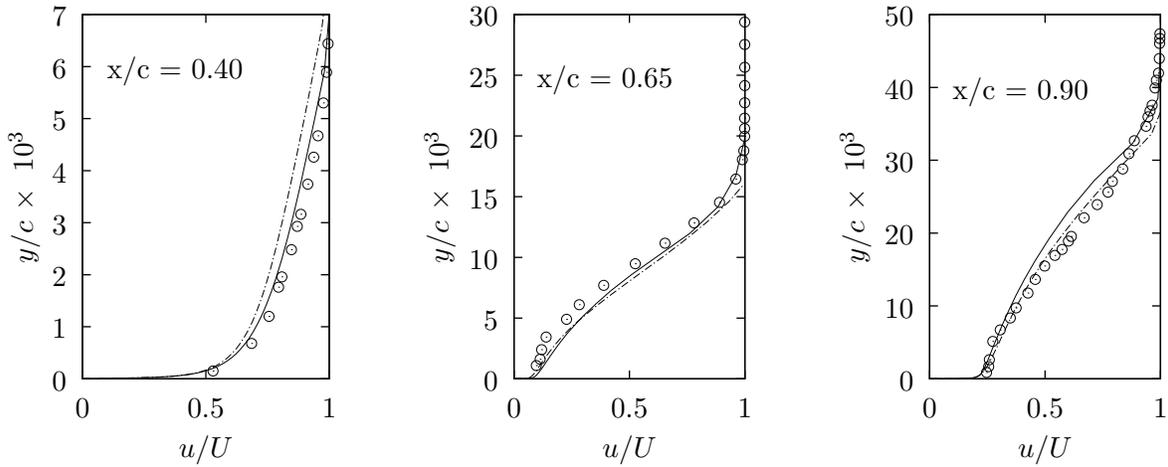
\begin{figure}[h!]
  \begin{center}
    \input{uprofile40_}
    \input{uprofile65_}
    \input{uprofile90_}
  \end{center}
\caption{Velocity profiles for the RAE2822 airfoil (Case 10 $Re = 6.2\times 10^6$, $M=0.745$,  angle of attack $\alpha =3.19^\circ$).
   (\solidrule)  Gas-kinetic scheme (GKS) on finest grid,
   (\protect\dotdashedrule) Navier-Stokes (Roe's approximate Riemann solver) on finest grid,
   (\protect\fewdots): experimental data from Cook \cite{cook1979aerofoil}.
 }
\label{fig:vel_profiles_case10}
\end{figure}

The distribution of pressure and skin friction coefficient obtained with the gas-kinetic scheme and a conventional Navier-Stokes scheme for case 9 and 10 are shown in Fig. \ref{fig:cpcfcase9} and \ref{fig:cpcfcase10}, respectively. 
In case 9 both schemes provide results which are in agreement with one another and with the experimental data. 
In case 10 the results obtained with the two schemes are very similar to one another, except for the shock area. The velocity profiles shown in Fig. \ref{fig:vel_profiles_case10} are reasonably close to each other and to experimental data. 
The results obtained from the gas-kinetic scheme  are more in line with the esperimental data, in terms of position and thickness of the shock, whereas the conventional scheme fails to capture the position of the shock correctly.
The relatively poor prediction of separated flows by the $k$-$\omega$ model in Navier-Stokes schemes is well-known: the errors found in this study from  the Navier-Stokes scheme are in line with literature (refer for instance to Wallin {\sl et al.} \cite{wallin2000explicit} and references therein). 
The predictions obtained from the gas-kinetic scheme show an accuracy comparable to the one provided by more sophisticated, higher-order turbulence models (refer for instance to Wallin {\sl et al.} \cite{wallin2000explicit}), although the concepts behind the modeling of the turbulent stress tensor are different. 

\begin{figure}[h!]
  \begin{center}
\includegraphics[width=142mm]{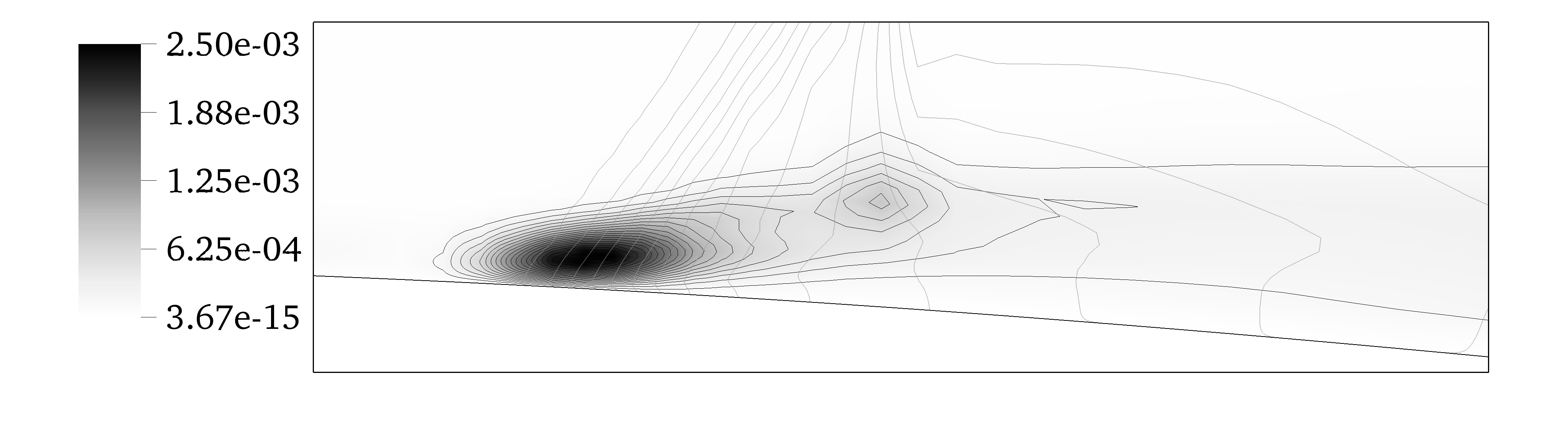}
\includegraphics[width=120mm]{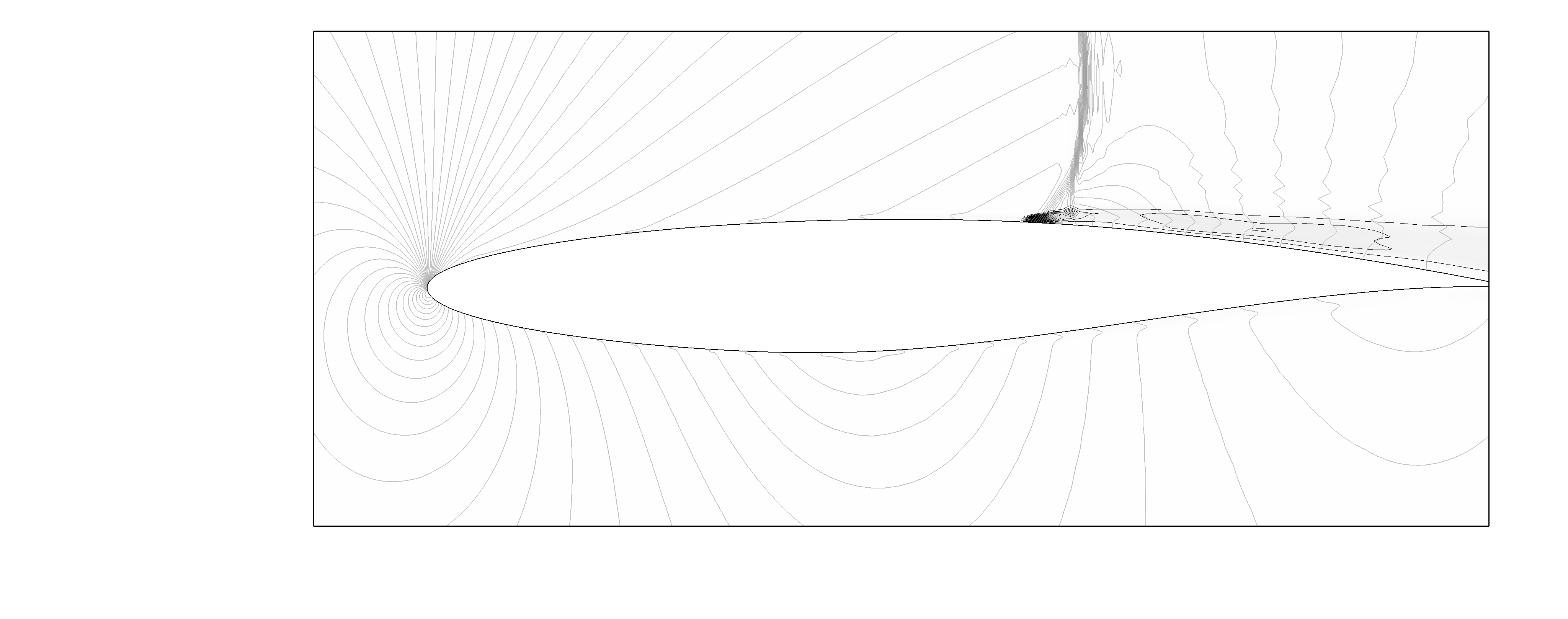}
  \end{center}
\caption{Airfoil RAE2822, Case 10. Degree of rarefaction expressed in term of $\epsilon^{k\omega}$ expressed in Eq. \ref{eq:tautchenadjusted}. 20 static pressure  isolines have been added for reference.}
\label{fig:kncontours}
\end{figure}

Fig. \ref{fig:kncontours} show the distribution of the degree of rarefaction calculated according to Eq. \ref{eq:tautchenadjusted}. In the shock region - where the gas-kinetic scheme works differently from the conventional scheme, the degree of rarefaction reaches values which are normally considered in the transitional regime.

%
%
%
%

\subsection{Transonic flow around a NACA 0012  airfoil in supercritical conditions}
\label{subsec:naca0012}

The NACA 0012 airfoil has been the object of several experimental investigations, including transonic and separated flow conditions. Two flow cases 
are included in the  experimental investigation by Harris \cite{harris1981two}: 
in the first one the flow at Mach $M=0.799$ and angle of attack   $\alpha =2.86^\circ$ causes an incipient separation of the boundary layer immediately downstream of the shock. In the second flow case, the flow at Mach $M=0.74$ and angle of attack   $\alpha =4.86^\circ$ causes the boundary layer to separate downstream of the shock and generate a large separated region. 
Reynolds number is 9 million in both cases. 

Two versions of each grid have been generated, with slightly different clustering around the expected shock wave position. 
As no special freestream conditions have been applied to keep into account the vorticity created, the domain has been extended to $50$ chords and the angle of attack has been chosen in order to match the pressure distribution upstream the shock. In all cases the angle of attack has resulted to be slightly smaller than the measured value. 
In the experiments, the boundary layer has been tripped at $5\%$ of the airfoil chord on the upper and lower sides; the calculation is fully turbulent downstream of this point.

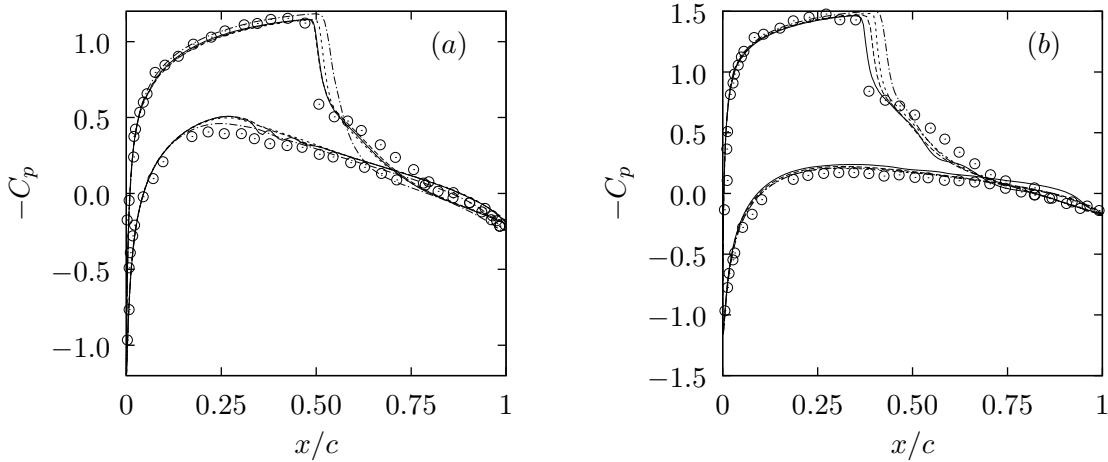
\begin{figure}[h]
  \begin{center}
    \input{cp0012m08_}
    \input{cp0012m074_}
  \end{center}
  \caption{
Wall pressure and skin friction coefficients for the NACA 0012  airfoil; (a) $Re = 9.0\times 10^6$, $M=0.799$,  angle of attack $\alpha =2.86^\circ$; (b) $Re = 9.0\times 10^6$, $M=0.74$,  angle of attack $\alpha =4.86^\circ$.
   (\solidrule)  Gas-kinetic scheme (GKS) on finest grid,
   (\protect\thindashedrule) GKS on medium grid, 
   (\protect\verythindashedrule) GKS on coarsest grid,  
   (\protect\dotdashedrule) Navier-Stokes  on finest grid,
   (\protect\fewdots): experimental data from Harris \cite{harris1981two}.  
}
\label{fig:cp0012m08}
\end{figure}

Fig. \ref{fig:cp0012m08} shows the pressure coefficient measured on the airfoil  and calculated with the gas-kinetic scheme and a conventional scheme. In both cases, the gas-kinetic scheme's predictions indicate a shock  position more in line with the experimental data \cite{harris1981two}, as the Navier-Stokes scheme delays separation and fails to position the shock  accurately in both cases. 
The degree of rarefaction calculated according to Eq. \ref{eq:tautchenadjusted} reaches values around $0.0025$ in the vicinity of the shocklayer, similarly to what happens around the RAE2822 airfoil. 


\subsection{Transonic flow in D\' elery bump channel  }
\label{subsec:delery}

The experiment  transonic bump flow (Case C), experimentally investigated by D\'elery \cite{delery1983experimental} was designed to produce a strong shock - boundary layer interaction leading to a large flow separation. 
The  Mach $0.615$  duct flow impacts a ramp-semicircular bump mounted on one side of the channel, reaching approximately Mach $1.45$ before the shock. The shock - boundary layer interaction generates the typical $\lambda$ structure, with the separation starting  at the foot  of the first leg.     
It is well known that predicting the position of the separation point and the size of the separated area are challenging for standard, linear two-equation models as they  tend to delay separation and underpredict the extension of the separation.

In order to match the experimental position of the shock, the outlet pressure is heuristically adjusted. The two solvers  therefore use slightly different values of outlet pressure. The calculation is fully turbulent. 

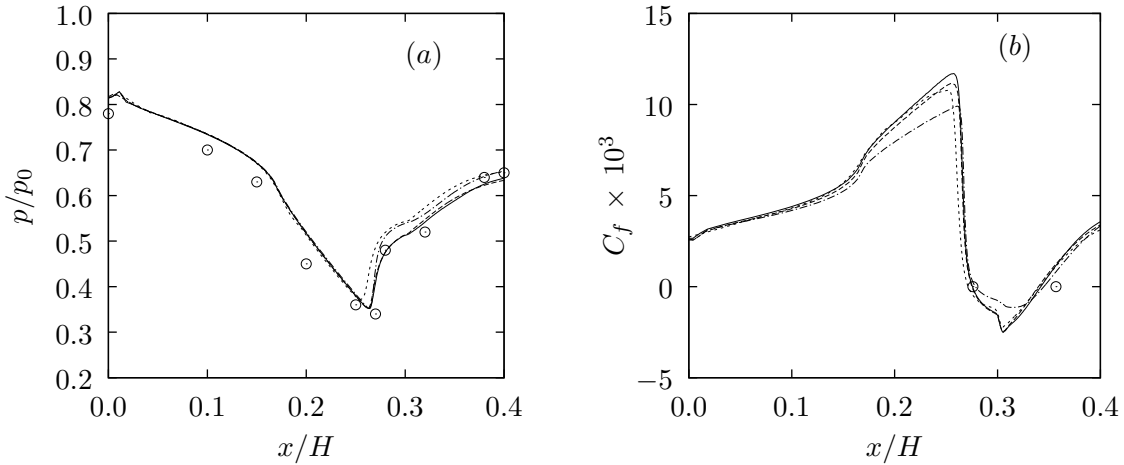
\begin{figure}[h]
  \begin{center}
    \input{pdelery_}
    \input{cfdelery_}
  \end{center}
  \caption{Pressure (a) and skin friction coefficient (b) for the D\'elery bump channel flow. 
   (\solidrule)  Gas-kinetic scheme (GKS) on finest grid,
   (\protect\thindashedrule) GKS on medium grid, 
   (\protect\verythindashedrule) GKS on coarsest grid,  
   (\protect\dotdashedrule) Navier-Stokes (Roe's approximate Riemann solver) on finest grid,
   (\protect\fewdots): experimental data from D\'elery \cite{delery1983experimental}.  }
\label{fig:cpdelery}
\end{figure}

\begin{figure}[h!]
\includegraphics[width=160mm]{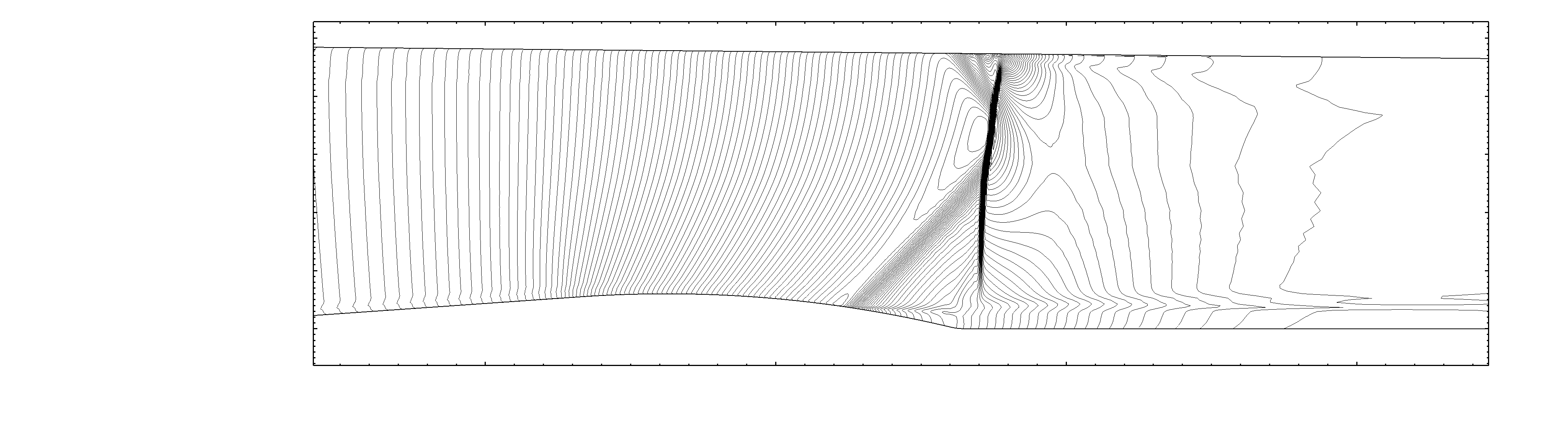}
\includegraphics[width=160mm]{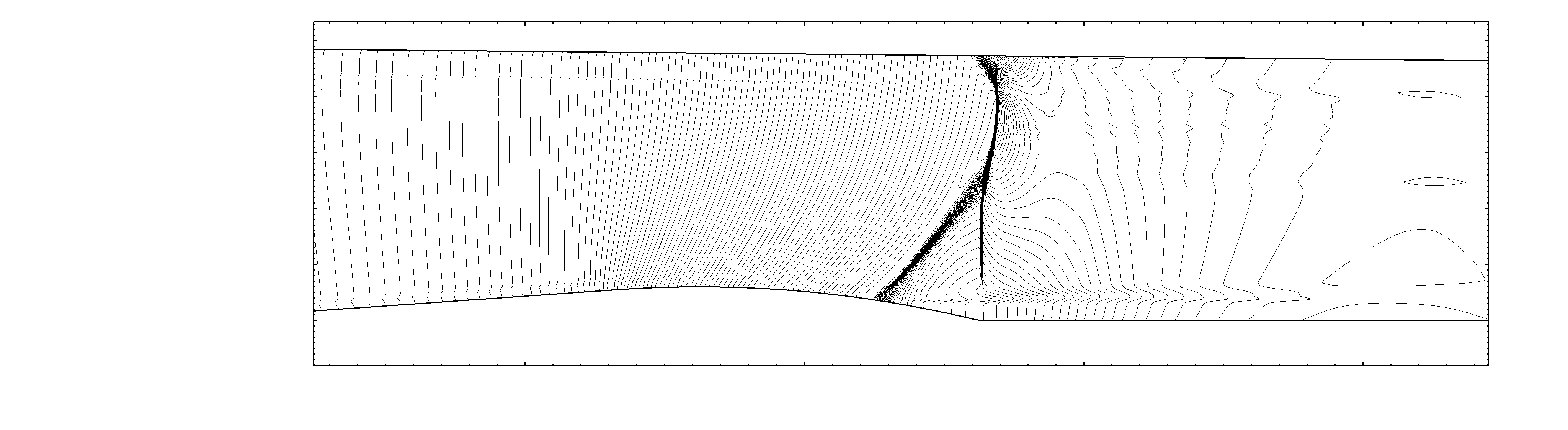}
\caption{D\'elery bump channel flow. Gas-kinetic scheme: plot above, conventional scheme: plot below. 100 static pressure isolines.   }
\label{fig:deleryshocksystem}
\end{figure}

\begin{figure}[h!]
\includegraphics[width=80mm]{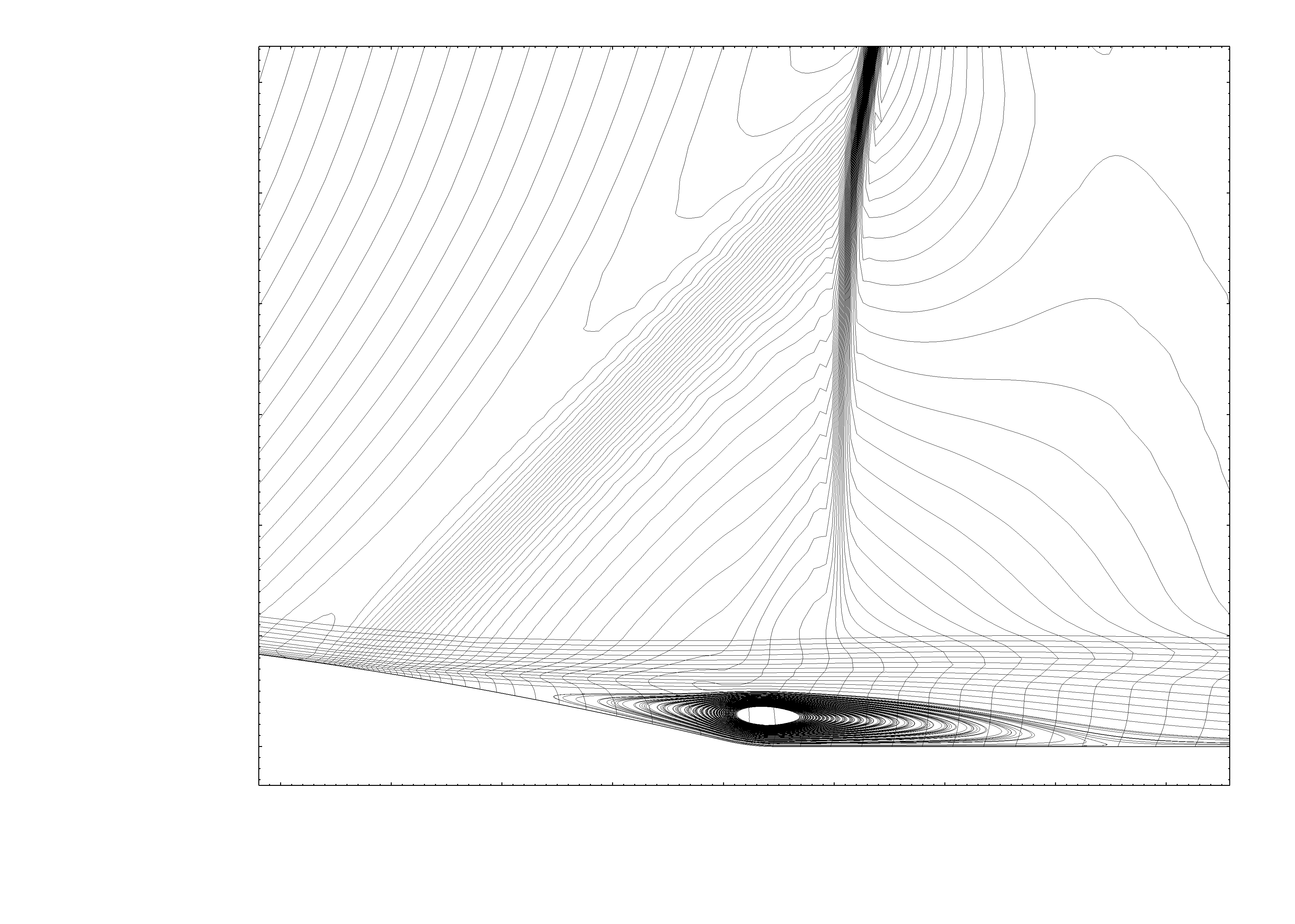}
\includegraphics[width=80mm]{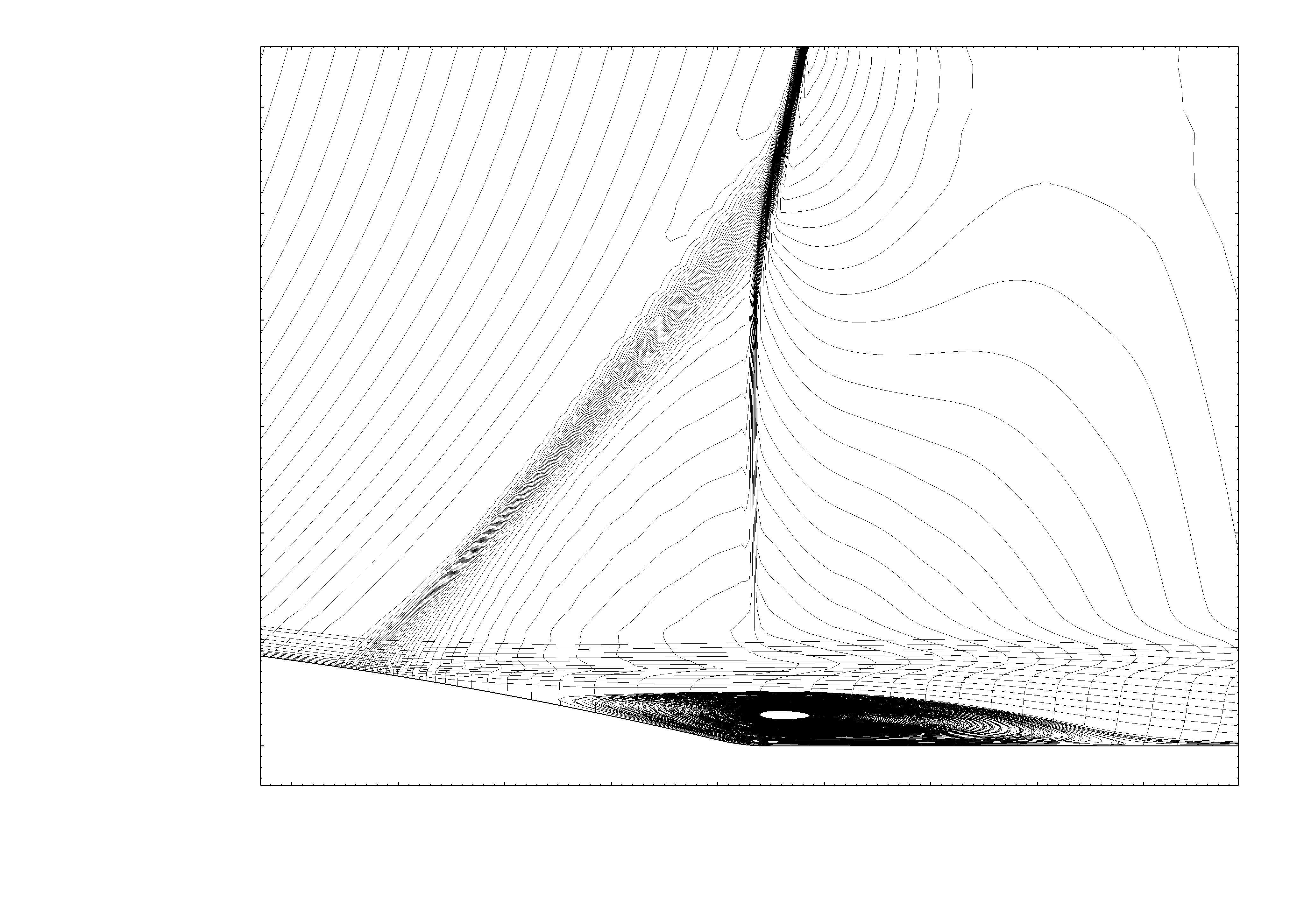}
\caption{D\'elery bump channel flow. Gas-kinetic scheme: left plot, conventional scheme: right plot. 100 static pressure isolines plus 22 streamlines.   }
\label{fig:deleryseparationbubbles}
\end{figure}

\begin{figure}[h!]
\includegraphics[width=160mm]{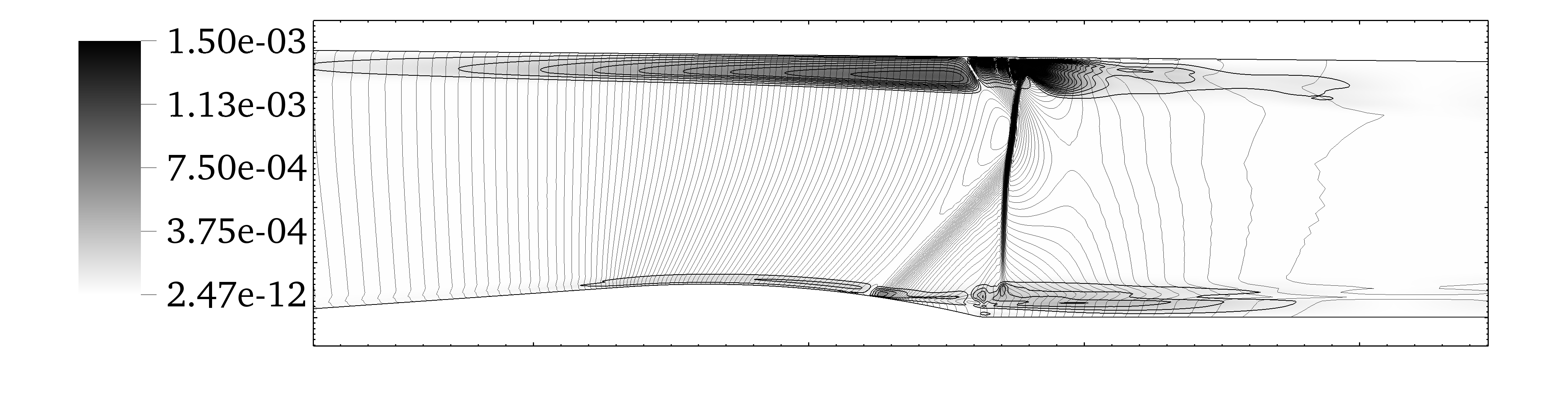}
\caption{D\'elery bump channel flow. Degree of rarefaction expressed in term of $\epsilon^{k\omega}$ expressed in Eq. \ref{eq:tautchenadjusted}. 100 static pressure  isolines have been added for reference. }
\label{fig:deleryrarefactionlevel}
\end{figure}

Fig. \ref{fig:cpdelery} shows static pressure and skin friction coefficient. The predicted extension of the separation region is in good agreement with experimental data for both schemes, but the behavior of static pressure downstream of the shock is more accurately predicted by the gas-kinetic scheme. As a matter of fact, the predicted shock structure is quite different between the two schemes. 
The different reconstruction of the shock system, generated by the gas-kinetic and the conventional scheme, can be observed in Fig. \ref{fig:deleryshocksystem}, whereas a detail view on the separation region is provided in Fig. \ref{fig:deleryseparationbubbles}. The different thickness of the oblique leg of the $\lambda$ shock system is evident; yet is the level of eddy viscosity (not shown here) in the two cases comparable. The separation predicted by the gas-kinetic scheme is more extended and its positioning vis-a-vis the shocks is more similar to the textbook sketches (refer for instance to D\'elery {\sl et al.}  \cite{delery1986shock}). The degree of rarefaction  is shown in Fig. \ref{fig:deleryrarefactionlevel}: high values, in the same order as the ones observed in the airfoil flow cases, appear in the proximity of the shocks and might be at the origin of the different behavior of the two schemes.

\subsection{Shock-separated supersonic turbulent boundary layer at a compression corner}
\label{subsec:rampsettles}

The supersonic flow impinging a compression corner may separate at the shock foot and produce a large separate region, depending on Reynolds number and geometry. 
Settles \cite{settles1976details,settles1979detailed} has investigated a flow at $Re_\theta = 23\,000$ (where $\theta$ is the momentum thickness of the incoming boundary layer)
 with a corner of $8^\circ$, $16^\circ$, $20^\circ$ and $24^\circ$. 
The $8^\circ$ corner does not separate the flow and the $16^\circ$ one generates only an incipient separation. The time-averaged separation at $20^\circ$ spans about $1\delta$ and 
the one at $24^\circ$ about $2\delta$ (where $\delta$ is the incoming boundary layer thickness).

Results from conventional Navier-Stokes schemes are easily found in the literature, e.g. in Golberg {\sl et al.} \cite{goldberg1998application} and Menter {\sl et al.} \cite{menter1994assessment}.  In general, conventional scheme fail to predict the right shock position and separation length.

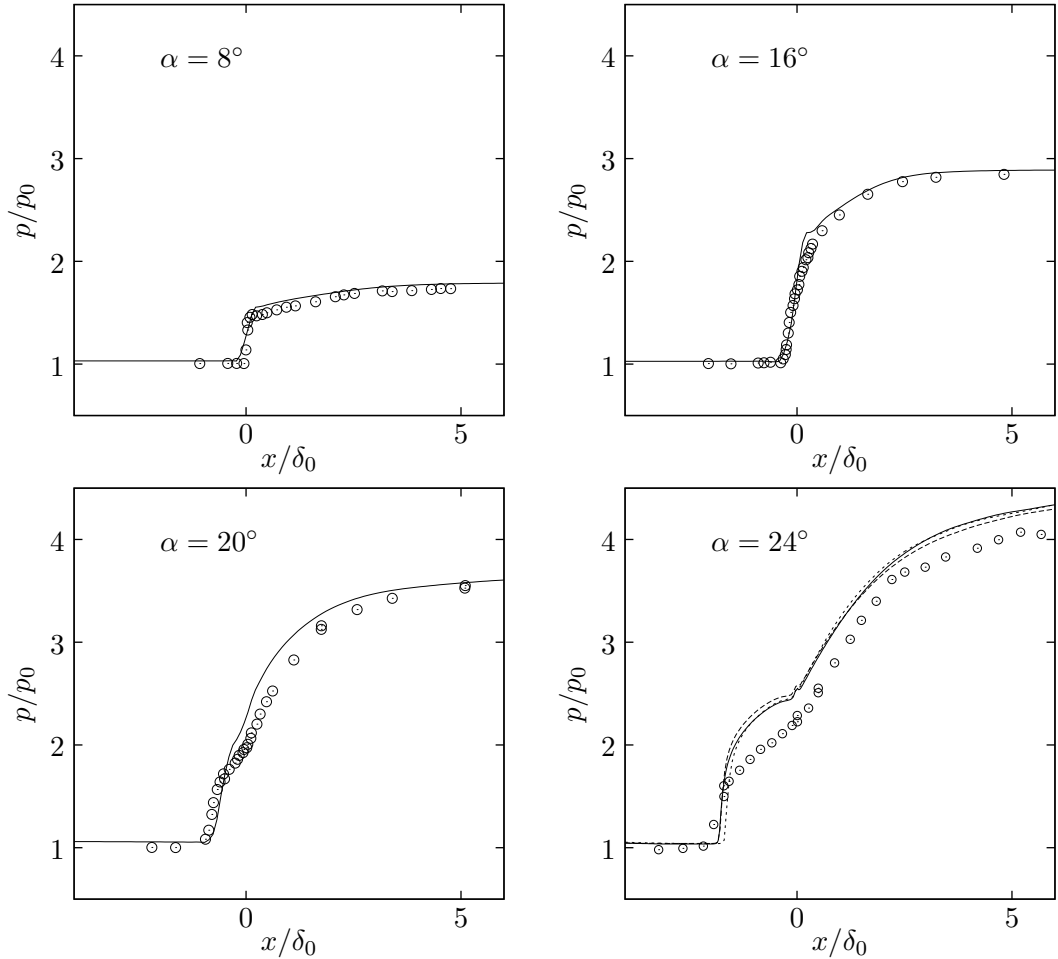
\begin{figure}[h]
  \begin{center}
    \input{cpramp8_}
    \input{cpramp16_}
    \input{cpramp20_}
    \input{cpramp24gridconv_}
  \end{center}
  \caption{Pressure calculated for four different compression corner flows, characterized by angles values of $8^\circ$, $16^\circ$, $20^\circ$ and $24^\circ$ (freestream conditions: $M = 2.85$, $Re = 7.0\times 10^7$ per length unit, $\delta_0 = 0.023\,m$).
   (\solidrule)  Gas-kinetic scheme (GKS) on finest grid,
   (\protect\thindashedrule) GKS on medium grid (only shown for $\alpha = 24^\circ$), 
   (\protect\verythindashedrule) GKS on coarsest grid (only shown for $\alpha = 24^\circ$),  
   (\protect\fewdots): experimental data from Settles \cite{settles1979detailed}.  
  }
\label{fig:cpallramps}
\end{figure}

\begin{figure}[h]
  \begin{center}
    \input{cframp8_}
    \input{cframp16_}
    \input{cframp20_}
    \input{cframp24gridconv_}
  \end{center}
  \caption{Skin friction coefficient calculated for four different compression corner flows, characterized by angles values of $8^\circ$, $16^\circ$, $20^\circ$ and $24^\circ$ (freestream conditions: $M = 2.85$, $Re = 7.0\times 10^7$ per length unit, $\delta_0 = 0.023\,m$).
   (\solidrule)  Gas-kinetic scheme (GKS) on finest grid,
   (\protect\thindashedrule) GKS on medium grid (only shown for $\alpha = 24^\circ$), 
   (\protect\verythindashedrule) GKS on coarsest grid (only shown for $\alpha = 24^\circ$),  
   (\protect\fewdots): experimental data from Settles \cite{settles1979detailed}. 
   }
\label{fig:cfallramps}
\end{figure}
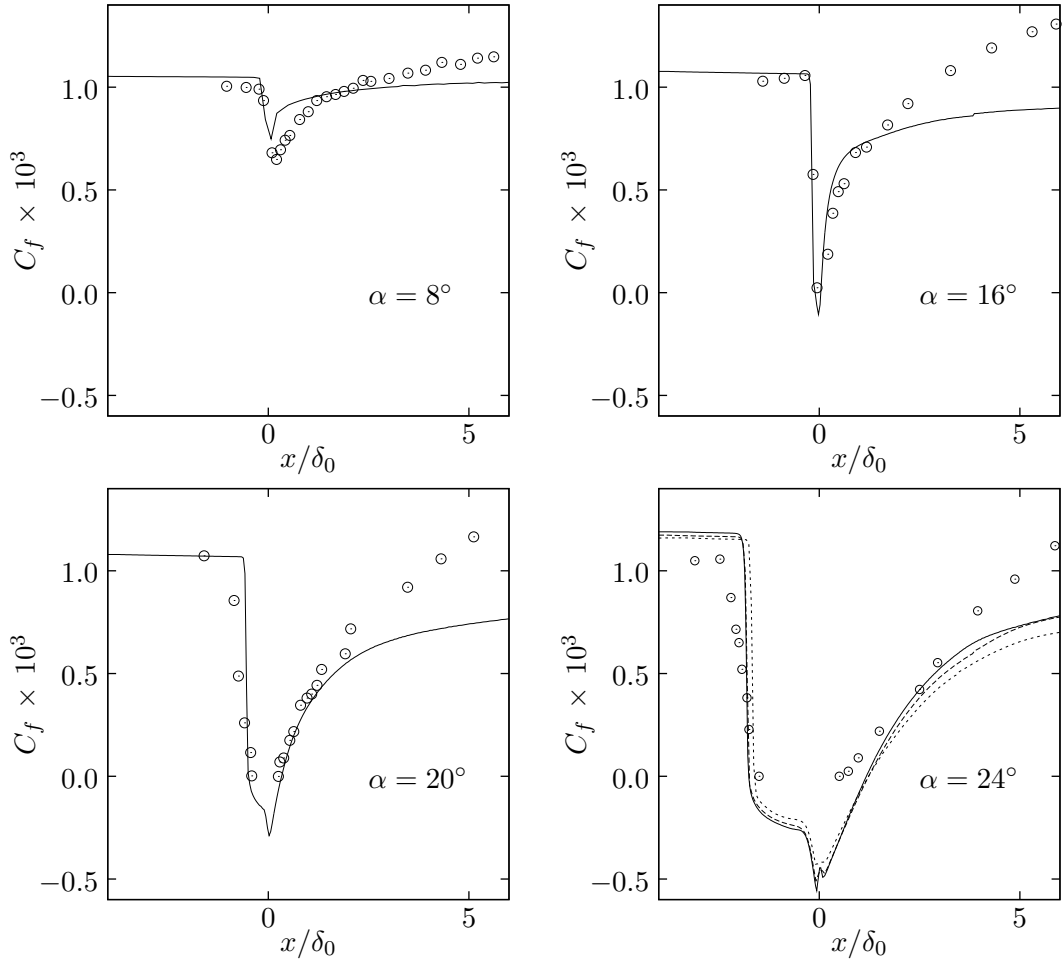

\begin{figure}[h!]
\includegraphics[width=120mm]{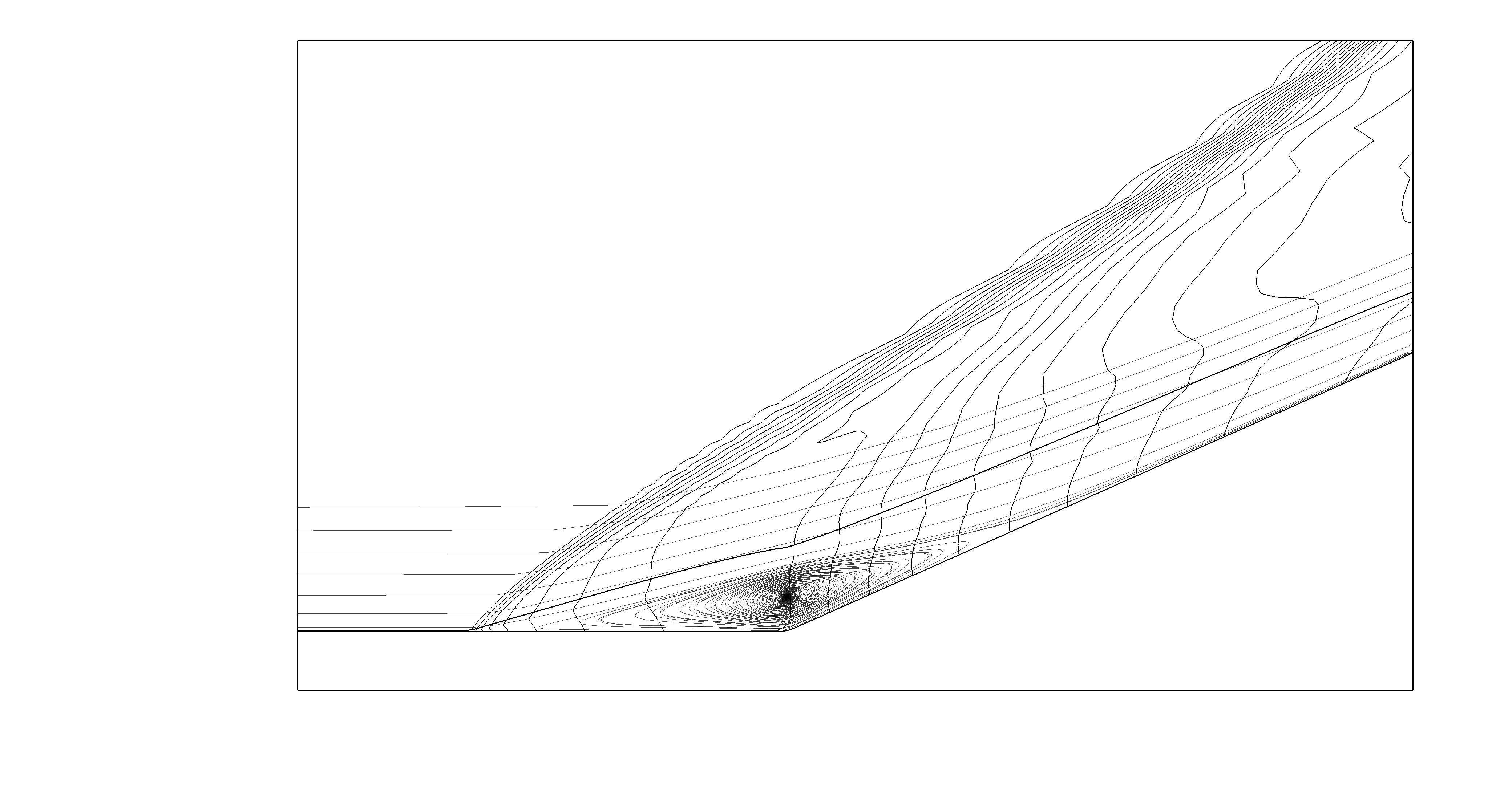}
\caption{Compression corner $ M=2.85$. Shock system represented with 20 static pressure isolines, sonic line (Mach=1 isoline) and 22 streamlines inside the boundary layer have bben added.  }
\label{fig:ramp24shockstructure}
\end{figure}

\begin{figure}[h!]
\includegraphics[width=120mm]{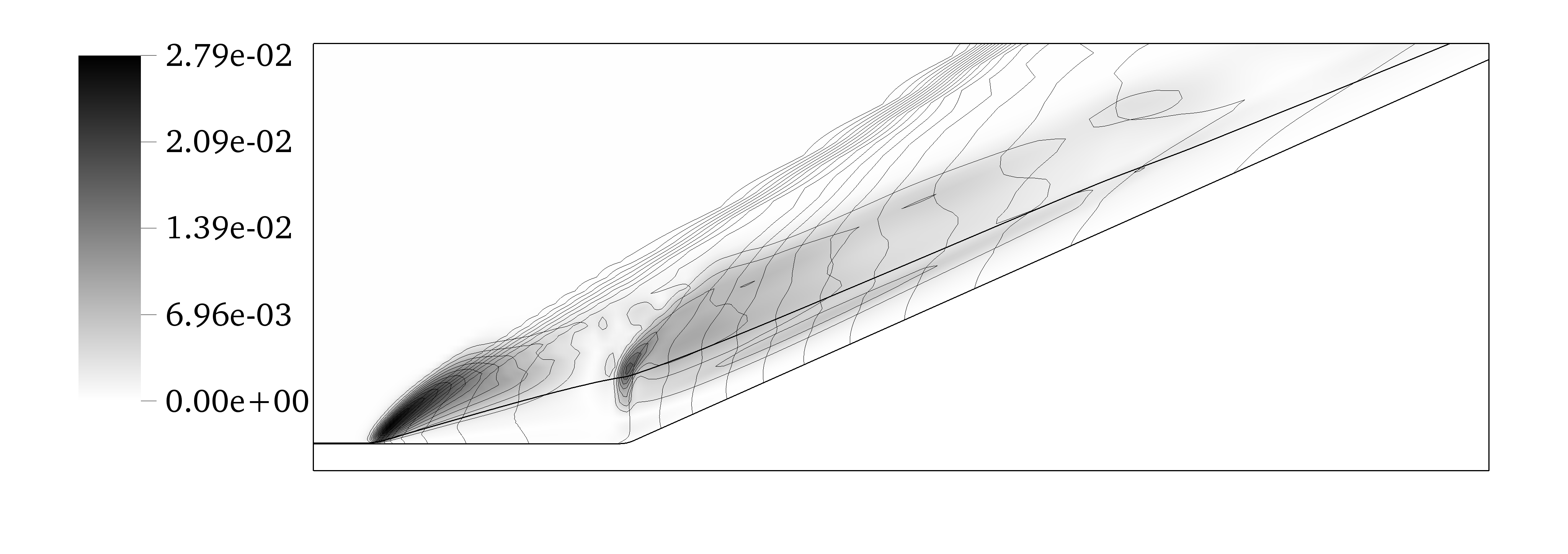}
\caption{Compression corner $ M=2.85$. Degree of rarefaction expressed in term of $\epsilon^{k\omega}$ expressed in Eq. \ref{eq:tautchenadjusted}. 20 static pressure  isolines have been added for reference.  }
\label{fig:ramp24knt}
\end{figure}

Figures \ref{fig:cpallramps} and \ref{fig:cfallramps} show the reasonably good agreement of predictions in all four cases in terms of wall pressure  and skin friction coefficient.
The shock system is shown in Fig. \ref{fig:ramp24shockstructure}. A comparison with textbook sketches in \cite{delery1986shock} and \cite{settles1979detailed} 
reveals though an inaccurate behavior in the re-attachment region and downstream, where the flow seems too slow in re-accelerating and reduce boundary layer thickness.  
Fig. \ref{fig:ramp24knt} shows the distribution of the degree of  rarefaction, which reaches in the proximity of the shocks a maximum of around $0.03$, which is  ten times higher than in the transonic flow cases. 

A number of simulations at different Reynolds number have been conducted with the $24^\circ$ corner in order to assess the dependence of the separation length from the Reynolds number. Results are summarized in Table \ref{tab:settlesreynolds}; they are in good agreement with the empirical law derived by Settles \cite{delery1986shock}: 

\begin{equation}
L_0/\delta_0 = 0.9 e^{0.23 \alpha},
\label{eq:empiricalsettles}
\end{equation}

\noindent where $\delta_0$ is the thickness of the incoming boundary layer, and $L_0$ the total length of the separation.

\begin{table}[h!]
\begin{center}
\begin{tabular}{rrrr}
\hline \hline
$Re_{\delta_0}$     & {$L_0/\delta_0$}   & {$\left( \frac{L_0}{\delta_0} \right) \, Re_{\delta_0}^{1/3}$}   &  Error ($\%$) \\ \hline
$    740,000$ & 2.38           &   216.74   &  3.53   \\ 
$1,694,000$ & 1.98           &   236.34   &  5.19     \\ 
$2,120,000$ & 1.70           &   218.15   &  2.90    \\ 
$2,912,000$ & 1.63           &   233.42   &  3.90     \\ 
\hline \hline
\end{tabular}
\end{center}
\caption{Compression corner, $M=2.85$, $\alpha = 24^\circ$. 
Summary of the interaction lengths obtained at different values of Reynolds number.  These values are compared to the empirical correlation found by Settles  $0.9 e^{0.23 \alpha}$  \cite{delery1986shock} }.  
\label{tab:settlesreynolds}
\end{table}

\clearpage

\section{Conclusions}
\label{sec:conclusions}





%
The turbulent gas-kinetic scheme models the effects of unresolved scales through a turbulent relaxation time, 
generating dissipation and kinetic effects.
Numerical experiments have so far provided encouraging results, showing that in the selected 2D flow cases, the turbulent gas-kinetic scheme systematically provides predictions in terms of pressure and viscous stresses distributions which are more in line with experiments than conventional schemes. 
Moreover the shock structures predicted by the gas-kinetic scheme seem to be consistent with the sketches reconstructed from experimental results by Settles {\sl et al.} \cite{settles1979detailed} and D\'elery {\sl et al.} \cite{delery1983experimental,delery1986shock}.
Additional validation is of course necessary to consolidate these findings, but it is significant that the gas-kinetic scheme does deviate from the conventional scheme even when using the same turbulence model.

This work is a first, exploratory step - limited to the steady RANS approach and 2D geometries, focusing mainly on the ability of a gas-kinetic scheme to exploit kinetic effects in turbulence. 
Additional investigations may concern a broader set of benchmark cases, and numerical approaches (Finite-Elements, spectral methods, different preconditioning), additional types of allied turbulence model (other two-equation models, algebraic stress models). 
Additionally, the truncation level in the Chapman-Enskog expansion for $f_0$, Eq. \ref{eq:initialf0} can be raised to the second level as in Eq. \ref{eq:chapman-enskog-burnett}, in order to add a  non-linear component to the assumed turbulent stress tensor. 
Finally, the gas-kinetic scheme could be implemented in unsteady simulations, following the (U)RANS or LES approaches. The application to LES would even be more meaningful than in RANS, as the gas-kinetic scheme activates kinetic effects in a resolution-dependent way. 

Gas-kinetic schemes still are computationally more expensive than conventional schemes, mainly because of the evaluation of the moments of distribution functions. 
Xuan {\sl et al.} \cite{xuan2012new} and Luo {\sl et al.} \cite{luo2013comparison} have recently proposed a new approach resulting in a much higher computational efficiency which could be exploited in turbulent simulations as well. 
The size of the time step, hence the time-stepping technique, may have a strong influence on the way gas-kinetic scheme works;  
in this work, this issue has driven the replacement of $\tau/\Delta t$ with $\tau/\widehat{\tau}$).


Finally, a gas-kinetic scheme could be developed in order to keep into account the dependence of $\tau_t$ on the microscopic degrees of freedom - in the spirit of the studies conducted by Succi et al. \cite{succi2002towards}, which would provide {\sl multiscale} modeling with an infinite number of scales.  

\bibliography{../bibliobgk}

\end{document}

%% file: cpcase9_.tex
\begingroup
  \fontfamily{Helvetica}%
  \selectfont
  \makeatletter
  \providecommand\color[2][]{%
    \GenericError{(gnuplot) \space\space\space\@spaces}{%
      Package color not loaded in conjunction with
      terminal option `colourtext'%
    }{See the gnuplot documentation for explanation.%
    }{Either use 'blacktext' in gnuplot or load the package
      color.sty in LaTeX.}%
    \renewcommand\color[2][]{}%
  }%
  \providecommand\includegraphics[2][]{%
    \GenericError{(gnuplot) \space\space\space\@spaces}{%
      Package graphicx or graphics not loaded%
    }{See the gnuplot documentation for explanation.%
    }{The gnuplot epslatex terminal needs graphicx.sty or graphics.sty.}%
    \renewcommand\includegraphics[2][]{}%
  }%
  \providecommand\rotatebox[2]{#2}%
  \@ifundefined{ifGPcolor}{%
    \newif\ifGPcolor
    \GPcolorfalse
  }{}%
  \@ifundefined{ifGPblacktext}{%
    \newif\ifGPblacktext
    \GPblacktexttrue
  }{}%
  \let\gplgaddtomacro\g@addto@macro
  \gdef\gplbacktext{}%
  \gdef\gplfronttext{}%
  \makeatother
  \ifGPblacktext
    \def\colorrgb#1{}%
    \def\colorgray#1{}%
  \else
    \ifGPcolor
      \def\colorrgb#1{\color[rgb]{#1}}%
      \def\colorgray#1{\color[gray]{#1}}%
      \expandafter\def\csname LTw\endcsname{\color{white}}%
      \expandafter\def\csname LTb\endcsname{\color{black}}%
      \expandafter\def\csname LTa\endcsname{\color{black}}%
      \expandafter\def\csname LT0\endcsname{\color[rgb]{1,0,0}}%
      \expandafter\def\csname LT1\endcsname{\color[rgb]{0,1,0}}%
      \expandafter\def\csname LT2\endcsname{\color[rgb]{0,0,1}}%
      \expandafter\def\csname LT3\endcsname{\color[rgb]{1,0,1}}%
      \expandafter\def\csname LT4\endcsname{\color[rgb]{0,1,1}}%
      \expandafter\def\csname LT5\endcsname{\color[rgb]{1,1,0}}%
      \expandafter\def\csname LT6\endcsname{\color[rgb]{0,0,0}}%
      \expandafter\def\csname LT7\endcsname{\color[rgb]{1,0.3,0}}%
      \expandafter\def\csname LT8\endcsname{\color[rgb]{0.5,0.5,0.5}}%
    \else
      \def\colorrgb#1{\color{black}}%
      \def\colorgray#1{\color[gray]{#1}}%
      \expandafter\def\csname LTw\endcsname{\color{white}}%
      \expandafter\def\csname LTb\endcsname{\color{black}}%
      \expandafter\def\csname LTa\endcsname{\color{black}}%
      \expandafter\def\csname LT0\endcsname{\color{black}}%
      \expandafter\def\csname LT1\endcsname{\color{black}}%
      \expandafter\def\csname LT2\endcsname{\color{black}}%
      \expandafter\def\csname LT3\endcsname{\color{black}}%
      \expandafter\def\csname LT4\endcsname{\color{black}}%
      \expandafter\def\csname LT5\endcsname{\color{black}}%
      \expandafter\def\csname LT6\endcsname{\color{black}}%
      \expandafter\def\csname LT7\endcsname{\color{black}}%
      \expandafter\def\csname LT8\endcsname{\color{black}}%
    \fi
  \fi
  \setlength{\unitlength}{0.0500bp}%
  \begin{picture}(4422.00,3628.00)%
    \gplgaddtomacro\gplbacktext{%
      \csname LTb\endcsname%
      \put(912,512){\makebox(0,0)[r]{\strut{}$-1.5$}}%
      \put(912,999){\makebox(0,0)[r]{\strut{}$-1.0$}}%
      \put(912,1487){\makebox(0,0)[r]{\strut{}$-0.5$}}%
      \put(912,1974){\makebox(0,0)[r]{\strut{}$0.0$}}%
      \put(912,2461){\makebox(0,0)[r]{\strut{}$0.5$}}%
      \put(912,2949){\makebox(0,0)[r]{\strut{}$1.0$}}%
      \put(912,3436){\makebox(0,0)[r]{\strut{}$1.5$}}%
      \put(1008,352){\makebox(0,0){\strut{}0}}%
      \put(1802,352){\makebox(0,0){\strut{}$0.25$}}%
      \put(2595,352){\makebox(0,0){\strut{}$0.50$}}%
      \put(3389,352){\makebox(0,0){\strut{}$0.75$}}%
      \put(4182,352){\makebox(0,0){\strut{}1}}%
      \put(352,1974){\rotatebox{90}{\makebox(0,0){\strut{}$-C_p$}}}%
      \put(2595,112){\makebox(0,0){\strut{}$x/c$}}%
      \put(3547,3192){\makebox(0,0)[l]{\strut{}$(a)$}}%
    }%
    \gplgaddtomacro\gplfronttext{%
    }%
    \gplbacktext
    \put(0,0){\includegraphics{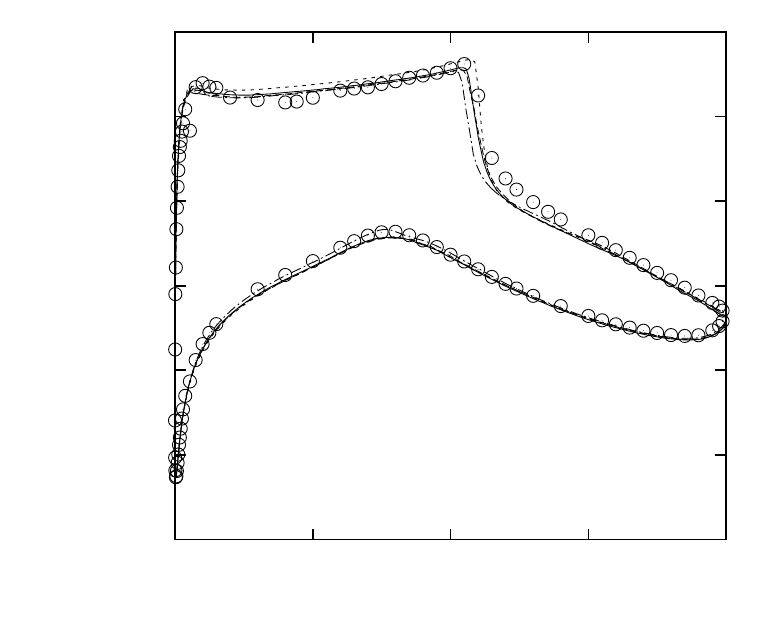}}%
    \gplfronttext
  \end{picture}%
\endgroup

%% file: cfcase9_.tex
\begingroup
  \fontfamily{Helvetica}%
  \selectfont
  \makeatletter
  \providecommand\color[2][]{%
    \GenericError{(gnuplot) \space\space\space\@spaces}{%
      Package color not loaded in conjunction with
      terminal option `colourtext'%
    }{See the gnuplot documentation for explanation.%
    }{Either use 'blacktext' in gnuplot or load the package
      color.sty in LaTeX.}%
    \renewcommand\color[2][]{}%
  }%
  \providecommand\includegraphics[2][]{%
    \GenericError{(gnuplot) \space\space\space\@spaces}{%
      Package graphicx or graphics not loaded%
    }{See the gnuplot documentation for explanation.%
    }{The gnuplot epslatex terminal needs graphicx.sty or graphics.sty.}%
    \renewcommand\includegraphics[2][]{}%
  }%
  \providecommand\rotatebox[2]{#2}%
  \@ifundefined{ifGPcolor}{%
    \newif\ifGPcolor
    \GPcolorfalse
  }{}%
  \@ifundefined{ifGPblacktext}{%
    \newif\ifGPblacktext
    \GPblacktexttrue
  }{}%
  \let\gplgaddtomacro\g@addto@macro
  \gdef\gplbacktext{}%
  \gdef\gplfronttext{}%
  \makeatother
  \ifGPblacktext
    \def\colorrgb#1{}%
    \def\colorgray#1{}%
  \else
    \ifGPcolor
      \def\colorrgb#1{\color[rgb]{#1}}%
      \def\colorgray#1{\color[gray]{#1}}%
      \expandafter\def\csname LTw\endcsname{\color{white}}%
      \expandafter\def\csname LTb\endcsname{\color{black}}%
      \expandafter\def\csname LTa\endcsname{\color{black}}%
      \expandafter\def\csname LT0\endcsname{\color[rgb]{1,0,0}}%
      \expandafter\def\csname LT1\endcsname{\color[rgb]{0,1,0}}%
      \expandafter\def\csname LT2\endcsname{\color[rgb]{0,0,1}}%
      \expandafter\def\csname LT3\endcsname{\color[rgb]{1,0,1}}%
      \expandafter\def\csname LT4\endcsname{\color[rgb]{0,1,1}}%
      \expandafter\def\csname LT5\endcsname{\color[rgb]{1,1,0}}%
      \expandafter\def\csname LT6\endcsname{\color[rgb]{0,0,0}}%
      \expandafter\def\csname LT7\endcsname{\color[rgb]{1,0.3,0}}%
      \expandafter\def\csname LT8\endcsname{\color[rgb]{0.5,0.5,0.5}}%
    \else
      \def\colorrgb#1{\color{black}}%
      \def\colorgray#1{\color[gray]{#1}}%
      \expandafter\def\csname LTw\endcsname{\color{white}}%
      \expandafter\def\csname LTb\endcsname{\color{black}}%
      \expandafter\def\csname LTa\endcsname{\color{black}}%
      \expandafter\def\csname LT0\endcsname{\color{black}}%
      \expandafter\def\csname LT1\endcsname{\color{black}}%
      \expandafter\def\csname LT2\endcsname{\color{black}}%
      \expandafter\def\csname LT3\endcsname{\color{black}}%
      \expandafter\def\csname LT4\endcsname{\color{black}}%
      \expandafter\def\csname LT5\endcsname{\color{black}}%
      \expandafter\def\csname LT6\endcsname{\color{black}}%
      \expandafter\def\csname LT7\endcsname{\color{black}}%
      \expandafter\def\csname LT8\endcsname{\color{black}}%
    \fi
  \fi
  \setlength{\unitlength}{0.0500bp}%
  \begin{picture}(4422.00,3628.00)%
    \gplgaddtomacro\gplbacktext{%
      \csname LTb\endcsname%
      \put(880,512){\makebox(0,0)[r]{\strut{}$-6$}}%
      \put(880,837){\makebox(0,0)[r]{\strut{}$-4$}}%
      \put(880,1162){\makebox(0,0)[r]{\strut{}$-2$}}%
      \put(880,1487){\makebox(0,0)[r]{\strut{}$0$}}%
      \put(880,1812){\makebox(0,0)[r]{\strut{}$2$}}%
      \put(880,2136){\makebox(0,0)[r]{\strut{}$4$}}%
      \put(880,2461){\makebox(0,0)[r]{\strut{}$6$}}%
      \put(880,2786){\makebox(0,0)[r]{\strut{}$8$}}%
      \put(880,3111){\makebox(0,0)[r]{\strut{}$10$}}%
      \put(880,3436){\makebox(0,0)[r]{\strut{}$12$}}%
      \put(976,352){\makebox(0,0){\strut{}0}}%
      \put(1778,352){\makebox(0,0){\strut{}$0.25$}}%
      \put(2579,352){\makebox(0,0){\strut{}$0.50$}}%
      \put(3381,352){\makebox(0,0){\strut{}$0.75$}}%
      \put(4182,352){\makebox(0,0){\strut{}1}}%
      \put(320,1974){\rotatebox{90}{\makebox(0,0){\strut{}$C_f \, \times \, 10^3$}}}%
      \put(2579,112){\makebox(0,0){\strut{}$x/c$}}%
      \put(3541,3192){\makebox(0,0)[l]{\strut{}$(b)$}}%
    }%
    \gplgaddtomacro\gplfronttext{%
    }%
    \gplbacktext
    \put(0,0){\includegraphics{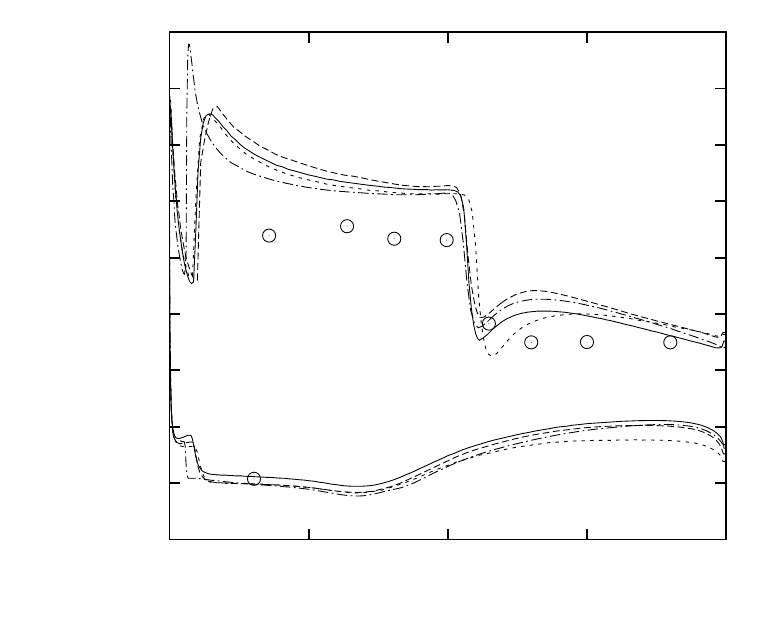}}%
    \gplfronttext
  \end{picture}%
\endgroup

%% file: cpcase10_.tex
\begingroup
  \fontfamily{Helvetica}%
  \selectfont
  \makeatletter
  \providecommand\color[2][]{%
    \GenericError{(gnuplot) \space\space\space\@spaces}{%
      Package color not loaded in conjunction with
      terminal option `colourtext'%
    }{See the gnuplot documentation for explanation.%
    }{Either use 'blacktext' in gnuplot or load the package
      color.sty in LaTeX.}%
    \renewcommand\color[2][]{}%
  }%
  \providecommand\includegraphics[2][]{%
    \GenericError{(gnuplot) \space\space\space\@spaces}{%
      Package graphicx or graphics not loaded%
    }{See the gnuplot documentation for explanation.%
    }{The gnuplot epslatex terminal needs graphicx.sty or graphics.sty.}%
    \renewcommand\includegraphics[2][]{}%
  }%
  \providecommand\rotatebox[2]{#2}%
  \@ifundefined{ifGPcolor}{%
    \newif\ifGPcolor
    \GPcolorfalse
  }{}%
  \@ifundefined{ifGPblacktext}{%
    \newif\ifGPblacktext
    \GPblacktexttrue
  }{}%
  \let\gplgaddtomacro\g@addto@macro
  \gdef\gplbacktext{}%
  \gdef\gplfronttext{}%
  \makeatother
  \ifGPblacktext
    \def\colorrgb#1{}%
    \def\colorgray#1{}%
  \else
    \ifGPcolor
      \def\colorrgb#1{\color[rgb]{#1}}%
      \def\colorgray#1{\color[gray]{#1}}%
      \expandafter\def\csname LTw\endcsname{\color{white}}%
      \expandafter\def\csname LTb\endcsname{\color{black}}%
      \expandafter\def\csname LTa\endcsname{\color{black}}%
      \expandafter\def\csname LT0\endcsname{\color[rgb]{1,0,0}}%
      \expandafter\def\csname LT1\endcsname{\color[rgb]{0,1,0}}%
      \expandafter\def\csname LT2\endcsname{\color[rgb]{0,0,1}}%
      \expandafter\def\csname LT3\endcsname{\color[rgb]{1,0,1}}%
      \expandafter\def\csname LT4\endcsname{\color[rgb]{0,1,1}}%
      \expandafter\def\csname LT5\endcsname{\color[rgb]{1,1,0}}%
      \expandafter\def\csname LT6\endcsname{\color[rgb]{0,0,0}}%
      \expandafter\def\csname LT7\endcsname{\color[rgb]{1,0.3,0}}%
      \expandafter\def\csname LT8\endcsname{\color[rgb]{0.5,0.5,0.5}}%
    \else
      \def\colorrgb#1{\color{black}}%
      \def\colorgray#1{\color[gray]{#1}}%
      \expandafter\def\csname LTw\endcsname{\color{white}}%
      \expandafter\def\csname LTb\endcsname{\color{black}}%
      \expandafter\def\csname LTa\endcsname{\color{black}}%
      \expandafter\def\csname LT0\endcsname{\color{black}}%
      \expandafter\def\csname LT1\endcsname{\color{black}}%
      \expandafter\def\csname LT2\endcsname{\color{black}}%
      \expandafter\def\csname LT3\endcsname{\color{black}}%
      \expandafter\def\csname LT4\endcsname{\color{black}}%
      \expandafter\def\csname LT5\endcsname{\color{black}}%
      \expandafter\def\csname LT6\endcsname{\color{black}}%
      \expandafter\def\csname LT7\endcsname{\color{black}}%
      \expandafter\def\csname LT8\endcsname{\color{black}}%
    \fi
  \fi
  \setlength{\unitlength}{0.0500bp}%
  \begin{picture}(4422.00,3628.00)%
    \gplgaddtomacro\gplbacktext{%
      \csname LTb\endcsname%
      \put(912,512){\makebox(0,0)[r]{\strut{}$-1.5$}}%
      \put(912,999){\makebox(0,0)[r]{\strut{}$-1.0$}}%
      \put(912,1487){\makebox(0,0)[r]{\strut{}$-0.5$}}%
      \put(912,1974){\makebox(0,0)[r]{\strut{}$0.0$}}%
      \put(912,2461){\makebox(0,0)[r]{\strut{}$0.5$}}%
      \put(912,2949){\makebox(0,0)[r]{\strut{}$1.0$}}%
      \put(912,3436){\makebox(0,0)[r]{\strut{}$1.5$}}%
      \put(1008,352){\makebox(0,0){\strut{}0}}%
      \put(1802,352){\makebox(0,0){\strut{}$0.25$}}%
      \put(2595,352){\makebox(0,0){\strut{}$0.50$}}%
      \put(3389,352){\makebox(0,0){\strut{}$0.75$}}%
      \put(4182,352){\makebox(0,0){\strut{}1}}%
      \put(352,1974){\rotatebox{90}{\makebox(0,0){\strut{}$-C_p$}}}%
      \put(2595,112){\makebox(0,0){\strut{}$x/c$}}%
      \put(3547,3192){\makebox(0,0)[l]{\strut{}$(a)$}}%
    }%
    \gplgaddtomacro\gplfronttext{%
    }%
    \gplbacktext
    \put(0,0){\includegraphics{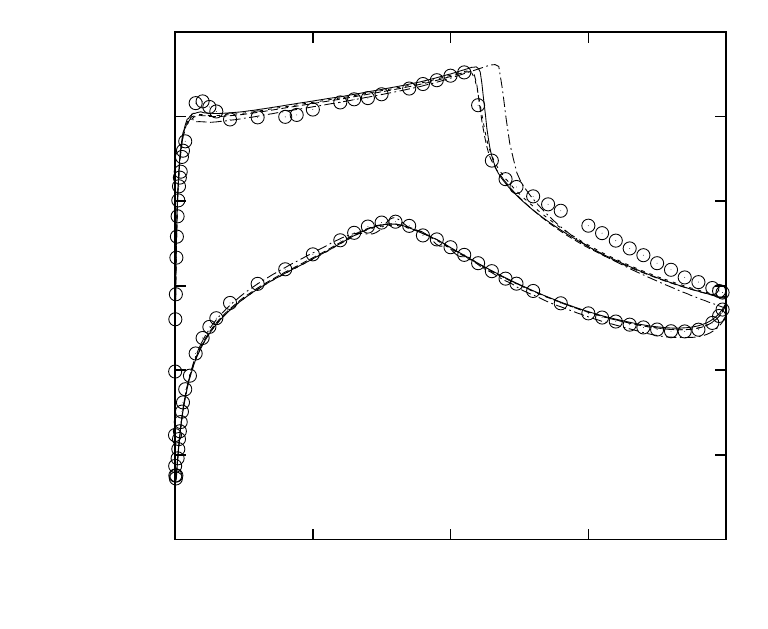}}%
    \gplfronttext
  \end{picture}%
\endgroup

%% file: cfcase10_.tex
\begingroup
  \fontfamily{Helvetica}%
  \selectfont
  \makeatletter
  \providecommand\color[2][]{%
    \GenericError{(gnuplot) \space\space\space\@spaces}{%
      Package color not loaded in conjunction with
      terminal option `colourtext'%
    }{See the gnuplot documentation for explanation.%
    }{Either use 'blacktext' in gnuplot or load the package
      color.sty in LaTeX.}%
    \renewcommand\color[2][]{}%
  }%
  \providecommand\includegraphics[2][]{%
    \GenericError{(gnuplot) \space\space\space\@spaces}{%
      Package graphicx or graphics not loaded%
    }{See the gnuplot documentation for explanation.%
    }{The gnuplot epslatex terminal needs graphicx.sty or graphics.sty.}%
    \renewcommand\includegraphics[2][]{}%
  }%
  \providecommand\rotatebox[2]{#2}%
  \@ifundefined{ifGPcolor}{%
    \newif\ifGPcolor
    \GPcolorfalse
  }{}%
  \@ifundefined{ifGPblacktext}{%
    \newif\ifGPblacktext
    \GPblacktexttrue
  }{}%
  \let\gplgaddtomacro\g@addto@macro
  \gdef\gplbacktext{}%
  \gdef\gplfronttext{}%
  \makeatother
  \ifGPblacktext
    \def\colorrgb#1{}%
    \def\colorgray#1{}%
  \else
    \ifGPcolor
      \def\colorrgb#1{\color[rgb]{#1}}%
      \def\colorgray#1{\color[gray]{#1}}%
      \expandafter\def\csname LTw\endcsname{\color{white}}%
      \expandafter\def\csname LTb\endcsname{\color{black}}%
      \expandafter\def\csname LTa\endcsname{\color{black}}%
      \expandafter\def\csname LT0\endcsname{\color[rgb]{1,0,0}}%
      \expandafter\def\csname LT1\endcsname{\color[rgb]{0,1,0}}%
      \expandafter\def\csname LT2\endcsname{\color[rgb]{0,0,1}}%
      \expandafter\def\csname LT3\endcsname{\color[rgb]{1,0,1}}%
      \expandafter\def\csname LT4\endcsname{\color[rgb]{0,1,1}}%
      \expandafter\def\csname LT5\endcsname{\color[rgb]{1,1,0}}%
      \expandafter\def\csname LT6\endcsname{\color[rgb]{0,0,0}}%
      \expandafter\def\csname LT7\endcsname{\color[rgb]{1,0.3,0}}%
      \expandafter\def\csname LT8\endcsname{\color[rgb]{0.5,0.5,0.5}}%
    \else
      \def\colorrgb#1{\color{black}}%
      \def\colorgray#1{\color[gray]{#1}}%
      \expandafter\def\csname LTw\endcsname{\color{white}}%
      \expandafter\def\csname LTb\endcsname{\color{black}}%
      \expandafter\def\csname LTa\endcsname{\color{black}}%
      \expandafter\def\csname LT0\endcsname{\color{black}}%
      \expandafter\def\csname LT1\endcsname{\color{black}}%
      \expandafter\def\csname LT2\endcsname{\color{black}}%
      \expandafter\def\csname LT3\endcsname{\color{black}}%
      \expandafter\def\csname LT4\endcsname{\color{black}}%
      \expandafter\def\csname LT5\endcsname{\color{black}}%
      \expandafter\def\csname LT6\endcsname{\color{black}}%
      \expandafter\def\csname LT7\endcsname{\color{black}}%
      \expandafter\def\csname LT8\endcsname{\color{black}}%
    \fi
  \fi
  \setlength{\unitlength}{0.0500bp}%
  \begin{picture}(4422.00,3628.00)%
    \gplgaddtomacro\gplbacktext{%
      \csname LTb\endcsname%
      \put(880,512){\makebox(0,0)[r]{\strut{}$-6$}}%
      \put(880,837){\makebox(0,0)[r]{\strut{}$-4$}}%
      \put(880,1162){\makebox(0,0)[r]{\strut{}$-2$}}%
      \put(880,1487){\makebox(0,0)[r]{\strut{}$0$}}%
      \put(880,1812){\makebox(0,0)[r]{\strut{}$2$}}%
      \put(880,2136){\makebox(0,0)[r]{\strut{}$4$}}%
      \put(880,2461){\makebox(0,0)[r]{\strut{}$6$}}%
      \put(880,2786){\makebox(0,0)[r]{\strut{}$8$}}%
      \put(880,3111){\makebox(0,0)[r]{\strut{}$10$}}%
      \put(880,3436){\makebox(0,0)[r]{\strut{}$12$}}%
      \put(976,352){\makebox(0,0){\strut{}0}}%
      \put(1778,352){\makebox(0,0){\strut{}$0.25$}}%
      \put(2579,352){\makebox(0,0){\strut{}$0.50$}}%
      \put(3381,352){\makebox(0,0){\strut{}$0.75$}}%
      \put(4182,352){\makebox(0,0){\strut{}1}}%
      \put(320,1974){\rotatebox{90}{\makebox(0,0){\strut{}$C_f \, \times \, 10^3$}}}%
      \put(2579,112){\makebox(0,0){\strut{}$x/c$}}%
      \put(3541,3192){\makebox(0,0)[l]{\strut{}$(b)$}}%
    }%
    \gplgaddtomacro\gplfronttext{%
    }%
    \gplbacktext
    \put(0,0){\includegraphics{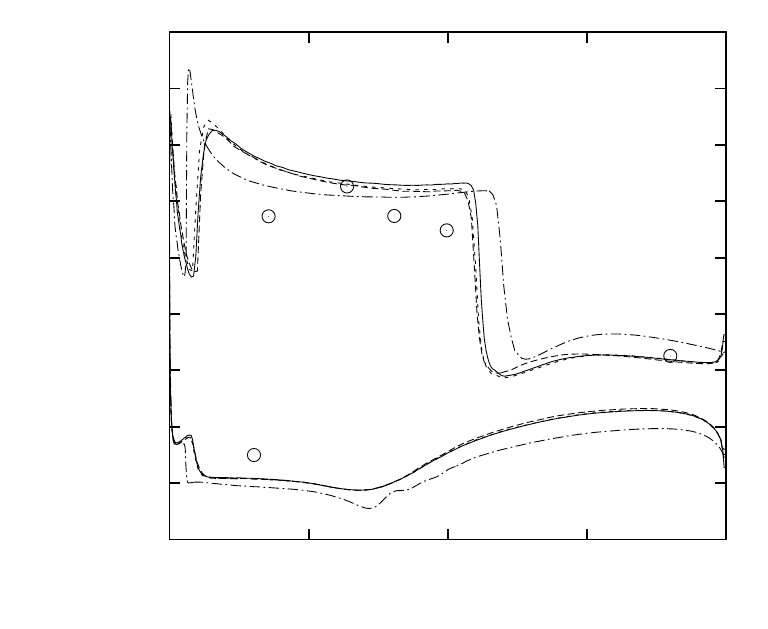}}%
    \gplfronttext
  \end{picture}%
\endgroup

%% file: uprofile40_.tex
\begingroup
  \fontfamily{Helvetica}%
  \selectfont
  \makeatletter
  \providecommand\color[2][]{%
    \GenericError{(gnuplot) \space\space\space\@spaces}{%
      Package color not loaded in conjunction with
      terminal option `colourtext'%
    }{See the gnuplot documentation for explanation.%
    }{Either use 'blacktext' in gnuplot or load the package
      color.sty in LaTeX.}%
    \renewcommand\color[2][]{}%
  }%
  \providecommand\includegraphics[2][]{%
    \GenericError{(gnuplot) \space\space\space\@spaces}{%
      Package graphicx or graphics not loaded%
    }{See the gnuplot documentation for explanation.%
    }{The gnuplot epslatex terminal needs graphicx.sty or graphics.sty.}%
    \renewcommand\includegraphics[2][]{}%
  }%
  \providecommand\rotatebox[2]{#2}%
  \@ifundefined{ifGPcolor}{%
    \newif\ifGPcolor
    \GPcolorfalse
  }{}%
  \@ifundefined{ifGPblacktext}{%
    \newif\ifGPblacktext
    \GPblacktexttrue
  }{}%
  \let\gplgaddtomacro\g@addto@macro
  \gdef\gplbacktext{}%
  \gdef\gplfronttext{}%
  \makeatother
  \ifGPblacktext
    \def\colorrgb#1{}%
    \def\colorgray#1{}%
  \else
    \ifGPcolor
      \def\colorrgb#1{\color[rgb]{#1}}%
      \def\colorgray#1{\color[gray]{#1}}%
      \expandafter\def\csname LTw\endcsname{\color{white}}%
      \expandafter\def\csname LTb\endcsname{\color{black}}%
      \expandafter\def\csname LTa\endcsname{\color{black}}%
      \expandafter\def\csname LT0\endcsname{\color[rgb]{1,0,0}}%
      \expandafter\def\csname LT1\endcsname{\color[rgb]{0,1,0}}%
      \expandafter\def\csname LT2\endcsname{\color[rgb]{0,0,1}}%
      \expandafter\def\csname LT3\endcsname{\color[rgb]{1,0,1}}%
      \expandafter\def\csname LT4\endcsname{\color[rgb]{0,1,1}}%
      \expandafter\def\csname LT5\endcsname{\color[rgb]{1,1,0}}%
      \expandafter\def\csname LT6\endcsname{\color[rgb]{0,0,0}}%
      \expandafter\def\csname LT7\endcsname{\color[rgb]{1,0.3,0}}%
      \expandafter\def\csname LT8\endcsname{\color[rgb]{0.5,0.5,0.5}}%
    \else
      \def\colorrgb#1{\color{black}}%
      \def\colorgray#1{\color[gray]{#1}}%
      \expandafter\def\csname LTw\endcsname{\color{white}}%
      \expandafter\def\csname LTb\endcsname{\color{black}}%
      \expandafter\def\csname LTa\endcsname{\color{black}}%
      \expandafter\def\csname LT0\endcsname{\color{black}}%
      \expandafter\def\csname LT1\endcsname{\color{black}}%
      \expandafter\def\csname LT2\endcsname{\color{black}}%
      \expandafter\def\csname LT3\endcsname{\color{black}}%
      \expandafter\def\csname LT4\endcsname{\color{black}}%
      \expandafter\def\csname LT5\endcsname{\color{black}}%
      \expandafter\def\csname LT6\endcsname{\color{black}}%
      \expandafter\def\csname LT7\endcsname{\color{black}}%
      \expandafter\def\csname LT8\endcsname{\color{black}}%
    \fi
  \fi
  \setlength{\unitlength}{0.0500bp}%
  \begin{picture}(3060.00,3628.00)%
    \gplgaddtomacro\gplbacktext{%
      \csname LTb\endcsname%
      \put(780,640){\makebox(0,0)[r]{\strut{}$0$}}%
      \put(780,1033){\makebox(0,0)[r]{\strut{}$1$}}%
      \put(780,1425){\makebox(0,0)[r]{\strut{}$2$}}%
      \put(780,1818){\makebox(0,0)[r]{\strut{}$3$}}%
      \put(780,2210){\makebox(0,0)[r]{\strut{}$4$}}%
      \put(780,2603){\makebox(0,0)[r]{\strut{}$5$}}%
      \put(780,2995){\makebox(0,0)[r]{\strut{}$6$}}%
      \put(780,3388){\makebox(0,0)[r]{\strut{}$7$}}%
      \put(900,440){\makebox(0,0){\strut{}0}}%
      \put(1830,440){\makebox(0,0){\strut{}$0.5$}}%
      \put(2760,440){\makebox(0,0){\strut{}1}}%
      \put(440,2014){\rotatebox{90}{\makebox(0,0){\strut{}$y/c \, \times \ 10^3$}}}%
      \put(1830,140){\makebox(0,0){\strut{}$u/U$}}%
      \put(1086,2995){\makebox(0,0)[l]{\strut{}x/c = 0.40}}%
    }%
    \gplgaddtomacro\gplfronttext{%
    }%
    \gplbacktext
    \put(0,0){\includegraphics{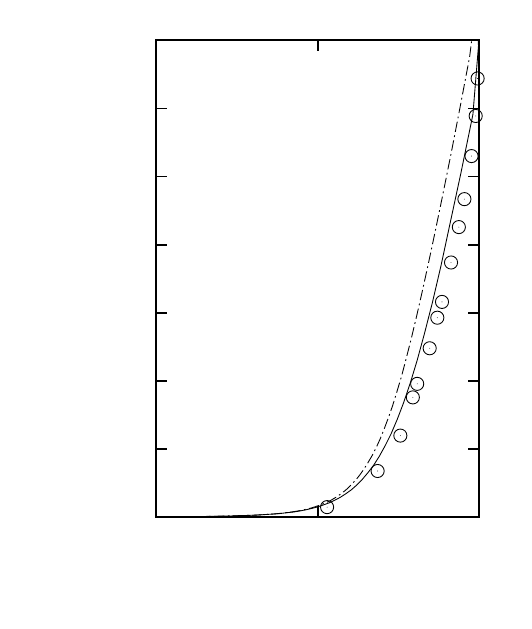}}%
    \gplfronttext
  \end{picture}%
\endgroup

%% file: uprofile65_.tex
\begingroup
  \fontfamily{Helvetica}%
  \selectfont
  \makeatletter
  \providecommand\color[2][]{%
    \GenericError{(gnuplot) \space\space\space\@spaces}{%
      Package color not loaded in conjunction with
      terminal option `colourtext'%
    }{See the gnuplot documentation for explanation.%
    }{Either use 'blacktext' in gnuplot or load the package
      color.sty in LaTeX.}%
    \renewcommand\color[2][]{}%
  }%
  \providecommand\includegraphics[2][]{%
    \GenericError{(gnuplot) \space\space\space\@spaces}{%
      Package graphicx or graphics not loaded%
    }{See the gnuplot documentation for explanation.%
    }{The gnuplot epslatex terminal needs graphicx.sty or graphics.sty.}%
    \renewcommand\includegraphics[2][]{}%
  }%
  \providecommand\rotatebox[2]{#2}%
  \@ifundefined{ifGPcolor}{%
    \newif\ifGPcolor
    \GPcolorfalse
  }{}%
  \@ifundefined{ifGPblacktext}{%
    \newif\ifGPblacktext
    \GPblacktexttrue
  }{}%
  \let\gplgaddtomacro\g@addto@macro
  \gdef\gplbacktext{}%
  \gdef\gplfronttext{}%
  \makeatother
  \ifGPblacktext
    \def\colorrgb#1{}%
    \def\colorgray#1{}%
  \else
    \ifGPcolor
      \def\colorrgb#1{\color[rgb]{#1}}%
      \def\colorgray#1{\color[gray]{#1}}%
      \expandafter\def\csname LTw\endcsname{\color{white}}%
      \expandafter\def\csname LTb\endcsname{\color{black}}%
      \expandafter\def\csname LTa\endcsname{\color{black}}%
      \expandafter\def\csname LT0\endcsname{\color[rgb]{1,0,0}}%
      \expandafter\def\csname LT1\endcsname{\color[rgb]{0,1,0}}%
      \expandafter\def\csname LT2\endcsname{\color[rgb]{0,0,1}}%
      \expandafter\def\csname LT3\endcsname{\color[rgb]{1,0,1}}%
      \expandafter\def\csname LT4\endcsname{\color[rgb]{0,1,1}}%
      \expandafter\def\csname LT5\endcsname{\color[rgb]{1,1,0}}%
      \expandafter\def\csname LT6\endcsname{\color[rgb]{0,0,0}}%
      \expandafter\def\csname LT7\endcsname{\color[rgb]{1,0.3,0}}%
      \expandafter\def\csname LT8\endcsname{\color[rgb]{0.5,0.5,0.5}}%
    \else
      \def\colorrgb#1{\color{black}}%
      \def\colorgray#1{\color[gray]{#1}}%
      \expandafter\def\csname LTw\endcsname{\color{white}}%
      \expandafter\def\csname LTb\endcsname{\color{black}}%
      \expandafter\def\csname LTa\endcsname{\color{black}}%
      \expandafter\def\csname LT0\endcsname{\color{black}}%
      \expandafter\def\csname LT1\endcsname{\color{black}}%
      \expandafter\def\csname LT2\endcsname{\color{black}}%
      \expandafter\def\csname LT3\endcsname{\color{black}}%
      \expandafter\def\csname LT4\endcsname{\color{black}}%
      \expandafter\def\csname LT5\endcsname{\color{black}}%
      \expandafter\def\csname LT6\endcsname{\color{black}}%
      \expandafter\def\csname LT7\endcsname{\color{black}}%
      \expandafter\def\csname LT8\endcsname{\color{black}}%
    \fi
  \fi
  \setlength{\unitlength}{0.0500bp}%
  \begin{picture}(3060.00,3628.00)%
    \gplgaddtomacro\gplbacktext{%
      \csname LTb\endcsname%
      \put(900,640){\makebox(0,0)[r]{\strut{}$0$}}%
      \put(900,1098){\makebox(0,0)[r]{\strut{}$5$}}%
      \put(900,1556){\makebox(0,0)[r]{\strut{}$10$}}%
      \put(900,2014){\makebox(0,0)[r]{\strut{}$15$}}%
      \put(900,2472){\makebox(0,0)[r]{\strut{}$20$}}%
      \put(900,2930){\makebox(0,0)[r]{\strut{}$25$}}%
      \put(900,3388){\makebox(0,0)[r]{\strut{}$30$}}%
      \put(1020,440){\makebox(0,0){\strut{}0}}%
      \put(1890,440){\makebox(0,0){\strut{}$0.5$}}%
      \put(2760,440){\makebox(0,0){\strut{}1}}%
      \put(440,2014){\rotatebox{90}{\makebox(0,0){\strut{}$y/c \, \times \ 10^3$}}}%
      \put(1890,140){\makebox(0,0){\strut{}$u/U$}}%
      \put(1194,2930){\makebox(0,0)[l]{\strut{}x/c = 0.65}}%
    }%
    \gplgaddtomacro\gplfronttext{%
    }%
    \gplbacktext
    \put(0,0){\includegraphics{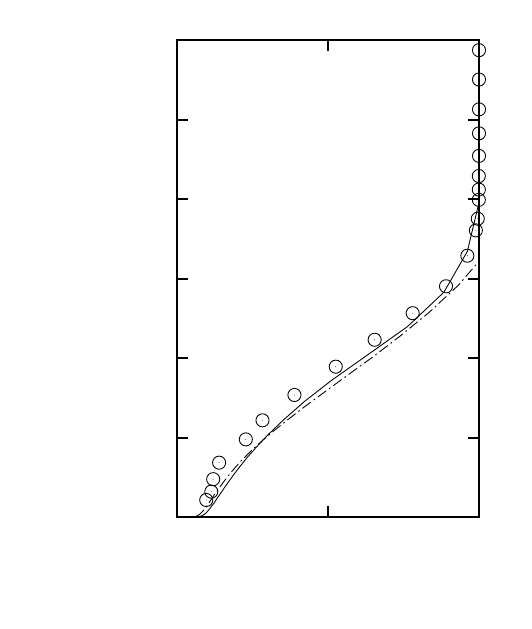}}%
    \gplfronttext
  \end{picture}%
\endgroup

%% file: uprofile90_.tex
\begingroup
  \fontfamily{Helvetica}%
  \selectfont
  \makeatletter
  \providecommand\color[2][]{%
    \GenericError{(gnuplot) \space\space\space\@spaces}{%
      Package color not loaded in conjunction with
      terminal option `colourtext'%
    }{See the gnuplot documentation for explanation.%
    }{Either use 'blacktext' in gnuplot or load the package
      color.sty in LaTeX.}%
    \renewcommand\color[2][]{}%
  }%
  \providecommand\includegraphics[2][]{%
    \GenericError{(gnuplot) \space\space\space\@spaces}{%
      Package graphicx or graphics not loaded%
    }{See the gnuplot documentation for explanation.%
    }{The gnuplot epslatex terminal needs graphicx.sty or graphics.sty.}%
    \renewcommand\includegraphics[2][]{}%
  }%
  \providecommand\rotatebox[2]{#2}%
  \@ifundefined{ifGPcolor}{%
    \newif\ifGPcolor
    \GPcolorfalse
  }{}%
  \@ifundefined{ifGPblacktext}{%
    \newif\ifGPblacktext
    \GPblacktexttrue
  }{}%
  \let\gplgaddtomacro\g@addto@macro
  \gdef\gplbacktext{}%
  \gdef\gplfronttext{}%
  \makeatother
  \ifGPblacktext
    \def\colorrgb#1{}%
    \def\colorgray#1{}%
  \else
    \ifGPcolor
      \def\colorrgb#1{\color[rgb]{#1}}%
      \def\colorgray#1{\color[gray]{#1}}%
      \expandafter\def\csname LTw\endcsname{\color{white}}%
      \expandafter\def\csname LTb\endcsname{\color{black}}%
      \expandafter\def\csname LTa\endcsname{\color{black}}%
      \expandafter\def\csname LT0\endcsname{\color[rgb]{1,0,0}}%
      \expandafter\def\csname LT1\endcsname{\color[rgb]{0,1,0}}%
      \expandafter\def\csname LT2\endcsname{\color[rgb]{0,0,1}}%
      \expandafter\def\csname LT3\endcsname{\color[rgb]{1,0,1}}%
      \expandafter\def\csname LT4\endcsname{\color[rgb]{0,1,1}}%
      \expandafter\def\csname LT5\endcsname{\color[rgb]{1,1,0}}%
      \expandafter\def\csname LT6\endcsname{\color[rgb]{0,0,0}}%
      \expandafter\def\csname LT7\endcsname{\color[rgb]{1,0.3,0}}%
      \expandafter\def\csname LT8\endcsname{\color[rgb]{0.5,0.5,0.5}}%
    \else
      \def\colorrgb#1{\color{black}}%
      \def\colorgray#1{\color[gray]{#1}}%
      \expandafter\def\csname LTw\endcsname{\color{white}}%
      \expandafter\def\csname LTb\endcsname{\color{black}}%
      \expandafter\def\csname LTa\endcsname{\color{black}}%
      \expandafter\def\csname LT0\endcsname{\color{black}}%
      \expandafter\def\csname LT1\endcsname{\color{black}}%
      \expandafter\def\csname LT2\endcsname{\color{black}}%
      \expandafter\def\csname LT3\endcsname{\color{black}}%
      \expandafter\def\csname LT4\endcsname{\color{black}}%
      \expandafter\def\csname LT5\endcsname{\color{black}}%
      \expandafter\def\csname LT6\endcsname{\color{black}}%
      \expandafter\def\csname LT7\endcsname{\color{black}}%
      \expandafter\def\csname LT8\endcsname{\color{black}}%
    \fi
  \fi
  \setlength{\unitlength}{0.0500bp}%
  \begin{picture}(3060.00,3628.00)%
    \gplgaddtomacro\gplbacktext{%
      \csname LTb\endcsname%
      \put(900,640){\makebox(0,0)[r]{\strut{}$0$}}%
      \put(900,1190){\makebox(0,0)[r]{\strut{}$10$}}%
      \put(900,1739){\makebox(0,0)[r]{\strut{}$20$}}%
      \put(900,2289){\makebox(0,0)[r]{\strut{}$30$}}%
      \put(900,2838){\makebox(0,0)[r]{\strut{}$40$}}%
      \put(900,3388){\makebox(0,0)[r]{\strut{}$50$}}%
      \put(1020,440){\makebox(0,0){\strut{}0}}%
      \put(1890,440){\makebox(0,0){\strut{}$0.5$}}%
      \put(2760,440){\makebox(0,0){\strut{}1}}%
      \put(440,2014){\rotatebox{90}{\makebox(0,0){\strut{}$y/c \, \times \ 10^3$}}}%
      \put(1890,140){\makebox(0,0){\strut{}$u/U$}}%
      \put(1194,2838){\makebox(0,0)[l]{\strut{}x/c = 0.90}}%
    }%
    \gplgaddtomacro\gplfronttext{%
    }%
    \gplbacktext
    \put(0,0){\includegraphics{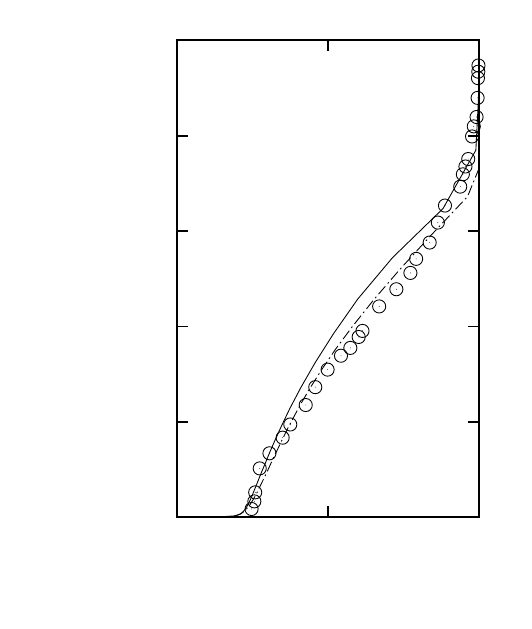}}%
    \gplfronttext
  \end{picture}%
\endgroup

%% file: cp0012m08_.tex
\begingroup
  \fontfamily{Helvetica}%
  \selectfont
  \makeatletter
  \providecommand\color[2][]{%
    \GenericError{(gnuplot) \space\space\space\@spaces}{%
      Package color not loaded in conjunction with
      terminal option `colourtext'%
    }{See the gnuplot documentation for explanation.%
    }{Either use 'blacktext' in gnuplot or load the package
      color.sty in LaTeX.}%
    \renewcommand\color[2][]{}%
  }%
  \providecommand\includegraphics[2][]{%
    \GenericError{(gnuplot) \space\space\space\@spaces}{%
      Package graphicx or graphics not loaded%
    }{See the gnuplot documentation for explanation.%
    }{The gnuplot epslatex terminal needs graphicx.sty or graphics.sty.}%
    \renewcommand\includegraphics[2][]{}%
  }%
  \providecommand\rotatebox[2]{#2}%
  \@ifundefined{ifGPcolor}{%
    \newif\ifGPcolor
    \GPcolorfalse
  }{}%
  \@ifundefined{ifGPblacktext}{%
    \newif\ifGPblacktext
    \GPblacktexttrue
  }{}%
  \let\gplgaddtomacro\g@addto@macro
  \gdef\gplbacktext{}%
  \gdef\gplfronttext{}%
  \makeatother
  \ifGPblacktext
    \def\colorrgb#1{}%
    \def\colorgray#1{}%
  \else
    \ifGPcolor
      \def\colorrgb#1{\color[rgb]{#1}}%
      \def\colorgray#1{\color[gray]{#1}}%
      \expandafter\def\csname LTw\endcsname{\color{white}}%
      \expandafter\def\csname LTb\endcsname{\color{black}}%
      \expandafter\def\csname LTa\endcsname{\color{black}}%
      \expandafter\def\csname LT0\endcsname{\color[rgb]{1,0,0}}%
      \expandafter\def\csname LT1\endcsname{\color[rgb]{0,1,0}}%
      \expandafter\def\csname LT2\endcsname{\color[rgb]{0,0,1}}%
      \expandafter\def\csname LT3\endcsname{\color[rgb]{1,0,1}}%
      \expandafter\def\csname LT4\endcsname{\color[rgb]{0,1,1}}%
      \expandafter\def\csname LT5\endcsname{\color[rgb]{1,1,0}}%
      \expandafter\def\csname LT6\endcsname{\color[rgb]{0,0,0}}%
      \expandafter\def\csname LT7\endcsname{\color[rgb]{1,0.3,0}}%
      \expandafter\def\csname LT8\endcsname{\color[rgb]{0.5,0.5,0.5}}%
    \else
      \def\colorrgb#1{\color{black}}%
      \def\colorgray#1{\color[gray]{#1}}%
      \expandafter\def\csname LTw\endcsname{\color{white}}%
      \expandafter\def\csname LTb\endcsname{\color{black}}%
      \expandafter\def\csname LTa\endcsname{\color{black}}%
      \expandafter\def\csname LT0\endcsname{\color{black}}%
      \expandafter\def\csname LT1\endcsname{\color{black}}%
      \expandafter\def\csname LT2\endcsname{\color{black}}%
      \expandafter\def\csname LT3\endcsname{\color{black}}%
      \expandafter\def\csname LT4\endcsname{\color{black}}%
      \expandafter\def\csname LT5\endcsname{\color{black}}%
      \expandafter\def\csname LT6\endcsname{\color{black}}%
      \expandafter\def\csname LT7\endcsname{\color{black}}%
      \expandafter\def\csname LT8\endcsname{\color{black}}%
    \fi
  \fi
  \setlength{\unitlength}{0.0500bp}%
  \begin{picture}(4422.00,3628.00)%
    \gplgaddtomacro\gplbacktext{%
      \csname LTb\endcsname%
      \put(1140,869){\makebox(0,0)[r]{\strut{}$-1.0$}}%
      \put(1140,1442){\makebox(0,0)[r]{\strut{}$-0.5$}}%
      \put(1140,2014){\makebox(0,0)[r]{\strut{}$0.0$}}%
      \put(1140,2587){\makebox(0,0)[r]{\strut{}$0.5$}}%
      \put(1140,3159){\makebox(0,0)[r]{\strut{}$1.0$}}%
      \put(1260,440){\makebox(0,0){\strut{}0}}%
      \put(1976,440){\makebox(0,0){\strut{}$0.25$}}%
      \put(2691,440){\makebox(0,0){\strut{}$0.50$}}%
      \put(3407,440){\makebox(0,0){\strut{}$0.75$}}%
      \put(4122,440){\makebox(0,0){\strut{}1}}%
      \put(440,2014){\rotatebox{90}{\makebox(0,0){\strut{}$-C_p$}}}%
      \put(2691,140){\makebox(0,0){\strut{}$x/c$}}%
      \put(3550,14037){\makebox(0,0)[l]{\strut{}$(b)$}}%
      \put(3550,3159){\makebox(0,0)[l]{\strut{}$(a)$}}%
    }%
    \gplgaddtomacro\gplfronttext{%
    }%
    \gplbacktext
    \put(0,0){\includegraphics{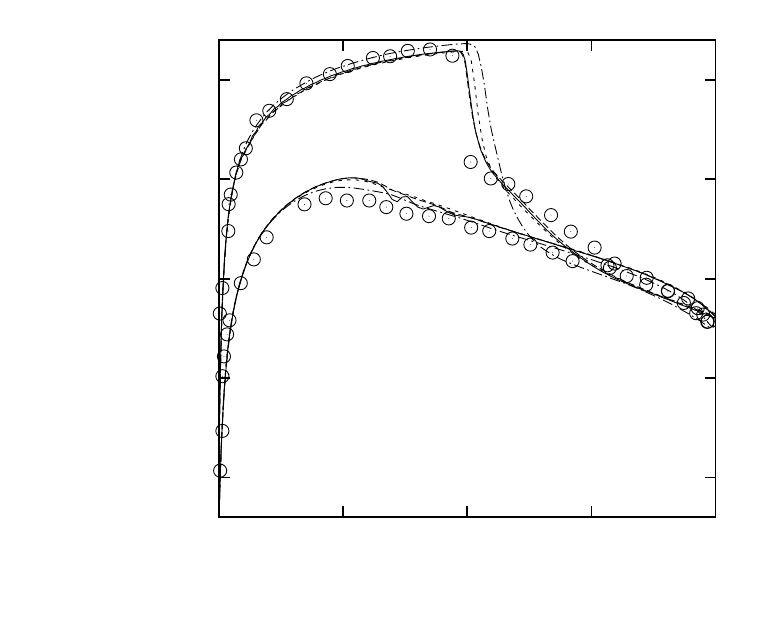}}%
    \gplfronttext
  \end{picture}%
\endgroup

%% file: cp0012m074_.tex
\begingroup
  \fontfamily{Helvetica}%
  \selectfont
  \makeatletter
  \providecommand\color[2][]{%
    \GenericError{(gnuplot) \space\space\space\@spaces}{%
      Package color not loaded in conjunction with
      terminal option `colourtext'%
    }{See the gnuplot documentation for explanation.%
    }{Either use 'blacktext' in gnuplot or load the package
      color.sty in LaTeX.}%
    \renewcommand\color[2][]{}%
  }%
  \providecommand\includegraphics[2][]{%
    \GenericError{(gnuplot) \space\space\space\@spaces}{%
      Package graphicx or graphics not loaded%
    }{See the gnuplot documentation for explanation.%
    }{The gnuplot epslatex terminal needs graphicx.sty or graphics.sty.}%
    \renewcommand\includegraphics[2][]{}%
  }%
  \providecommand\rotatebox[2]{#2}%
  \@ifundefined{ifGPcolor}{%
    \newif\ifGPcolor
    \GPcolorfalse
  }{}%
  \@ifundefined{ifGPblacktext}{%
    \newif\ifGPblacktext
    \GPblacktexttrue
  }{}%
  \let\gplgaddtomacro\g@addto@macro
  \gdef\gplbacktext{}%
  \gdef\gplfronttext{}%
  \makeatother
  \ifGPblacktext
    \def\colorrgb#1{}%
    \def\colorgray#1{}%
  \else
    \ifGPcolor
      \def\colorrgb#1{\color[rgb]{#1}}%
      \def\colorgray#1{\color[gray]{#1}}%
      \expandafter\def\csname LTw\endcsname{\color{white}}%
      \expandafter\def\csname LTb\endcsname{\color{black}}%
      \expandafter\def\csname LTa\endcsname{\color{black}}%
      \expandafter\def\csname LT0\endcsname{\color[rgb]{1,0,0}}%
      \expandafter\def\csname LT1\endcsname{\color[rgb]{0,1,0}}%
      \expandafter\def\csname LT2\endcsname{\color[rgb]{0,0,1}}%
      \expandafter\def\csname LT3\endcsname{\color[rgb]{1,0,1}}%
      \expandafter\def\csname LT4\endcsname{\color[rgb]{0,1,1}}%
      \expandafter\def\csname LT5\endcsname{\color[rgb]{1,1,0}}%
      \expandafter\def\csname LT6\endcsname{\color[rgb]{0,0,0}}%
      \expandafter\def\csname LT7\endcsname{\color[rgb]{1,0.3,0}}%
      \expandafter\def\csname LT8\endcsname{\color[rgb]{0.5,0.5,0.5}}%
    \else
      \def\colorrgb#1{\color{black}}%
      \def\colorgray#1{\color[gray]{#1}}%
      \expandafter\def\csname LTw\endcsname{\color{white}}%
      \expandafter\def\csname LTb\endcsname{\color{black}}%
      \expandafter\def\csname LTa\endcsname{\color{black}}%
      \expandafter\def\csname LT0\endcsname{\color{black}}%
      \expandafter\def\csname LT1\endcsname{\color{black}}%
      \expandafter\def\csname LT2\endcsname{\color{black}}%
      \expandafter\def\csname LT3\endcsname{\color{black}}%
      \expandafter\def\csname LT4\endcsname{\color{black}}%
      \expandafter\def\csname LT5\endcsname{\color{black}}%
      \expandafter\def\csname LT6\endcsname{\color{black}}%
      \expandafter\def\csname LT7\endcsname{\color{black}}%
      \expandafter\def\csname LT8\endcsname{\color{black}}%
    \fi
  \fi
  \setlength{\unitlength}{0.0500bp}%
  \begin{picture}(4422.00,3628.00)%
    \gplgaddtomacro\gplbacktext{%
      \csname LTb\endcsname%
      \put(1140,640){\makebox(0,0)[r]{\strut{}$-1.5$}}%
      \put(1140,1098){\makebox(0,0)[r]{\strut{}$-1.0$}}%
      \put(1140,1556){\makebox(0,0)[r]{\strut{}$-0.5$}}%
      \put(1140,2014){\makebox(0,0)[r]{\strut{}$0.0$}}%
      \put(1140,2472){\makebox(0,0)[r]{\strut{}$0.5$}}%
      \put(1140,2930){\makebox(0,0)[r]{\strut{}$1.0$}}%
      \put(1140,3388){\makebox(0,0)[r]{\strut{}$1.5$}}%
      \put(1260,440){\makebox(0,0){\strut{}0}}%
      \put(1976,440){\makebox(0,0){\strut{}$0.25$}}%
      \put(2691,440){\makebox(0,0){\strut{}$0.50$}}%
      \put(3407,440){\makebox(0,0){\strut{}$0.75$}}%
      \put(4122,440){\makebox(0,0){\strut{}1}}%
      \put(440,2014){\rotatebox{90}{\makebox(0,0){\strut{}$-C_p$}}}%
      \put(2691,140){\makebox(0,0){\strut{}$x/c$}}%
      \put(3550,3159){\makebox(0,0)[l]{\strut{}$(b)$}}%
    }%
    \gplgaddtomacro\gplfronttext{%
    }%
    \gplbacktext
    \put(0,0){\includegraphics{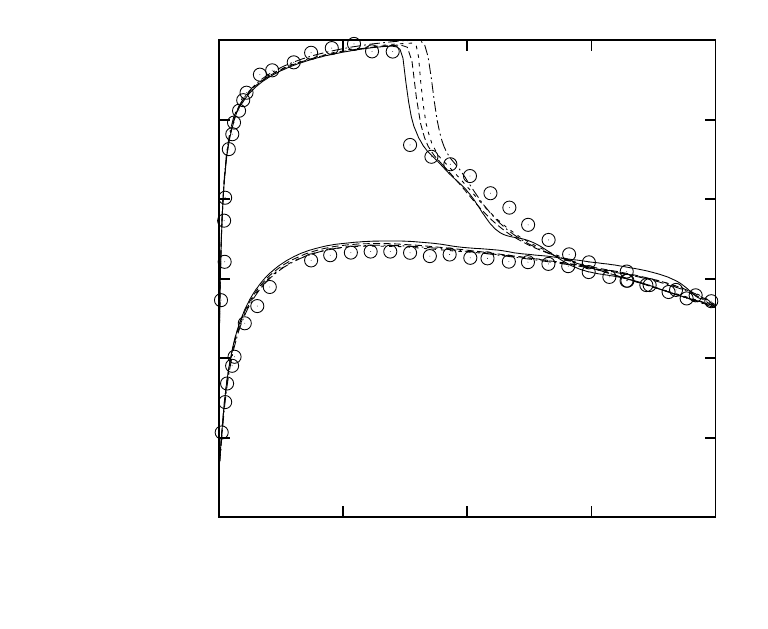}}%
    \gplfronttext
  \end{picture}%
\endgroup

%% file: pdelery_.tex
\begingroup
  \fontfamily{Helvetica}%
  \selectfont
  \makeatletter
  \providecommand\color[2][]{%
    \GenericError{(gnuplot) \space\space\space\@spaces}{%
      Package color not loaded in conjunction with
      terminal option `colourtext'%
    }{See the gnuplot documentation for explanation.%
    }{Either use 'blacktext' in gnuplot or load the package
      color.sty in LaTeX.}%
    \renewcommand\color[2][]{}%
  }%
  \providecommand\includegraphics[2][]{%
    \GenericError{(gnuplot) \space\space\space\@spaces}{%
      Package graphicx or graphics not loaded%
    }{See the gnuplot documentation for explanation.%
    }{The gnuplot epslatex terminal needs graphicx.sty or graphics.sty.}%
    \renewcommand\includegraphics[2][]{}%
  }%
  \providecommand\rotatebox[2]{#2}%
  \@ifundefined{ifGPcolor}{%
    \newif\ifGPcolor
    \GPcolorfalse
  }{}%
  \@ifundefined{ifGPblacktext}{%
    \newif\ifGPblacktext
    \GPblacktexttrue
  }{}%
  \let\gplgaddtomacro\g@addto@macro
  \gdef\gplbacktext{}%
  \gdef\gplfronttext{}%
  \makeatother
  \ifGPblacktext
    \def\colorrgb#1{}%
    \def\colorgray#1{}%
  \else
    \ifGPcolor
      \def\colorrgb#1{\color[rgb]{#1}}%
      \def\colorgray#1{\color[gray]{#1}}%
      \expandafter\def\csname LTw\endcsname{\color{white}}%
      \expandafter\def\csname LTb\endcsname{\color{black}}%
      \expandafter\def\csname LTa\endcsname{\color{black}}%
      \expandafter\def\csname LT0\endcsname{\color[rgb]{1,0,0}}%
      \expandafter\def\csname LT1\endcsname{\color[rgb]{0,1,0}}%
      \expandafter\def\csname LT2\endcsname{\color[rgb]{0,0,1}}%
      \expandafter\def\csname LT3\endcsname{\color[rgb]{1,0,1}}%
      \expandafter\def\csname LT4\endcsname{\color[rgb]{0,1,1}}%
      \expandafter\def\csname LT5\endcsname{\color[rgb]{1,1,0}}%
      \expandafter\def\csname LT6\endcsname{\color[rgb]{0,0,0}}%
      \expandafter\def\csname LT7\endcsname{\color[rgb]{1,0.3,0}}%
      \expandafter\def\csname LT8\endcsname{\color[rgb]{0.5,0.5,0.5}}%
    \else
      \def\colorrgb#1{\color{black}}%
      \def\colorgray#1{\color[gray]{#1}}%
      \expandafter\def\csname LTw\endcsname{\color{white}}%
      \expandafter\def\csname LTb\endcsname{\color{black}}%
      \expandafter\def\csname LTa\endcsname{\color{black}}%
      \expandafter\def\csname LT0\endcsname{\color{black}}%
      \expandafter\def\csname LT1\endcsname{\color{black}}%
      \expandafter\def\csname LT2\endcsname{\color{black}}%
      \expandafter\def\csname LT3\endcsname{\color{black}}%
      \expandafter\def\csname LT4\endcsname{\color{black}}%
      \expandafter\def\csname LT5\endcsname{\color{black}}%
      \expandafter\def\csname LT6\endcsname{\color{black}}%
      \expandafter\def\csname LT7\endcsname{\color{black}}%
      \expandafter\def\csname LT8\endcsname{\color{black}}%
    \fi
  \fi
  \setlength{\unitlength}{0.0500bp}%
  \begin{picture}(4422.00,3628.00)%
    \gplgaddtomacro\gplbacktext{%
      \csname LTb\endcsname%
      \put(1020,640){\makebox(0,0)[r]{\strut{}$0.2$}}%
      \put(1020,984){\makebox(0,0)[r]{\strut{}$0.3$}}%
      \put(1020,1327){\makebox(0,0)[r]{\strut{}$0.4$}}%
      \put(1020,1671){\makebox(0,0)[r]{\strut{}$0.5$}}%
      \put(1020,2014){\makebox(0,0)[r]{\strut{}$0.6$}}%
      \put(1020,2358){\makebox(0,0)[r]{\strut{}$0.7$}}%
      \put(1020,2701){\makebox(0,0)[r]{\strut{}$0.8$}}%
      \put(1020,3045){\makebox(0,0)[r]{\strut{}$0.9$}}%
      \put(1020,3388){\makebox(0,0)[r]{\strut{}$1.0$}}%
      \put(1140,440){\makebox(0,0){\strut{}$0.0$}}%
      \put(1886,440){\makebox(0,0){\strut{}$0.1$}}%
      \put(2631,440){\makebox(0,0){\strut{}$0.2$}}%
      \put(3377,440){\makebox(0,0){\strut{}$0.3$}}%
      \put(4122,440){\makebox(0,0){\strut{}$0.4$}}%
      \put(440,2014){\rotatebox{90}{\makebox(0,0){\strut{}$p/p_0$}}}%
      \put(2631,140){\makebox(0,0){\strut{}$x/H$}}%
      \put(3377,3102){\makebox(0,0)[l]{\strut{}$(a)$}}%
    }%
    \gplgaddtomacro\gplfronttext{%
    }%
    \gplbacktext
    \put(0,0){\includegraphics{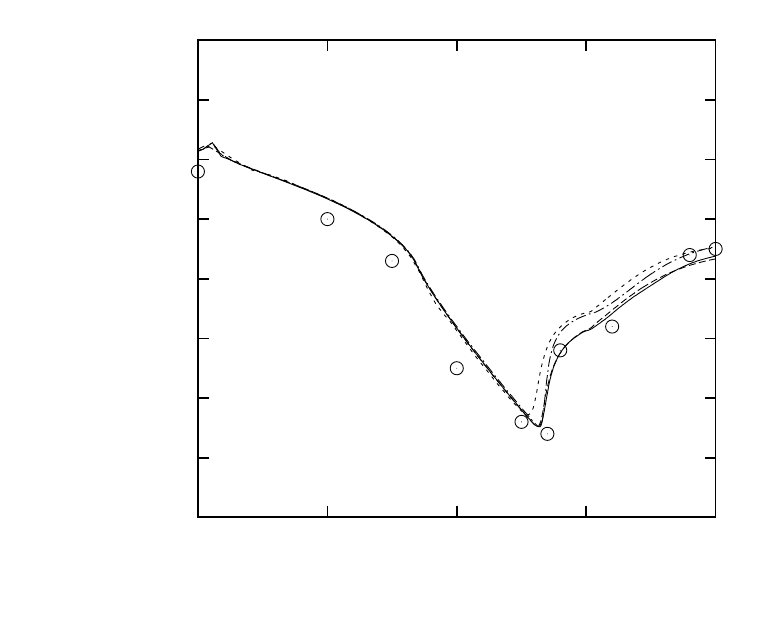}}%
    \gplfronttext
  \end{picture}%
\endgroup

%% file: cfdelery_.tex
\begingroup
  \fontfamily{Helvetica}%
  \selectfont
  \makeatletter
  \providecommand\color[2][]{%
    \GenericError{(gnuplot) \space\space\space\@spaces}{%
      Package color not loaded in conjunction with
      terminal option `colourtext'%
    }{See the gnuplot documentation for explanation.%
    }{Either use 'blacktext' in gnuplot or load the package
      color.sty in LaTeX.}%
    \renewcommand\color[2][]{}%
  }%
  \providecommand\includegraphics[2][]{%
    \GenericError{(gnuplot) \space\space\space\@spaces}{%
      Package graphicx or graphics not loaded%
    }{See the gnuplot documentation for explanation.%
    }{The gnuplot epslatex terminal needs graphicx.sty or graphics.sty.}%
    \renewcommand\includegraphics[2][]{}%
  }%
  \providecommand\rotatebox[2]{#2}%
  \@ifundefined{ifGPcolor}{%
    \newif\ifGPcolor
    \GPcolorfalse
  }{}%
  \@ifundefined{ifGPblacktext}{%
    \newif\ifGPblacktext
    \GPblacktexttrue
  }{}%
  \let\gplgaddtomacro\g@addto@macro
  \gdef\gplbacktext{}%
  \gdef\gplfronttext{}%
  \makeatother
  \ifGPblacktext
    \def\colorrgb#1{}%
    \def\colorgray#1{}%
  \else
    \ifGPcolor
      \def\colorrgb#1{\color[rgb]{#1}}%
      \def\colorgray#1{\color[gray]{#1}}%
      \expandafter\def\csname LTw\endcsname{\color{white}}%
      \expandafter\def\csname LTb\endcsname{\color{black}}%
      \expandafter\def\csname LTa\endcsname{\color{black}}%
      \expandafter\def\csname LT0\endcsname{\color[rgb]{1,0,0}}%
      \expandafter\def\csname LT1\endcsname{\color[rgb]{0,1,0}}%
      \expandafter\def\csname LT2\endcsname{\color[rgb]{0,0,1}}%
      \expandafter\def\csname LT3\endcsname{\color[rgb]{1,0,1}}%
      \expandafter\def\csname LT4\endcsname{\color[rgb]{0,1,1}}%
      \expandafter\def\csname LT5\endcsname{\color[rgb]{1,1,0}}%
      \expandafter\def\csname LT6\endcsname{\color[rgb]{0,0,0}}%
      \expandafter\def\csname LT7\endcsname{\color[rgb]{1,0.3,0}}%
      \expandafter\def\csname LT8\endcsname{\color[rgb]{0.5,0.5,0.5}}%
    \else
      \def\colorrgb#1{\color{black}}%
      \def\colorgray#1{\color[gray]{#1}}%
      \expandafter\def\csname LTw\endcsname{\color{white}}%
      \expandafter\def\csname LTb\endcsname{\color{black}}%
      \expandafter\def\csname LTa\endcsname{\color{black}}%
      \expandafter\def\csname LT0\endcsname{\color{black}}%
      \expandafter\def\csname LT1\endcsname{\color{black}}%
      \expandafter\def\csname LT2\endcsname{\color{black}}%
      \expandafter\def\csname LT3\endcsname{\color{black}}%
      \expandafter\def\csname LT4\endcsname{\color{black}}%
      \expandafter\def\csname LT5\endcsname{\color{black}}%
      \expandafter\def\csname LT6\endcsname{\color{black}}%
      \expandafter\def\csname LT7\endcsname{\color{black}}%
      \expandafter\def\csname LT8\endcsname{\color{black}}%
    \fi
  \fi
  \setlength{\unitlength}{0.0500bp}%
  \begin{picture}(4422.00,3628.00)%
    \gplgaddtomacro\gplbacktext{%
      \csname LTb\endcsname%
      \put(900,640){\makebox(0,0)[r]{\strut{}$-5$}}%
      \put(900,1327){\makebox(0,0)[r]{\strut{}$0$}}%
      \put(900,2014){\makebox(0,0)[r]{\strut{}$5$}}%
      \put(900,2701){\makebox(0,0)[r]{\strut{}$10$}}%
      \put(900,3388){\makebox(0,0)[r]{\strut{}$15$}}%
      \put(1020,440){\makebox(0,0){\strut{}$0.0$}}%
      \put(1796,440){\makebox(0,0){\strut{}$0.1$}}%
      \put(2571,440){\makebox(0,0){\strut{}$0.2$}}%
      \put(3347,440){\makebox(0,0){\strut{}$0.3$}}%
      \put(4122,440){\makebox(0,0){\strut{}$0.4$}}%
      \put(440,2014){\rotatebox{90}{\makebox(0,0){\strut{}$C_f \, \times \, 10^3$}}}%
      \put(2571,140){\makebox(0,0){\strut{}$x/H$}}%
      \put(3347,3159){\makebox(0,0)[l]{\strut{}$(b)$}}%
    }%
    \gplgaddtomacro\gplfronttext{%
    }%
    \gplbacktext
    \put(0,0){\includegraphics{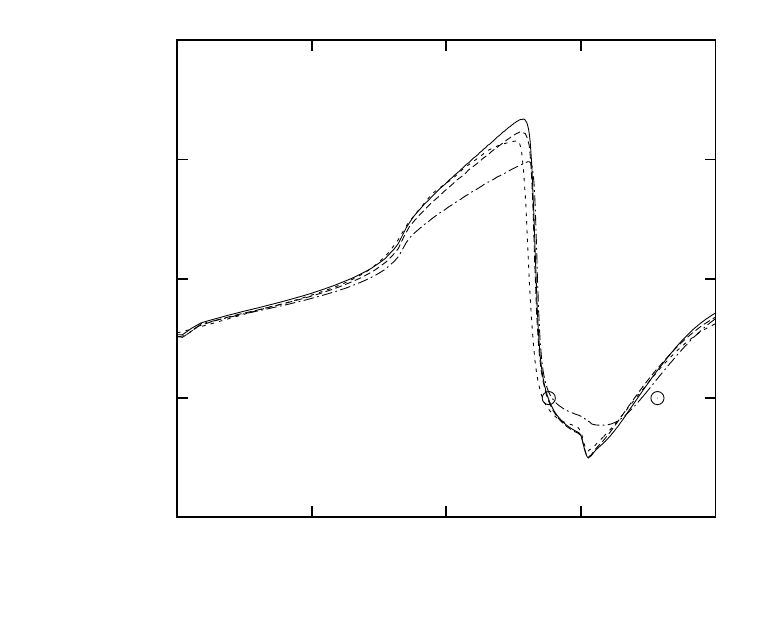}}%
    \gplfronttext
  \end{picture}%
\endgroup

%% file: cpramp8_.tex
\begingroup
  \fontfamily{Helvetica}%
  \selectfont
  \makeatletter
  \providecommand\color[2][]{%
    \GenericError{(gnuplot) \space\space\space\@spaces}{%
      Package color not loaded in conjunction with
      terminal option `colourtext'%
    }{See the gnuplot documentation for explanation.%
    }{Either use 'blacktext' in gnuplot or load the package
      color.sty in LaTeX.}%
    \renewcommand\color[2][]{}%
  }%
  \providecommand\includegraphics[2][]{%
    \GenericError{(gnuplot) \space\space\space\@spaces}{%
      Package graphicx or graphics not loaded%
    }{See the gnuplot documentation for explanation.%
    }{The gnuplot epslatex terminal needs graphicx.sty or graphics.sty.}%
    \renewcommand\includegraphics[2][]{}%
  }%
  \providecommand\rotatebox[2]{#2}%
  \@ifundefined{ifGPcolor}{%
    \newif\ifGPcolor
    \GPcolorfalse
  }{}%
  \@ifundefined{ifGPblacktext}{%
    \newif\ifGPblacktext
    \GPblacktexttrue
  }{}%
  \let\gplgaddtomacro\g@addto@macro
  \gdef\gplbacktext{}%
  \gdef\gplfronttext{}%
  \makeatother
  \ifGPblacktext
    \def\colorrgb#1{}%
    \def\colorgray#1{}%
  \else
    \ifGPcolor
      \def\colorrgb#1{\color[rgb]{#1}}%
      \def\colorgray#1{\color[gray]{#1}}%
      \expandafter\def\csname LTw\endcsname{\color{white}}%
      \expandafter\def\csname LTb\endcsname{\color{black}}%
      \expandafter\def\csname LTa\endcsname{\color{black}}%
      \expandafter\def\csname LT0\endcsname{\color[rgb]{1,0,0}}%
      \expandafter\def\csname LT1\endcsname{\color[rgb]{0,1,0}}%
      \expandafter\def\csname LT2\endcsname{\color[rgb]{0,0,1}}%
      \expandafter\def\csname LT3\endcsname{\color[rgb]{1,0,1}}%
      \expandafter\def\csname LT4\endcsname{\color[rgb]{0,1,1}}%
      \expandafter\def\csname LT5\endcsname{\color[rgb]{1,1,0}}%
      \expandafter\def\csname LT6\endcsname{\color[rgb]{0,0,0}}%
      \expandafter\def\csname LT7\endcsname{\color[rgb]{1,0.3,0}}%
      \expandafter\def\csname LT8\endcsname{\color[rgb]{0.5,0.5,0.5}}%
    \else
      \def\colorrgb#1{\color{black}}%
      \def\colorgray#1{\color[gray]{#1}}%
      \expandafter\def\csname LTw\endcsname{\color{white}}%
      \expandafter\def\csname LTb\endcsname{\color{black}}%
      \expandafter\def\csname LTa\endcsname{\color{black}}%
      \expandafter\def\csname LT0\endcsname{\color{black}}%
      \expandafter\def\csname LT1\endcsname{\color{black}}%
      \expandafter\def\csname LT2\endcsname{\color{black}}%
      \expandafter\def\csname LT3\endcsname{\color{black}}%
      \expandafter\def\csname LT4\endcsname{\color{black}}%
      \expandafter\def\csname LT5\endcsname{\color{black}}%
      \expandafter\def\csname LT6\endcsname{\color{black}}%
      \expandafter\def\csname LT7\endcsname{\color{black}}%
      \expandafter\def\csname LT8\endcsname{\color{black}}%
    \fi
  \fi
  \setlength{\unitlength}{0.0500bp}%
  \begin{picture}(4080.00,3628.00)%
    \gplgaddtomacro\gplbacktext{%
      \csname LTb\endcsname%
      \put(588,772){\makebox(0,0)[r]{\strut{}$1$}}%
      \put(588,1547){\makebox(0,0)[r]{\strut{}$2$}}%
      \put(588,2322){\makebox(0,0)[r]{\strut{}$3$}}%
      \put(588,3097){\makebox(0,0)[r]{\strut{}$4$}}%
      \put(1956,264){\makebox(0,0){\strut{}$0$}}%
      \put(3576,264){\makebox(0,0){\strut{}$5$}}%
      \put(240,1934){\rotatebox{90}{\makebox(0,0){\strut{}$p/p_0$}}}%
      \put(2280,84){\makebox(0,0){\strut{}$x/{\delta}_0$}}%
      \put(1308,3097){\makebox(0,0)[l]{\strut{}${\alpha} = 8^{\circ}$}}%
    }%
    \gplgaddtomacro\gplfronttext{%
    }%
    \gplbacktext
    \put(0,0){\includegraphics{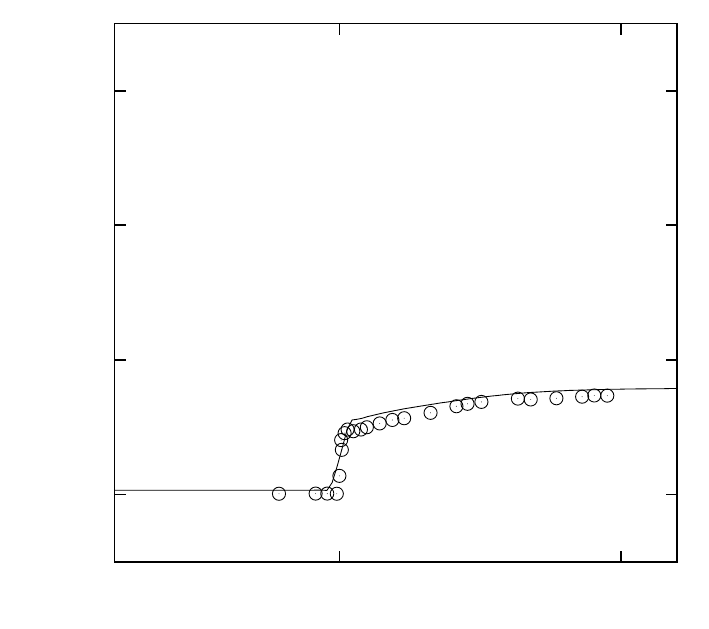}}%
    \gplfronttext
  \end{picture}%
\endgroup

%% file: cpramp16_.tex
\begingroup
  \fontfamily{Helvetica}%
  \selectfont
  \makeatletter
  \providecommand\color[2][]{%
    \GenericError{(gnuplot) \space\space\space\@spaces}{%
      Package color not loaded in conjunction with
      terminal option `colourtext'%
    }{See the gnuplot documentation for explanation.%
    }{Either use 'blacktext' in gnuplot or load the package
      color.sty in LaTeX.}%
    \renewcommand\color[2][]{}%
  }%
  \providecommand\includegraphics[2][]{%
    \GenericError{(gnuplot) \space\space\space\@spaces}{%
      Package graphicx or graphics not loaded%
    }{See the gnuplot documentation for explanation.%
    }{The gnuplot epslatex terminal needs graphicx.sty or graphics.sty.}%
    \renewcommand\includegraphics[2][]{}%
  }%
  \providecommand\rotatebox[2]{#2}%
  \@ifundefined{ifGPcolor}{%
    \newif\ifGPcolor
    \GPcolorfalse
  }{}%
  \@ifundefined{ifGPblacktext}{%
    \newif\ifGPblacktext
    \GPblacktexttrue
  }{}%
  \let\gplgaddtomacro\g@addto@macro
  \gdef\gplbacktext{}%
  \gdef\gplfronttext{}%
  \makeatother
  \ifGPblacktext
    \def\colorrgb#1{}%
    \def\colorgray#1{}%
  \else
    \ifGPcolor
      \def\colorrgb#1{\color[rgb]{#1}}%
      \def\colorgray#1{\color[gray]{#1}}%
      \expandafter\def\csname LTw\endcsname{\color{white}}%
      \expandafter\def\csname LTb\endcsname{\color{black}}%
      \expandafter\def\csname LTa\endcsname{\color{black}}%
      \expandafter\def\csname LT0\endcsname{\color[rgb]{1,0,0}}%
      \expandafter\def\csname LT1\endcsname{\color[rgb]{0,1,0}}%
      \expandafter\def\csname LT2\endcsname{\color[rgb]{0,0,1}}%
      \expandafter\def\csname LT3\endcsname{\color[rgb]{1,0,1}}%
      \expandafter\def\csname LT4\endcsname{\color[rgb]{0,1,1}}%
      \expandafter\def\csname LT5\endcsname{\color[rgb]{1,1,0}}%
      \expandafter\def\csname LT6\endcsname{\color[rgb]{0,0,0}}%
      \expandafter\def\csname LT7\endcsname{\color[rgb]{1,0.3,0}}%
      \expandafter\def\csname LT8\endcsname{\color[rgb]{0.5,0.5,0.5}}%
    \else
      \def\colorrgb#1{\color{black}}%
      \def\colorgray#1{\color[gray]{#1}}%
      \expandafter\def\csname LTw\endcsname{\color{white}}%
      \expandafter\def\csname LTb\endcsname{\color{black}}%
      \expandafter\def\csname LTa\endcsname{\color{black}}%
      \expandafter\def\csname LT0\endcsname{\color{black}}%
      \expandafter\def\csname LT1\endcsname{\color{black}}%
      \expandafter\def\csname LT2\endcsname{\color{black}}%
      \expandafter\def\csname LT3\endcsname{\color{black}}%
      \expandafter\def\csname LT4\endcsname{\color{black}}%
      \expandafter\def\csname LT5\endcsname{\color{black}}%
      \expandafter\def\csname LT6\endcsname{\color{black}}%
      \expandafter\def\csname LT7\endcsname{\color{black}}%
      \expandafter\def\csname LT8\endcsname{\color{black}}%
    \fi
  \fi
  \setlength{\unitlength}{0.0500bp}%
  \begin{picture}(4080.00,3628.00)%
    \gplgaddtomacro\gplbacktext{%
      \csname LTb\endcsname%
      \put(588,772){\makebox(0,0)[r]{\strut{}$1$}}%
      \put(588,1547){\makebox(0,0)[r]{\strut{}$2$}}%
      \put(588,2322){\makebox(0,0)[r]{\strut{}$3$}}%
      \put(588,3097){\makebox(0,0)[r]{\strut{}$4$}}%
      \put(1956,264){\makebox(0,0){\strut{}$0$}}%
      \put(3576,264){\makebox(0,0){\strut{}$5$}}%
      \put(240,1934){\rotatebox{90}{\makebox(0,0){\strut{}$p/p_0$}}}%
      \put(2280,84){\makebox(0,0){\strut{}$x/{\delta}_0$}}%
      \put(1308,3097){\makebox(0,0)[l]{\strut{}${\alpha} = 16^{\circ}$}}%
    }%
    \gplgaddtomacro\gplfronttext{%
    }%
    \gplbacktext
    \put(0,0){\includegraphics{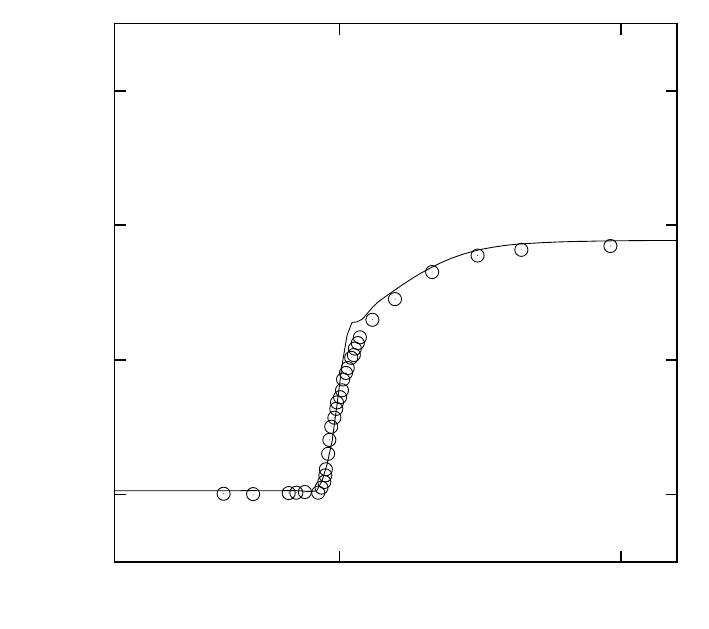}}%
    \gplfronttext
  \end{picture}%
\endgroup

%% file: cpramp20_.tex
\begingroup
  \fontfamily{Helvetica}%
  \selectfont
  \makeatletter
  \providecommand\color[2][]{%
    \GenericError{(gnuplot) \space\space\space\@spaces}{%
      Package color not loaded in conjunction with
      terminal option `colourtext'%
    }{See the gnuplot documentation for explanation.%
    }{Either use 'blacktext' in gnuplot or load the package
      color.sty in LaTeX.}%
    \renewcommand\color[2][]{}%
  }%
  \providecommand\includegraphics[2][]{%
    \GenericError{(gnuplot) \space\space\space\@spaces}{%
      Package graphicx or graphics not loaded%
    }{See the gnuplot documentation for explanation.%
    }{The gnuplot epslatex terminal needs graphicx.sty or graphics.sty.}%
    \renewcommand\includegraphics[2][]{}%
  }%
  \providecommand\rotatebox[2]{#2}%
  \@ifundefined{ifGPcolor}{%
    \newif\ifGPcolor
    \GPcolorfalse
  }{}%
  \@ifundefined{ifGPblacktext}{%
    \newif\ifGPblacktext
    \GPblacktexttrue
  }{}%
  \let\gplgaddtomacro\g@addto@macro
  \gdef\gplbacktext{}%
  \gdef\gplfronttext{}%
  \makeatother
  \ifGPblacktext
    \def\colorrgb#1{}%
    \def\colorgray#1{}%
  \else
    \ifGPcolor
      \def\colorrgb#1{\color[rgb]{#1}}%
      \def\colorgray#1{\color[gray]{#1}}%
      \expandafter\def\csname LTw\endcsname{\color{white}}%
      \expandafter\def\csname LTb\endcsname{\color{black}}%
      \expandafter\def\csname LTa\endcsname{\color{black}}%
      \expandafter\def\csname LT0\endcsname{\color[rgb]{1,0,0}}%
      \expandafter\def\csname LT1\endcsname{\color[rgb]{0,1,0}}%
      \expandafter\def\csname LT2\endcsname{\color[rgb]{0,0,1}}%
      \expandafter\def\csname LT3\endcsname{\color[rgb]{1,0,1}}%
      \expandafter\def\csname LT4\endcsname{\color[rgb]{0,1,1}}%
      \expandafter\def\csname LT5\endcsname{\color[rgb]{1,1,0}}%
      \expandafter\def\csname LT6\endcsname{\color[rgb]{0,0,0}}%
      \expandafter\def\csname LT7\endcsname{\color[rgb]{1,0.3,0}}%
      \expandafter\def\csname LT8\endcsname{\color[rgb]{0.5,0.5,0.5}}%
    \else
      \def\colorrgb#1{\color{black}}%
      \def\colorgray#1{\color[gray]{#1}}%
      \expandafter\def\csname LTw\endcsname{\color{white}}%
      \expandafter\def\csname LTb\endcsname{\color{black}}%
      \expandafter\def\csname LTa\endcsname{\color{black}}%
      \expandafter\def\csname LT0\endcsname{\color{black}}%
      \expandafter\def\csname LT1\endcsname{\color{black}}%
      \expandafter\def\csname LT2\endcsname{\color{black}}%
      \expandafter\def\csname LT3\endcsname{\color{black}}%
      \expandafter\def\csname LT4\endcsname{\color{black}}%
      \expandafter\def\csname LT5\endcsname{\color{black}}%
      \expandafter\def\csname LT6\endcsname{\color{black}}%
      \expandafter\def\csname LT7\endcsname{\color{black}}%
      \expandafter\def\csname LT8\endcsname{\color{black}}%
    \fi
  \fi
  \setlength{\unitlength}{0.0500bp}%
  \begin{picture}(4080.00,3628.00)%
    \gplgaddtomacro\gplbacktext{%
      \csname LTb\endcsname%
      \put(588,772){\makebox(0,0)[r]{\strut{}$1$}}%
      \put(588,1547){\makebox(0,0)[r]{\strut{}$2$}}%
      \put(588,2322){\makebox(0,0)[r]{\strut{}$3$}}%
      \put(588,3097){\makebox(0,0)[r]{\strut{}$4$}}%
      \put(1956,264){\makebox(0,0){\strut{}$0$}}%
      \put(3576,264){\makebox(0,0){\strut{}$5$}}%
      \put(240,1934){\rotatebox{90}{\makebox(0,0){\strut{}$p/p_0$}}}%
      \put(2280,84){\makebox(0,0){\strut{}$x/{\delta}_0$}}%
      \put(1308,3097){\makebox(0,0)[l]{\strut{}${\alpha} = 20^{\circ}$}}%
    }%
    \gplgaddtomacro\gplfronttext{%
    }%
    \gplbacktext
    \put(0,0){\includegraphics{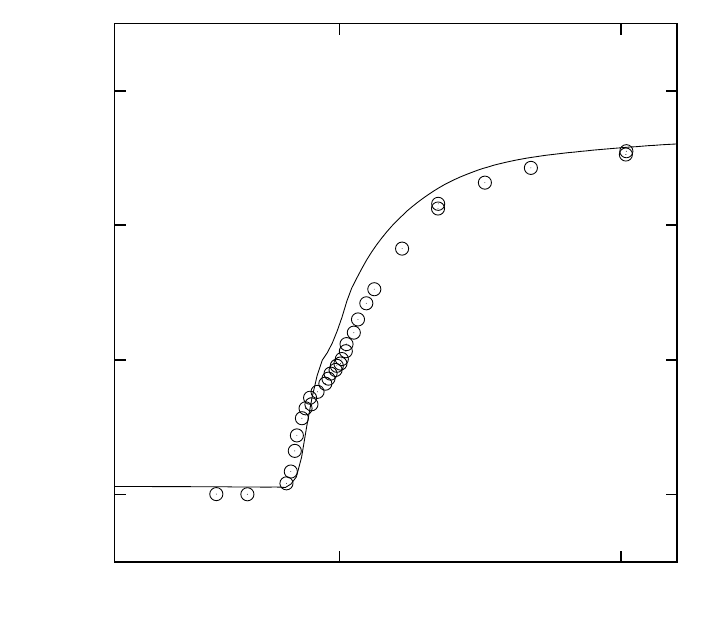}}%
    \gplfronttext
  \end{picture}%
\endgroup

%% file: cpramp24gridconv_.tex
\begingroup
  \fontfamily{Helvetica}%
  \selectfont
  \makeatletter
  \providecommand\color[2][]{%
    \GenericError{(gnuplot) \space\space\space\@spaces}{%
      Package color not loaded in conjunction with
      terminal option `colourtext'%
    }{See the gnuplot documentation for explanation.%
    }{Either use 'blacktext' in gnuplot or load the package
      color.sty in LaTeX.}%
    \renewcommand\color[2][]{}%
  }%
  \providecommand\includegraphics[2][]{%
    \GenericError{(gnuplot) \space\space\space\@spaces}{%
      Package graphicx or graphics not loaded%
    }{See the gnuplot documentation for explanation.%
    }{The gnuplot epslatex terminal needs graphicx.sty or graphics.sty.}%
    \renewcommand\includegraphics[2][]{}%
  }%
  \providecommand\rotatebox[2]{#2}%
  \@ifundefined{ifGPcolor}{%
    \newif\ifGPcolor
    \GPcolorfalse
  }{}%
  \@ifundefined{ifGPblacktext}{%
    \newif\ifGPblacktext
    \GPblacktexttrue
  }{}%
  \let\gplgaddtomacro\g@addto@macro
  \gdef\gplbacktext{}%
  \gdef\gplfronttext{}%
  \makeatother
  \ifGPblacktext
    \def\colorrgb#1{}%
    \def\colorgray#1{}%
  \else
    \ifGPcolor
      \def\colorrgb#1{\color[rgb]{#1}}%
      \def\colorgray#1{\color[gray]{#1}}%
      \expandafter\def\csname LTw\endcsname{\color{white}}%
      \expandafter\def\csname LTb\endcsname{\color{black}}%
      \expandafter\def\csname LTa\endcsname{\color{black}}%
      \expandafter\def\csname LT0\endcsname{\color[rgb]{1,0,0}}%
      \expandafter\def\csname LT1\endcsname{\color[rgb]{0,1,0}}%
      \expandafter\def\csname LT2\endcsname{\color[rgb]{0,0,1}}%
      \expandafter\def\csname LT3\endcsname{\color[rgb]{1,0,1}}%
      \expandafter\def\csname LT4\endcsname{\color[rgb]{0,1,1}}%
      \expandafter\def\csname LT5\endcsname{\color[rgb]{1,1,0}}%
      \expandafter\def\csname LT6\endcsname{\color[rgb]{0,0,0}}%
      \expandafter\def\csname LT7\endcsname{\color[rgb]{1,0.3,0}}%
      \expandafter\def\csname LT8\endcsname{\color[rgb]{0.5,0.5,0.5}}%
    \else
      \def\colorrgb#1{\color{black}}%
      \def\colorgray#1{\color[gray]{#1}}%
      \expandafter\def\csname LTw\endcsname{\color{white}}%
      \expandafter\def\csname LTb\endcsname{\color{black}}%
      \expandafter\def\csname LTa\endcsname{\color{black}}%
      \expandafter\def\csname LT0\endcsname{\color{black}}%
      \expandafter\def\csname LT1\endcsname{\color{black}}%
      \expandafter\def\csname LT2\endcsname{\color{black}}%
      \expandafter\def\csname LT3\endcsname{\color{black}}%
      \expandafter\def\csname LT4\endcsname{\color{black}}%
      \expandafter\def\csname LT5\endcsname{\color{black}}%
      \expandafter\def\csname LT6\endcsname{\color{black}}%
      \expandafter\def\csname LT7\endcsname{\color{black}}%
      \expandafter\def\csname LT8\endcsname{\color{black}}%
    \fi
  \fi
  \setlength{\unitlength}{0.0500bp}%
  \begin{picture}(4080.00,3628.00)%
    \gplgaddtomacro\gplbacktext{%
      \csname LTb\endcsname%
      \put(588,772){\makebox(0,0)[r]{\strut{}$1$}}%
      \put(588,1547){\makebox(0,0)[r]{\strut{}$2$}}%
      \put(588,2322){\makebox(0,0)[r]{\strut{}$3$}}%
      \put(588,3097){\makebox(0,0)[r]{\strut{}$4$}}%
      \put(1956,264){\makebox(0,0){\strut{}$0$}}%
      \put(3576,264){\makebox(0,0){\strut{}$5$}}%
      \put(240,1934){\rotatebox{90}{\makebox(0,0){\strut{}$p/p_0$}}}%
      \put(2280,84){\makebox(0,0){\strut{}$x/{\delta}_0$}}%
      \put(1308,3097){\makebox(0,0)[l]{\strut{}${\alpha} = 24^{\circ}$}}%
    }%
    \gplgaddtomacro\gplfronttext{%
    }%
    \gplbacktext
    \put(0,0){\includegraphics{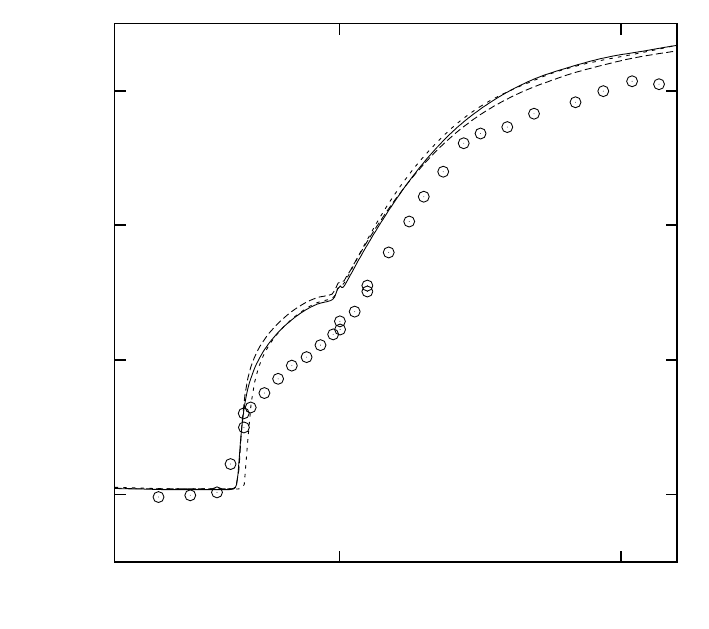}}%
    \gplfronttext
  \end{picture}%
\endgroup

%% file: cframp8_.tex
\begingroup
  \fontfamily{Helvetica}%
  \selectfont
  \makeatletter
  \providecommand\color[2][]{%
    \GenericError{(gnuplot) \space\space\space\@spaces}{%
      Package color not loaded in conjunction with
      terminal option `colourtext'%
    }{See the gnuplot documentation for explanation.%
    }{Either use 'blacktext' in gnuplot or load the package
      color.sty in LaTeX.}%
    \renewcommand\color[2][]{}%
  }%
  \providecommand\includegraphics[2][]{%
    \GenericError{(gnuplot) \space\space\space\@spaces}{%
      Package graphicx or graphics not loaded%
    }{See the gnuplot documentation for explanation.%
    }{The gnuplot epslatex terminal needs graphicx.sty or graphics.sty.}%
    \renewcommand\includegraphics[2][]{}%
  }%
  \providecommand\rotatebox[2]{#2}%
  \@ifundefined{ifGPcolor}{%
    \newif\ifGPcolor
    \GPcolorfalse
  }{}%
  \@ifundefined{ifGPblacktext}{%
    \newif\ifGPblacktext
    \GPblacktexttrue
  }{}%
  \let\gplgaddtomacro\g@addto@macro
  \gdef\gplbacktext{}%
  \gdef\gplfronttext{}%
  \makeatother
  \ifGPblacktext
    \def\colorrgb#1{}%
    \def\colorgray#1{}%
  \else
    \ifGPcolor
      \def\colorrgb#1{\color[rgb]{#1}}%
      \def\colorgray#1{\color[gray]{#1}}%
      \expandafter\def\csname LTw\endcsname{\color{white}}%
      \expandafter\def\csname LTb\endcsname{\color{black}}%
      \expandafter\def\csname LTa\endcsname{\color{black}}%
      \expandafter\def\csname LT0\endcsname{\color[rgb]{1,0,0}}%
      \expandafter\def\csname LT1\endcsname{\color[rgb]{0,1,0}}%
      \expandafter\def\csname LT2\endcsname{\color[rgb]{0,0,1}}%
      \expandafter\def\csname LT3\endcsname{\color[rgb]{1,0,1}}%
      \expandafter\def\csname LT4\endcsname{\color[rgb]{0,1,1}}%
      \expandafter\def\csname LT5\endcsname{\color[rgb]{1,1,0}}%
      \expandafter\def\csname LT6\endcsname{\color[rgb]{0,0,0}}%
      \expandafter\def\csname LT7\endcsname{\color[rgb]{1,0.3,0}}%
      \expandafter\def\csname LT8\endcsname{\color[rgb]{0.5,0.5,0.5}}%
    \else
      \def\colorrgb#1{\color{black}}%
      \def\colorgray#1{\color[gray]{#1}}%
      \expandafter\def\csname LTw\endcsname{\color{white}}%
      \expandafter\def\csname LTb\endcsname{\color{black}}%
      \expandafter\def\csname LTa\endcsname{\color{black}}%
      \expandafter\def\csname LT0\endcsname{\color{black}}%
      \expandafter\def\csname LT1\endcsname{\color{black}}%
      \expandafter\def\csname LT2\endcsname{\color{black}}%
      \expandafter\def\csname LT3\endcsname{\color{black}}%
      \expandafter\def\csname LT4\endcsname{\color{black}}%
      \expandafter\def\csname LT5\endcsname{\color{black}}%
      \expandafter\def\csname LT6\endcsname{\color{black}}%
      \expandafter\def\csname LT7\endcsname{\color{black}}%
      \expandafter\def\csname LT8\endcsname{\color{black}}%
    \fi
  \fi
  \setlength{\unitlength}{0.0500bp}%
  \begin{picture}(4080.00,3628.00)%
    \gplgaddtomacro\gplbacktext{%
      \csname LTb\endcsname%
      \put(804,539){\makebox(0,0)[r]{\strut{}$-0.5$}}%
      \put(804,1314){\makebox(0,0)[r]{\strut{}$0.0$}}%
      \put(804,2089){\makebox(0,0)[r]{\strut{}$0.5$}}%
      \put(804,2864){\makebox(0,0)[r]{\strut{}$1.0$}}%
      \put(2086,264){\makebox(0,0){\strut{}$0$}}%
      \put(3598,264){\makebox(0,0){\strut{}$5$}}%
      \put(240,1934){\rotatebox{90}{\makebox(0,0){\strut{}$C_f \, \times \, 10^3$}}}%
      \put(2388,84){\makebox(0,0){\strut{}$x/{\delta}_0$}}%
      \put(2842,1314){\makebox(0,0)[l]{\strut{}${\alpha} = 8^{\circ}$}}%
    }%
    \gplgaddtomacro\gplfronttext{%
    }%
    \gplbacktext
    \put(0,0){\includegraphics{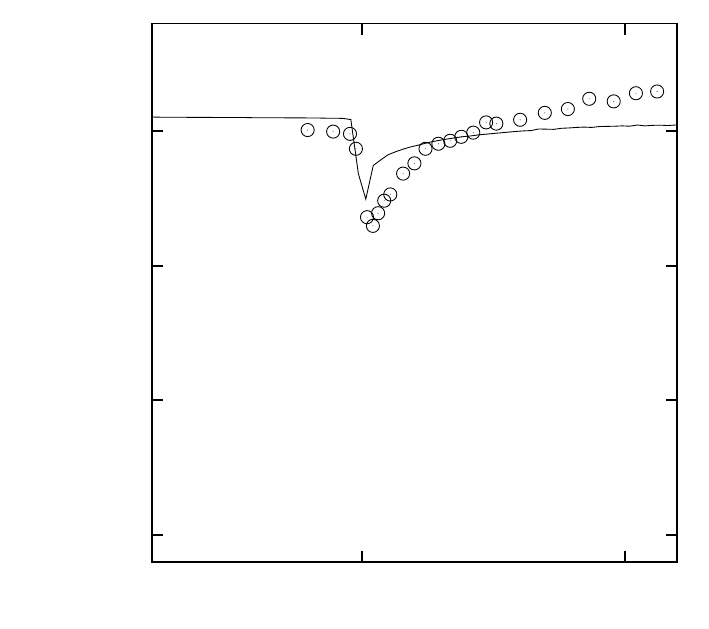}}%
    \gplfronttext
  \end{picture}%
\endgroup

%% file: cframp16_.tex
\begingroup
  \fontfamily{Helvetica}%
  \selectfont
  \makeatletter
  \providecommand\color[2][]{%
    \GenericError{(gnuplot) \space\space\space\@spaces}{%
      Package color not loaded in conjunction with
      terminal option `colourtext'%
    }{See the gnuplot documentation for explanation.%
    }{Either use 'blacktext' in gnuplot or load the package
      color.sty in LaTeX.}%
    \renewcommand\color[2][]{}%
  }%
  \providecommand\includegraphics[2][]{%
    \GenericError{(gnuplot) \space\space\space\@spaces}{%
      Package graphicx or graphics not loaded%
    }{See the gnuplot documentation for explanation.%
    }{The gnuplot epslatex terminal needs graphicx.sty or graphics.sty.}%
    \renewcommand\includegraphics[2][]{}%
  }%
  \providecommand\rotatebox[2]{#2}%
  \@ifundefined{ifGPcolor}{%
    \newif\ifGPcolor
    \GPcolorfalse
  }{}%
  \@ifundefined{ifGPblacktext}{%
    \newif\ifGPblacktext
    \GPblacktexttrue
  }{}%
  \let\gplgaddtomacro\g@addto@macro
  \gdef\gplbacktext{}%
  \gdef\gplfronttext{}%
  \makeatother
  \ifGPblacktext
    \def\colorrgb#1{}%
    \def\colorgray#1{}%
  \else
    \ifGPcolor
      \def\colorrgb#1{\color[rgb]{#1}}%
      \def\colorgray#1{\color[gray]{#1}}%
      \expandafter\def\csname LTw\endcsname{\color{white}}%
      \expandafter\def\csname LTb\endcsname{\color{black}}%
      \expandafter\def\csname LTa\endcsname{\color{black}}%
      \expandafter\def\csname LT0\endcsname{\color[rgb]{1,0,0}}%
      \expandafter\def\csname LT1\endcsname{\color[rgb]{0,1,0}}%
      \expandafter\def\csname LT2\endcsname{\color[rgb]{0,0,1}}%
      \expandafter\def\csname LT3\endcsname{\color[rgb]{1,0,1}}%
      \expandafter\def\csname LT4\endcsname{\color[rgb]{0,1,1}}%
      \expandafter\def\csname LT5\endcsname{\color[rgb]{1,1,0}}%
      \expandafter\def\csname LT6\endcsname{\color[rgb]{0,0,0}}%
      \expandafter\def\csname LT7\endcsname{\color[rgb]{1,0.3,0}}%
      \expandafter\def\csname LT8\endcsname{\color[rgb]{0.5,0.5,0.5}}%
    \else
      \def\colorrgb#1{\color{black}}%
      \def\colorgray#1{\color[gray]{#1}}%
      \expandafter\def\csname LTw\endcsname{\color{white}}%
      \expandafter\def\csname LTb\endcsname{\color{black}}%
      \expandafter\def\csname LTa\endcsname{\color{black}}%
      \expandafter\def\csname LT0\endcsname{\color{black}}%
      \expandafter\def\csname LT1\endcsname{\color{black}}%
      \expandafter\def\csname LT2\endcsname{\color{black}}%
      \expandafter\def\csname LT3\endcsname{\color{black}}%
      \expandafter\def\csname LT4\endcsname{\color{black}}%
      \expandafter\def\csname LT5\endcsname{\color{black}}%
      \expandafter\def\csname LT6\endcsname{\color{black}}%
      \expandafter\def\csname LT7\endcsname{\color{black}}%
      \expandafter\def\csname LT8\endcsname{\color{black}}%
    \fi
  \fi
  \setlength{\unitlength}{0.0500bp}%
  \begin{picture}(4080.00,3628.00)%
    \gplgaddtomacro\gplbacktext{%
      \csname LTb\endcsname%
      \put(804,539){\makebox(0,0)[r]{\strut{}$-0.5$}}%
      \put(804,1314){\makebox(0,0)[r]{\strut{}$0.0$}}%
      \put(804,2089){\makebox(0,0)[r]{\strut{}$0.5$}}%
      \put(804,2864){\makebox(0,0)[r]{\strut{}$1.0$}}%
      \put(2086,264){\makebox(0,0){\strut{}$0$}}%
      \put(3598,264){\makebox(0,0){\strut{}$5$}}%
      \put(240,1934){\rotatebox{90}{\makebox(0,0){\strut{}$C_f \, \times \, 10^3$}}}%
      \put(2388,84){\makebox(0,0){\strut{}$x/{\delta}_0$}}%
      \put(2842,1314){\makebox(0,0)[l]{\strut{}${\alpha} = 16^{\circ}$}}%
    }%
    \gplgaddtomacro\gplfronttext{%
    }%
    \gplbacktext
    \put(0,0){\includegraphics{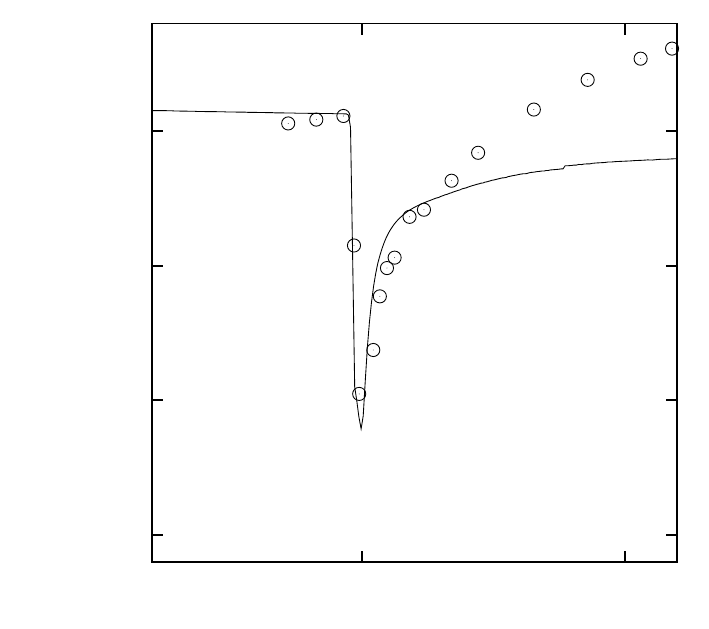}}%
    \gplfronttext
  \end{picture}%
\endgroup

%% file: cframp20_.tex
\begingroup
  \fontfamily{Helvetica}%
  \selectfont
  \makeatletter
  \providecommand\color[2][]{%
    \GenericError{(gnuplot) \space\space\space\@spaces}{%
      Package color not loaded in conjunction with
      terminal option `colourtext'%
    }{See the gnuplot documentation for explanation.%
    }{Either use 'blacktext' in gnuplot or load the package
      color.sty in LaTeX.}%
    \renewcommand\color[2][]{}%
  }%
  \providecommand\includegraphics[2][]{%
    \GenericError{(gnuplot) \space\space\space\@spaces}{%
      Package graphicx or graphics not loaded%
    }{See the gnuplot documentation for explanation.%
    }{The gnuplot epslatex terminal needs graphicx.sty or graphics.sty.}%
    \renewcommand\includegraphics[2][]{}%
  }%
  \providecommand\rotatebox[2]{#2}%
  \@ifundefined{ifGPcolor}{%
    \newif\ifGPcolor
    \GPcolorfalse
  }{}%
  \@ifundefined{ifGPblacktext}{%
    \newif\ifGPblacktext
    \GPblacktexttrue
  }{}%
  \let\gplgaddtomacro\g@addto@macro
  \gdef\gplbacktext{}%
  \gdef\gplfronttext{}%
  \makeatother
  \ifGPblacktext
    \def\colorrgb#1{}%
    \def\colorgray#1{}%
  \else
    \ifGPcolor
      \def\colorrgb#1{\color[rgb]{#1}}%
      \def\colorgray#1{\color[gray]{#1}}%
      \expandafter\def\csname LTw\endcsname{\color{white}}%
      \expandafter\def\csname LTb\endcsname{\color{black}}%
      \expandafter\def\csname LTa\endcsname{\color{black}}%
      \expandafter\def\csname LT0\endcsname{\color[rgb]{1,0,0}}%
      \expandafter\def\csname LT1\endcsname{\color[rgb]{0,1,0}}%
      \expandafter\def\csname LT2\endcsname{\color[rgb]{0,0,1}}%
      \expandafter\def\csname LT3\endcsname{\color[rgb]{1,0,1}}%
      \expandafter\def\csname LT4\endcsname{\color[rgb]{0,1,1}}%
      \expandafter\def\csname LT5\endcsname{\color[rgb]{1,1,0}}%
      \expandafter\def\csname LT6\endcsname{\color[rgb]{0,0,0}}%
      \expandafter\def\csname LT7\endcsname{\color[rgb]{1,0.3,0}}%
      \expandafter\def\csname LT8\endcsname{\color[rgb]{0.5,0.5,0.5}}%
    \else
      \def\colorrgb#1{\color{black}}%
      \def\colorgray#1{\color[gray]{#1}}%
      \expandafter\def\csname LTw\endcsname{\color{white}}%
      \expandafter\def\csname LTb\endcsname{\color{black}}%
      \expandafter\def\csname LTa\endcsname{\color{black}}%
      \expandafter\def\csname LT0\endcsname{\color{black}}%
      \expandafter\def\csname LT1\endcsname{\color{black}}%
      \expandafter\def\csname LT2\endcsname{\color{black}}%
      \expandafter\def\csname LT3\endcsname{\color{black}}%
      \expandafter\def\csname LT4\endcsname{\color{black}}%
      \expandafter\def\csname LT5\endcsname{\color{black}}%
      \expandafter\def\csname LT6\endcsname{\color{black}}%
      \expandafter\def\csname LT7\endcsname{\color{black}}%
      \expandafter\def\csname LT8\endcsname{\color{black}}%
    \fi
  \fi
  \setlength{\unitlength}{0.0500bp}%
  \begin{picture}(4080.00,3628.00)%
    \gplgaddtomacro\gplbacktext{%
      \csname LTb\endcsname%
      \put(804,539){\makebox(0,0)[r]{\strut{}$-0.5$}}%
      \put(804,1314){\makebox(0,0)[r]{\strut{}$0.0$}}%
      \put(804,2089){\makebox(0,0)[r]{\strut{}$0.5$}}%
      \put(804,2864){\makebox(0,0)[r]{\strut{}$1.0$}}%
      \put(2086,264){\makebox(0,0){\strut{}$0$}}%
      \put(3598,264){\makebox(0,0){\strut{}$5$}}%
      \put(240,1934){\rotatebox{90}{\makebox(0,0){\strut{}$C_f \, \times \, 10^3$}}}%
      \put(2388,84){\makebox(0,0){\strut{}$x/{\delta}_0$}}%
      \put(2842,1314){\makebox(0,0)[l]{\strut{}${\alpha} = 20^{\circ}$}}%
    }%
    \gplgaddtomacro\gplfronttext{%
    }%
    \gplbacktext
    \put(0,0){\includegraphics{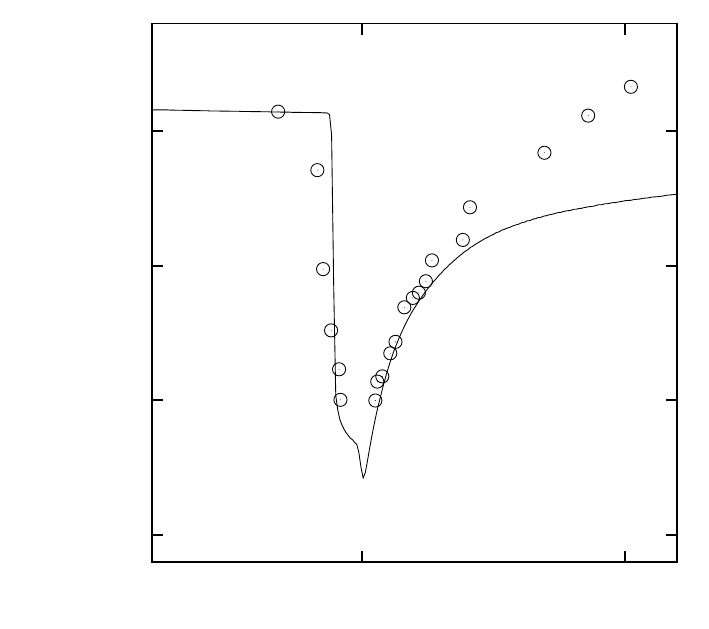}}%
    \gplfronttext
  \end{picture}%
\endgroup

%% file: cframp24gridconv_.tex
\begingroup
  \fontfamily{Helvetica}%
  \selectfont
  \makeatletter
  \providecommand\color[2][]{%
    \GenericError{(gnuplot) \space\space\space\@spaces}{%
      Package color not loaded in conjunction with
      terminal option `colourtext'%
    }{See the gnuplot documentation for explanation.%
    }{Either use 'blacktext' in gnuplot or load the package
      color.sty in LaTeX.}%
    \renewcommand\color[2][]{}%
  }%
  \providecommand\includegraphics[2][]{%
    \GenericError{(gnuplot) \space\space\space\@spaces}{%
      Package graphicx or graphics not loaded%
    }{See the gnuplot documentation for explanation.%
    }{The gnuplot epslatex terminal needs graphicx.sty or graphics.sty.}%
    \renewcommand\includegraphics[2][]{}%
  }%
  \providecommand\rotatebox[2]{#2}%
  \@ifundefined{ifGPcolor}{%
    \newif\ifGPcolor
    \GPcolorfalse
  }{}%
  \@ifundefined{ifGPblacktext}{%
    \newif\ifGPblacktext
    \GPblacktexttrue
  }{}%
  \let\gplgaddtomacro\g@addto@macro
  \gdef\gplbacktext{}%
  \gdef\gplfronttext{}%
  \makeatother
  \ifGPblacktext
    \def\colorrgb#1{}%
    \def\colorgray#1{}%
  \else
    \ifGPcolor
      \def\colorrgb#1{\color[rgb]{#1}}%
      \def\colorgray#1{\color[gray]{#1}}%
      \expandafter\def\csname LTw\endcsname{\color{white}}%
      \expandafter\def\csname LTb\endcsname{\color{black}}%
      \expandafter\def\csname LTa\endcsname{\color{black}}%
      \expandafter\def\csname LT0\endcsname{\color[rgb]{1,0,0}}%
      \expandafter\def\csname LT1\endcsname{\color[rgb]{0,1,0}}%
      \expandafter\def\csname LT2\endcsname{\color[rgb]{0,0,1}}%
      \expandafter\def\csname LT3\endcsname{\color[rgb]{1,0,1}}%
      \expandafter\def\csname LT4\endcsname{\color[rgb]{0,1,1}}%
      \expandafter\def\csname LT5\endcsname{\color[rgb]{1,1,0}}%
      \expandafter\def\csname LT6\endcsname{\color[rgb]{0,0,0}}%
      \expandafter\def\csname LT7\endcsname{\color[rgb]{1,0.3,0}}%
      \expandafter\def\csname LT8\endcsname{\color[rgb]{0.5,0.5,0.5}}%
    \else
      \def\colorrgb#1{\color{black}}%
      \def\colorgray#1{\color[gray]{#1}}%
      \expandafter\def\csname LTw\endcsname{\color{white}}%
      \expandafter\def\csname LTb\endcsname{\color{black}}%
      \expandafter\def\csname LTa\endcsname{\color{black}}%
      \expandafter\def\csname LT0\endcsname{\color{black}}%
      \expandafter\def\csname LT1\endcsname{\color{black}}%
      \expandafter\def\csname LT2\endcsname{\color{black}}%
      \expandafter\def\csname LT3\endcsname{\color{black}}%
      \expandafter\def\csname LT4\endcsname{\color{black}}%
      \expandafter\def\csname LT5\endcsname{\color{black}}%
      \expandafter\def\csname LT6\endcsname{\color{black}}%
      \expandafter\def\csname LT7\endcsname{\color{black}}%
      \expandafter\def\csname LT8\endcsname{\color{black}}%
    \fi
  \fi
  \setlength{\unitlength}{0.0500bp}%
  \begin{picture}(4080.00,3628.00)%
    \gplgaddtomacro\gplbacktext{%
      \csname LTb\endcsname%
      \put(804,539){\makebox(0,0)[r]{\strut{}$-0.5$}}%
      \put(804,1314){\makebox(0,0)[r]{\strut{}$0.0$}}%
      \put(804,2089){\makebox(0,0)[r]{\strut{}$0.5$}}%
      \put(804,2864){\makebox(0,0)[r]{\strut{}$1.0$}}%
      \put(2086,264){\makebox(0,0){\strut{}$0$}}%
      \put(3598,264){\makebox(0,0){\strut{}$5$}}%
      \put(240,1934){\rotatebox{90}{\makebox(0,0){\strut{}$C_f \, \times \, 10^3$}}}%
      \put(2388,84){\makebox(0,0){\strut{}$x/{\delta}_0$}}%
      \put(2842,1314){\makebox(0,0)[l]{\strut{}${\alpha} = 24^{\circ}$}}%
    }%
    \gplgaddtomacro\gplfronttext{%
    }%
    \gplbacktext
    \put(0,0){\includegraphics{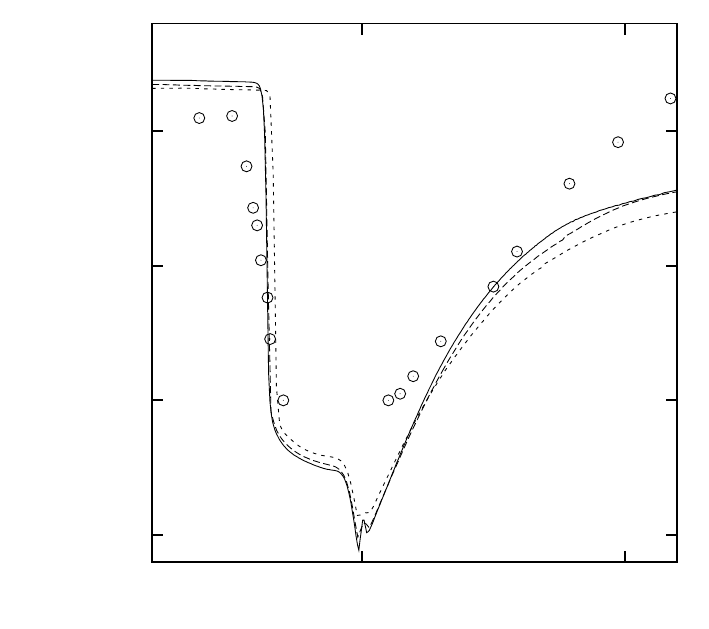}}%
    \gplfronttext
  \end{picture}%
\endgroup

%% file: righigks_rev1_arxiv.bbl
\begin{thebibliography}{35}%
\makeatletter
\providecommand \@ifxundefined [1]{%
 \@ifx{#1\undefined}
}%
\providecommand \@ifnum [1]{%
 \ifnum #1\expandafter \@firstoftwo
 \else \expandafter \@secondoftwo
 \fi
}%
\providecommand \@ifx [1]{%
 \ifx #1\expandafter \@firstoftwo
 \else \expandafter \@secondoftwo
 \fi
}%
\providecommand \natexlab [1]{#1}%
\providecommand \enquote  [1]{``#1''}%
\providecommand \bibnamefont  [1]{#1}%
\providecommand \bibfnamefont [1]{#1}%
\providecommand \citenamefont [1]{#1}%
\providecommand \href@noop [0]{\@secondoftwo}%
\providecommand \href [0]{\begingroup \@sanitize@url \@href}%
\providecommand \@href[1]{\@@startlink{#1}\@@href}%
\providecommand \@@href[1]{\endgroup#1\@@endlink}%
\providecommand \@sanitize@url [0]{\catcode `\\12\catcode `\$12\catcode
  `\&12\catcode `\#12\catcode `\^12\catcode `\_12\catcode `\%12\relax}%
\providecommand \@@startlink[1]{}%
\providecommand \@@endlink[0]{}%
\providecommand \url  [0]{\begingroup\@sanitize@url \@url }%
\providecommand \@url [1]{\endgroup\@href {#1}{\urlprefix }}%
\providecommand \urlprefix  [0]{URL }%
\providecommand \Eprint [0]{\href }%
\providecommand \doibase [0]{http://dx.doi.org/}%
\providecommand \selectlanguage [0]{\@gobble}%
\providecommand \bibinfo  [0]{\@secondoftwo}%
\providecommand \bibfield  [0]{\@secondoftwo}%
\providecommand \translation [1]{[#1]}%
\providecommand \BibitemOpen [0]{}%
\providecommand \bibitemStop [0]{}%
\providecommand \bibitemNoStop [0]{.\EOS\space}%
\providecommand \EOS [0]{\spacefactor3000\relax}%
\providecommand \BibitemShut  [1]{\csname bibitem#1\endcsname}%
\let\auto@bib@innerbib\@empty
\bibitem [{\citenamefont {Chen}\ \emph {et~al.}(2003)\citenamefont {Chen},
  \citenamefont {Kandasamy}, \citenamefont {Orszag}, \citenamefont {Shock},
  \citenamefont {Succi},\ and\ \citenamefont {Yakhot}}]{chen2003extended}%
  \BibitemOpen
  \bibfield  {author} {\bibinfo {author} {\bibfnamefont {H.}~\bibnamefont
  {Chen}}, \bibinfo {author} {\bibfnamefont {S.}~\bibnamefont {Kandasamy}},
  \bibinfo {author} {\bibfnamefont {S.}~\bibnamefont {Orszag}}, \bibinfo
  {author} {\bibfnamefont {R.}~\bibnamefont {Shock}}, \bibinfo {author}
  {\bibfnamefont {S.}~\bibnamefont {Succi}}, \ and\ \bibinfo {author}
  {\bibfnamefont {V.}~\bibnamefont {Yakhot}},\ }\bibfield  {title} {\enquote
  {\bibinfo {title} {{Extended Boltzmann kinetic equation for turbulent
  flows}},}\ }\href@noop {} {\bibfield  {journal} {\bibinfo  {journal}
  {Science}\ }\textbf {\bibinfo {volume} {301}},\ \bibinfo {pages} {633--636}
  (\bibinfo {year} {2003})}\BibitemShut {NoStop}%
\bibitem [{\citenamefont {Chen}\ \emph {et~al.}(2004)\citenamefont {Chen},
  \citenamefont {Orszag}, \citenamefont {Staroselsky},\ and\ \citenamefont
  {Succi}}]{chen2004expanded}%
  \BibitemOpen
  \bibfield  {author} {\bibinfo {author} {\bibfnamefont {H.}~\bibnamefont
  {Chen}}, \bibinfo {author} {\bibfnamefont {S.}~\bibnamefont {Orszag}},
  \bibinfo {author} {\bibfnamefont {I.}~\bibnamefont {Staroselsky}}, \ and\
  \bibinfo {author} {\bibfnamefont {S.}~\bibnamefont {Succi}},\ }\bibfield
  {title} {\enquote {\bibinfo {title} {{Expanded analogy between Boltzmann
  kinetic theory of fluids and turbulence}},}\ }\href@noop {} {\bibfield
  {journal} {\bibinfo  {journal} {J. Fluid Mech.}\ }\textbf {\bibinfo {volume}
  {519}},\ \bibinfo {pages} {301--314} (\bibinfo {year} {2004})}\BibitemShut
  {NoStop}%
\bibitem [{\citenamefont {Succi}\ \emph {et~al.}(2002)\citenamefont {Succi},
  \citenamefont {Filippova}, \citenamefont {Chen},\ and\ \citenamefont
  {Orszag}}]{succi2002towards}%
  \BibitemOpen
  \bibfield  {author} {\bibinfo {author} {\bibfnamefont {S.}~\bibnamefont
  {Succi}}, \bibinfo {author} {\bibfnamefont {O.}~\bibnamefont {Filippova}},
  \bibinfo {author} {\bibfnamefont {H.}~\bibnamefont {Chen}}, \ and\ \bibinfo
  {author} {\bibfnamefont {S.}~\bibnamefont {Orszag}},\ }\bibfield  {title}
  {\enquote {\bibinfo {title} {Towards a renormalized lattice boltzmann
  equation for fluid turbulence},}\ }\href@noop {} {\bibfield  {journal}
  {\bibinfo  {journal} {J. Stat. Phyis.}\ }\textbf {\bibinfo {volume} {107}},\
  \bibinfo {pages} {261--278} (\bibinfo {year} {2002})}\BibitemShut {NoStop}%
\bibitem [{\citenamefont {Xu}(2001)}]{xu2001gas}%
  \BibitemOpen
  \bibfield  {author} {\bibinfo {author} {\bibfnamefont {K.}~\bibnamefont
  {Xu}},\ }\bibfield  {title} {\enquote {\bibinfo {title} {{A gas-kinetic BGK
  scheme for the Navier--Stokes equations and its connection with artificial
  dissipation and Godunov method}},}\ }\href@noop {} {\bibfield  {journal}
  {\bibinfo  {journal} {J. Comput. Phys.}\ }\textbf {\bibinfo {volume} {171}},\
  \bibinfo {pages} {289--335} (\bibinfo {year} {2001})}\BibitemShut {NoStop}%
\bibitem [{\citenamefont {May}, \citenamefont {Srinivasan},\ and\ \citenamefont
  {Jameson}(2007)}]{may2007improved}%
  \BibitemOpen
  \bibfield  {author} {\bibinfo {author} {\bibfnamefont {G.}~\bibnamefont
  {May}}, \bibinfo {author} {\bibfnamefont {B.}~\bibnamefont {Srinivasan}}, \
  and\ \bibinfo {author} {\bibfnamefont {A.}~\bibnamefont {Jameson}},\
  }\bibfield  {title} {\enquote {\bibinfo {title} {{An improved gas-kinetic BGK
  finite-volume method for three-dimensional transonic flow}},}\ }\href@noop {}
  {\bibfield  {journal} {\bibinfo  {journal} {J. Comput. Phys.}\ }\textbf
  {\bibinfo {volume} {220}},\ \bibinfo {pages} {856--878} (\bibinfo {year}
  {2007})}\BibitemShut {NoStop}%
\bibitem [{\citenamefont {Mandal}\ and\ \citenamefont
  {Deshpande}(1994)}]{mandal1994kinetic}%
  \BibitemOpen
  \bibfield  {author} {\bibinfo {author} {\bibfnamefont {J.}~\bibnamefont
  {Mandal}}\ and\ \bibinfo {author} {\bibfnamefont {S.}~\bibnamefont
  {Deshpande}},\ }\bibfield  {title} {\enquote {\bibinfo {title} {{Kinetic flux
  vector splitting for Euler equations}},}\ }\href@noop {} {\bibfield
  {journal} {\bibinfo  {journal} {Comput. Fluids}\ }\textbf {\bibinfo {volume}
  {23}},\ \bibinfo {pages} {447--478} (\bibinfo {year} {1994})}\BibitemShut
  {NoStop}%
\bibitem [{\citenamefont {Chou}\ and\ \citenamefont
  {Baganoff}(1997)}]{chou1997kinetic}%
  \BibitemOpen
  \bibfield  {author} {\bibinfo {author} {\bibfnamefont {S.}~\bibnamefont
  {Chou}}\ and\ \bibinfo {author} {\bibfnamefont {D.}~\bibnamefont
  {Baganoff}},\ }\bibfield  {title} {\enquote {\bibinfo {title} {{Kinetic
  flux--vector splitting for the Navier--Stokes equations}},}\ }\href@noop {}
  {\bibfield  {journal} {\bibinfo  {journal} {J. Comput. Phys.}\ }\textbf
  {\bibinfo {volume} {130}},\ \bibinfo {pages} {217--230} (\bibinfo {year}
  {1997})}\BibitemShut {NoStop}%
\bibitem [{\citenamefont {Xu}\ and\ \citenamefont
  {Prendergast}(1994)}]{xu1994numerical}%
  \BibitemOpen
  \bibfield  {author} {\bibinfo {author} {\bibfnamefont {K.}~\bibnamefont
  {Xu}}\ and\ \bibinfo {author} {\bibfnamefont {K.}~\bibnamefont
  {Prendergast}},\ }\bibfield  {title} {\enquote {\bibinfo {title} {{Numerical
  Navier-Stokes solutions from gas kinetic theory}},}\ }\href@noop {}
  {\bibfield  {journal} {\bibinfo  {journal} {J. Comput. Phys.}\ }\textbf
  {\bibinfo {volume} {114}},\ \bibinfo {pages} {9--17} (\bibinfo {year}
  {1994})}\BibitemShut {NoStop}%
\bibitem [{\citenamefont {Xu}, \citenamefont {Mao},\ and\ \citenamefont
  {Tang}(2005)}]{xu2005multidimensional}%
  \BibitemOpen
  \bibfield  {author} {\bibinfo {author} {\bibfnamefont {K.}~\bibnamefont
  {Xu}}, \bibinfo {author} {\bibfnamefont {M.}~\bibnamefont {Mao}}, \ and\
  \bibinfo {author} {\bibfnamefont {L.}~\bibnamefont {Tang}},\ }\bibfield
  {title} {\enquote {\bibinfo {title} {{A multidimensional gas-kinetic BGK
  scheme for hypersonic viscous flow}},}\ }\href@noop {} {\bibfield  {journal}
  {\bibinfo  {journal} {J. Comput. Phys.}\ }\textbf {\bibinfo {volume} {203}},\
  \bibinfo {pages} {405--421} (\bibinfo {year} {2005})}\BibitemShut {NoStop}%
\bibitem [{\citenamefont {Li}, \citenamefont {Xu},\ and\ \citenamefont
  {Fu}(2010)}]{li2010high}%
  \BibitemOpen
  \bibfield  {author} {\bibinfo {author} {\bibfnamefont {Q.}~\bibnamefont
  {Li}}, \bibinfo {author} {\bibfnamefont {K.}~\bibnamefont {Xu}}, \ and\
  \bibinfo {author} {\bibfnamefont {S.}~\bibnamefont {Fu}},\ }\bibfield
  {title} {\enquote {\bibinfo {title} {{A high-order gas-kinetic Navier--Stokes
  flow solver}},}\ }\href@noop {} {\bibfield  {journal} {\bibinfo  {journal}
  {Journal of Computational Physics}\ }\textbf {\bibinfo {volume} {229}},\
  \bibinfo {pages} {6715--6731} (\bibinfo {year} {2010})}\BibitemShut {NoStop}%
\bibitem [{\citenamefont {Righi}(2013)}]{righi2012aeronautical}%
  \BibitemOpen
  \bibfield  {author} {\bibinfo {author} {\bibfnamefont {M.}~\bibnamefont
  {Righi}},\ }\bibfield  {title} {\enquote {\bibinfo {title} {{A finite-volume
  gas-kinetic method for the solution of the Navier-Stokes equations}},}\
  }\href@noop {} {\bibfield  {journal} {\bibinfo  {journal} {Royal Aeronautical
  Society, Aeronaut. J.}\ }\textbf {\bibinfo {volume} {(117)}} (\bibinfo {year}
  {2013})}\BibitemShut {NoStop}%
\bibitem [{\citenamefont {Liao}, \citenamefont {Luo},\ and\ \citenamefont
  {Xu}(2007)}]{liao2007gas}%
  \BibitemOpen
  \bibfield  {author} {\bibinfo {author} {\bibfnamefont {W.}~\bibnamefont
  {Liao}}, \bibinfo {author} {\bibfnamefont {L.}~\bibnamefont {Luo}}, \ and\
  \bibinfo {author} {\bibfnamefont {K.}~\bibnamefont {Xu}},\ }\bibfield
  {title} {\enquote {\bibinfo {title} {Gas-kinetic scheme for continuum and
  near-continuum hypersonic flows},}\ }\href@noop {} {\bibfield  {journal}
  {\bibinfo  {journal} {J. Spacecraft Rockets}\ }\textbf {\bibinfo {volume}
  {44}},\ \bibinfo {pages} {1232--1240} (\bibinfo {year} {2007})}\BibitemShut
  {NoStop}%
\bibitem [{\citenamefont {Xu}\ and\ \citenamefont
  {Huang}(2010)}]{xu2010unified}%
  \BibitemOpen
  \bibfield  {author} {\bibinfo {author} {\bibfnamefont {K.}~\bibnamefont
  {Xu}}\ and\ \bibinfo {author} {\bibfnamefont {J.}~\bibnamefont {Huang}},\
  }\bibfield  {title} {\enquote {\bibinfo {title} {A unified gas-kinetic scheme
  for continuum and rarefied flows},}\ }\href@noop {} {\bibfield  {journal}
  {\bibinfo  {journal} {J. Comput. Phys.}\ }\textbf {\bibinfo {volume} {229}},\
  \bibinfo {pages} {7747--7764} (\bibinfo {year} {2010})}\BibitemShut {NoStop}%
\bibitem [{\citenamefont {Bhatnagar}, \citenamefont {Gross},\ and\
  \citenamefont {Krook}(1954)}]{bhatnagar1954model}%
  \BibitemOpen
  \bibfield  {author} {\bibinfo {author} {\bibfnamefont {P.}~\bibnamefont
  {Bhatnagar}}, \bibinfo {author} {\bibfnamefont {E.}~\bibnamefont {Gross}}, \
  and\ \bibinfo {author} {\bibfnamefont {M.}~\bibnamefont {Krook}},\ }\bibfield
   {title} {\enquote {\bibinfo {title} {{A model for collision processes in
  gases. I. Small amplitude processes in charged and neutral one-component
  systems}},}\ }\href@noop {} {\bibfield  {journal} {\bibinfo  {journal} {Phys.
  Rev.}\ }\textbf {\bibinfo {volume} {94}},\ \bibinfo {pages} {511 -- 525}
  (\bibinfo {year} {1954})}\BibitemShut {NoStop}%
\bibitem [{\citenamefont {Cercignani}(1988)}]{cercignani1988boltzmann}%
  \BibitemOpen
  \bibfield  {author} {\bibinfo {author} {\bibfnamefont {C.}~\bibnamefont
  {Cercignani}},\ }\href@noop {} {\emph {\bibinfo {title} {{The Boltzmann
  equation and its applications}}}}\ (\bibinfo  {publisher} {Springer, New
  York},\ \bibinfo {year} {1988})\BibitemShut {NoStop}%
\bibitem [{\citenamefont {Xu}(1998)}]{xu1998gas}%
  \BibitemOpen
  \bibfield  {author} {\bibinfo {author} {\bibfnamefont {K.}~\bibnamefont
  {Xu}},\ }\bibfield  {title} {\enquote {\bibinfo {title} {{Gas-kinetic schemes
  for unsteady compressible flow simulations}},}\ }\href@noop {} {\bibfield
  {journal} {\bibinfo  {journal} {Von Karman Institute, Computational Fluid
  Dynamics, Annual Lecture Series, 29th, Rhode-Saint-Genese, Belgium}\ }
  (\bibinfo {year} {1998})}\BibitemShut {NoStop}%
\bibitem [{\citenamefont {Kogan}(1969)}]{kogan1969rarefied}%
  \BibitemOpen
  \bibfield  {author} {\bibinfo {author} {\bibfnamefont {M.~N.}\ \bibnamefont
  {Kogan}},\ }\href@noop {} {\emph {\bibinfo {title} {Rarefied gas dynamics}}}\
  (\bibinfo  {publisher} {Plenum Press, New York},\ \bibinfo {year}
  {1969})\BibitemShut {NoStop}%
\bibitem [{\citenamefont {Smagorinsky}(1963)}]{smagorinsky1963general}%
  \BibitemOpen
  \bibfield  {author} {\bibinfo {author} {\bibfnamefont {J.}~\bibnamefont
  {Smagorinsky}},\ }\bibfield  {title} {\enquote {\bibinfo {title} {General
  circulation experiments with the primitive equations},}\ }\href@noop {}
  {\bibfield  {journal} {\bibinfo  {journal} {Mon. Weather Rev.}\ }\textbf
  {\bibinfo {volume} {91}},\ \bibinfo {pages} {99--164} (\bibinfo {year}
  {1963})}\BibitemShut {NoStop}%
\bibitem [{\citenamefont {Germano}(1992)}]{germano1992turbulence}%
  \BibitemOpen
  \bibfield  {author} {\bibinfo {author} {\bibfnamefont {M.}~\bibnamefont
  {Germano}},\ }\bibfield  {title} {\enquote {\bibinfo {title} {{Turbulence:
  the filtering approach}},}\ }\href@noop {} {\bibfield  {journal} {\bibinfo
  {journal} {J. Fluid Mech.}\ }\textbf {\bibinfo {volume} {238}},\ \bibinfo
  {pages} {325--336} (\bibinfo {year} {1992})}\BibitemShut {NoStop}%
\bibitem [{\citenamefont {Ohwada}\ and\ \citenamefont
  {Xu}(2004)}]{ohwada2004kinetic}%
  \BibitemOpen
  \bibfield  {author} {\bibinfo {author} {\bibfnamefont {T.}~\bibnamefont
  {Ohwada}}\ and\ \bibinfo {author} {\bibfnamefont {K.}~\bibnamefont {Xu}},\
  }\bibfield  {title} {\enquote {\bibinfo {title} {{The kinetic scheme for the
  full-Burnett equations}},}\ }\href@noop {} {\bibfield  {journal} {\bibinfo
  {journal} {J. Comput. Phys.}\ }\textbf {\bibinfo {volume} {201}},\ \bibinfo
  {pages} {315--332} (\bibinfo {year} {2004})}\BibitemShut {NoStop}%
\bibitem [{\citenamefont {Righi}(2012)}]{righi2012rgd}%
  \BibitemOpen
  \bibfield  {author} {\bibinfo {author} {\bibfnamefont {M.}~\bibnamefont
  {Righi}},\ }\bibfield  {title} {\enquote {\bibinfo {title} {{A Gas-Kinetic
  Scheme For The Simulation Of Turbulent Flows}},}\ }in\ \href@noop {} {\emph
  {\bibinfo {booktitle} {Proceeding of the 28th Internaltional Symposium on
  Rarefied Gas Dynamics, Zaragoza}}},\ \bibinfo {editor} {edited by\ \bibinfo
  {editor} {\bibfnamefont {M.}~\bibnamefont {Mareschal}}\ and\ \bibinfo
  {editor} {\bibfnamefont {A.}~\bibnamefont {Santos}}}\ (\bibinfo  {publisher}
  {American Institute of Physics},\ \bibinfo {year} {2012})\ pp.\ \bibinfo
  {pages} {481--488}\BibitemShut {NoStop}%
\bibitem [{\citenamefont {Yoon}\ and\ \citenamefont
  {Jameson}(1988)}]{yoon1988lower}%
  \BibitemOpen
  \bibfield  {author} {\bibinfo {author} {\bibfnamefont {S.}~\bibnamefont
  {Yoon}}\ and\ \bibinfo {author} {\bibfnamefont {A.}~\bibnamefont {Jameson}},\
  }\bibfield  {title} {\enquote {\bibinfo {title} {{Lower-upper
  symmetric-Gauss-Seidel method for the Euler and Navier-Stokes equations}},}\
  }\href@noop {} {\bibfield  {journal} {\bibinfo  {journal} {AIAA J.}\ }\textbf
  {\bibinfo {volume} {26}},\ \bibinfo {pages} {1025--1026} (\bibinfo {year}
  {1988})}\BibitemShut {NoStop}%
\bibitem [{\citenamefont {Jameson}(1983)}]{jameson1983solution}%
  \BibitemOpen
  \bibfield  {author} {\bibinfo {author} {\bibfnamefont {A.}~\bibnamefont
  {Jameson}},\ }\bibfield  {title} {\enquote {\bibinfo {title} {{Solution of
  the Euler equations for two dimensional transonic flow by a multigrid
  method}},}\ }\href@noop {} {\bibfield  {journal} {\bibinfo  {journal} {Appl.
  Math. Comput.}\ }\textbf {\bibinfo {volume} {13}},\ \bibinfo {pages}
  {327--356} (\bibinfo {year} {1983})}\BibitemShut {NoStop}%
\bibitem [{\citenamefont {Wilcox}(2006)}]{wilcox2006}%
  \BibitemOpen
  \bibfield  {author} {\bibinfo {author} {\bibfnamefont {D.~C.}\ \bibnamefont
  {Wilcox}},\ }\href@noop {} {\emph {\bibinfo {title} {{Turbulence Modeling for
  CFD, 3rd edition}}}}\ (\bibinfo  {publisher} {DCW Industries, Inc., La Canada
  CA},\ \bibinfo {year} {2006})\BibitemShut {NoStop}%
\bibitem [{\citenamefont {Cook}, \citenamefont {McDonald},\ and\ \citenamefont
  {Firman}(1979)}]{cook1979aerofoil}%
  \BibitemOpen
  \bibfield  {author} {\bibinfo {author} {\bibfnamefont {P.}~\bibnamefont
  {Cook}}, \bibinfo {author} {\bibfnamefont {M.}~\bibnamefont {McDonald}}, \
  and\ \bibinfo {author} {\bibfnamefont {M.}~\bibnamefont {Firman}},\
  }\bibfield  {title} {\enquote {\bibinfo {title} {{Aerofoil RAE 2822--pressure
  distributions, and boundary layer andwake measurements. Experimental data
  base for computer program assessment}},}\ }\href@noop {} {\bibfield
  {journal} {\bibinfo  {journal} {AGARD Advisory}\ } (\bibinfo {year}
  {1979})}\BibitemShut {NoStop}%
\bibitem [{\citenamefont {Harris}(1981)}]{harris1981two}%
  \BibitemOpen
  \bibfield  {author} {\bibinfo {author} {\bibfnamefont {C.}~\bibnamefont
  {Harris}},\ }\bibfield  {title} {\enquote {\bibinfo {title} {{Two-dimensional
  aerodynamic characteristics of the NACA 0012 airfoil in the Langley 8 foot
  transonic pressure tunnel}},}\ }\href@noop {} {\bibfield  {journal} {\bibinfo
   {journal} {NASA Technical Memorandum 81-927}\ } (\bibinfo {year}
  {1981})}\BibitemShut {NoStop}%
\bibitem [{\citenamefont {D\'elery}(1983)}]{delery1983experimental}%
  \BibitemOpen
  \bibfield  {author} {\bibinfo {author} {\bibfnamefont {J.}~\bibnamefont
  {D\'elery}},\ }\bibfield  {title} {\enquote {\bibinfo {title} {{Experimental
  investigation of turbulence properties in transonic shock/boundary-layer
  interactions}},}\ }\href@noop {} {\bibfield  {journal} {\bibinfo  {journal}
  {AIAA J.}\ }\textbf {\bibinfo {volume} {21}},\ \bibinfo {pages} {180--185}
  (\bibinfo {year} {1983})}\BibitemShut {NoStop}%
\bibitem [{\citenamefont {Settles}, \citenamefont {Fitzpatrick},\ and\
  \citenamefont {Bogdonoff}(1979)}]{settles1979detailed}%
  \BibitemOpen
  \bibfield  {author} {\bibinfo {author} {\bibfnamefont {G.}~\bibnamefont
  {Settles}}, \bibinfo {author} {\bibfnamefont {T.}~\bibnamefont
  {Fitzpatrick}}, \ and\ \bibinfo {author} {\bibfnamefont {S.}~\bibnamefont
  {Bogdonoff}},\ }\bibfield  {title} {\enquote {\bibinfo {title} {{Detailed
  study of attached and separated compression corner flowfields in high
  Reynolds number supersonic flow}},}\ }\href@noop {} {\bibfield  {journal}
  {\bibinfo  {journal} {AIAA J.}\ }\textbf {\bibinfo {volume} {17}},\ \bibinfo
  {pages} {579--585} (\bibinfo {year} {1979})}\BibitemShut {NoStop}%
\bibitem [{\citenamefont {Wallin}\ and\ \citenamefont
  {Johansson}(2000)}]{wallin2000explicit}%
  \BibitemOpen
  \bibfield  {author} {\bibinfo {author} {\bibfnamefont {S.}~\bibnamefont
  {Wallin}}\ and\ \bibinfo {author} {\bibfnamefont {A.}~\bibnamefont
  {Johansson}},\ }\bibfield  {title} {\enquote {\bibinfo {title} {{An explicit
  algebraic Reynolds stress model for incompressible and compressible turbulent
  flows}},}\ }\href@noop {} {\bibfield  {journal} {\bibinfo  {journal} {J.
  Fluid Mech.}\ }\textbf {\bibinfo {volume} {403}},\ \bibinfo {pages} {89--132}
  (\bibinfo {year} {2000})}\BibitemShut {NoStop}%
\bibitem [{\citenamefont {D{\'e}lery}, \citenamefont {Marvin},\ and\
  \citenamefont {Reshotko}(1986)}]{delery1986shock}%
  \BibitemOpen
  \bibfield  {author} {\bibinfo {author} {\bibfnamefont {J.}~\bibnamefont
  {D{\'e}lery}}, \bibinfo {author} {\bibfnamefont {J.}~\bibnamefont {Marvin}},
  \ and\ \bibinfo {author} {\bibfnamefont {E.}~\bibnamefont {Reshotko}},\
  }\bibfield  {title} {\enquote {\bibinfo {title} {Shock-wave boundary layer
  interactions},}\ }\href@noop {} {\bibfield  {journal} {\bibinfo  {journal}
  {AGARDograph}\ }\textbf {\bibinfo {volume} {280}} (\bibinfo {year}
  {1986})}\BibitemShut {NoStop}%
\bibitem [{\citenamefont {Settles}, \citenamefont {Vas},\ and\ \citenamefont
  {Bogdonoff}(1976)}]{settles1976details}%
  \BibitemOpen
  \bibfield  {author} {\bibinfo {author} {\bibfnamefont {G.}~\bibnamefont
  {Settles}}, \bibinfo {author} {\bibfnamefont {I.}~\bibnamefont {Vas}}, \ and\
  \bibinfo {author} {\bibfnamefont {S.}~\bibnamefont {Bogdonoff}},\ }\bibfield
  {title} {\enquote {\bibinfo {title} {{Details of a shock-separated turbulent
  boundary layer at a compression corner}},}\ }\href@noop {} {\bibfield
  {journal} {\bibinfo  {journal} {AIAA J.}\ }\textbf {\bibinfo {volume} {14}},\
  \bibinfo {pages} {1709--1715} (\bibinfo {year} {1976})}\BibitemShut {NoStop}%
\bibitem [{\citenamefont {Goldberg}, \citenamefont {Peroomian},\ and\
  \citenamefont {Chakravarthy}(1998)}]{goldberg1998application}%
  \BibitemOpen
  \bibfield  {author} {\bibinfo {author} {\bibfnamefont {U.}~\bibnamefont
  {Goldberg}}, \bibinfo {author} {\bibfnamefont {O.}~\bibnamefont {Peroomian}},
  \ and\ \bibinfo {author} {\bibfnamefont {S.}~\bibnamefont {Chakravarthy}},\
  }\bibfield  {title} {\enquote {\bibinfo {title} {{Application of the k-e-R
  Turbulence Model to Wall-Bounded Compressive Flows}},}\ }\href@noop {}
  {\bibfield  {journal} {\bibinfo  {journal} {AIAA Paper No. 98-0323}\ }
  (\bibinfo {year} {1998})}\BibitemShut {NoStop}%
\bibitem [{\citenamefont {Menter}\ and\ \citenamefont
  {Rumsey}(1994)}]{menter1994assessment}%
  \BibitemOpen
  \bibfield  {author} {\bibinfo {author} {\bibfnamefont {F.}~\bibnamefont
  {Menter}}\ and\ \bibinfo {author} {\bibfnamefont {C.}~\bibnamefont
  {Rumsey}},\ }\bibfield  {title} {\enquote {\bibinfo {title} {{Assessment of
  two-equation turbulence models for transonic flows}},}\ }\href@noop {}
  {\bibfield  {journal} {\bibinfo  {journal} {AIAA Paper No. 94-2343}\ }
  (\bibinfo {year} {1994})}\BibitemShut {NoStop}%
\bibitem [{\citenamefont {Xuan}\ and\ \citenamefont {Xu}(2012)}]{xuan2012new}%
  \BibitemOpen
  \bibfield  {author} {\bibinfo {author} {\bibfnamefont {L.-J.}\ \bibnamefont
  {Xuan}}\ and\ \bibinfo {author} {\bibfnamefont {K.}~\bibnamefont {Xu}},\
  }\bibfield  {title} {\enquote {\bibinfo {title} {A new gas-kinetic scheme
  based on analytical solutions of the bgk equation},}\ }\href@noop {}
  {\bibfield  {journal} {\bibinfo  {journal} {J. of Comput. Phys.}\ } (\bibinfo
  {year} {2012})}\BibitemShut {NoStop}%
\bibitem [{\citenamefont {Luo}, \citenamefont {Xuan},\ and\ \citenamefont
  {Xu}(2013)}]{luo2013comparison}%
  \BibitemOpen
  \bibfield  {author} {\bibinfo {author} {\bibfnamefont {J.}~\bibnamefont
  {Luo}}, \bibinfo {author} {\bibfnamefont {L.}~\bibnamefont {Xuan}}, \ and\
  \bibinfo {author} {\bibfnamefont {K.}~\bibnamefont {Xu}},\ }\bibfield
  {title} {\enquote {\bibinfo {title} {{Comparison of Fifth-Order WENO Scheme
  and Finite Volume WENO-Gas-Kinetic Scheme for Inviscid and Viscous Flow
  Simulation}},}\ }\href@noop {} {\bibfield  {journal} {\bibinfo  {journal}
  {Commun. Comput. Phys.}\ } (\bibinfo {year} {2013})}\BibitemShut {NoStop}%
\end{thebibliography}%
